\documentclass[a4paper,fleqn,usenatbib]{mnras}

\usepackage{newtxtext,newtxmath}
\usepackage[T1]{fontenc}
\usepackage{ae,aecompl}

\usepackage{graphicx}	% Including figure files
\usepackage{amsmath}	% Advanced maths commands
\usepackage{csvsimple}
\usepackage{multirow}
\usepackage[normalem]{ulem}
\graphicspath{{Figures/}}
\newcommand{\todo}{\ifmmode \text{\color{red}\Huge{\(\bullet\)}} \else {\color{red}{\Huge$\bullet$}}\fi}
\newcommand{\tido}{\ifmmode {{\color{red}\bullet}} \else {\color{red}$\bullet$}\fi}

\newcommand{\E        }[1]{\ifmmode 10^{#1} \else $10^{#1}$\fi}
\newcommand{\tE        }[1]{\ifmmode \times10^{#1} \else $\times10^{#1}$\fi}
\newcommand{\til}{\ifmmode \sim \else $\sim$\fi}
\newcommand{\pc}	{\ifmmode {\rm pc} \else pc\fi}
\newcommand{\kpc}	{\ifmmode {\rm kpc} \else kpc\fi}
\newcommand{\ld}	{\ifmmode {\rm l.d.} \else l.d.\fi}
\newcommand{\kms}	{\ifmmode {\rm km\,s}^{-1} \else km\,s$^{-1}$\fi}
\newcommand{\cc}	{\ifmmode {\rm cm}^{-3}    \else cm$^{-3}$\fi}
\newcommand{\cmii}	{\ifmmode {\rm cm}^{-2}    \else cm$^{-2}$\fi}
\newcommand{\ergss}	{\ifmmode {\rm erg\,s}^{-1} \else erg s$^{-1}$\fi}
\newcommand{\ergcms}	{\ifmmode {\rm erg\,cm}^{-2}\,{\rm s}^{-1} \else erg\,cm$^{-2}$\,s$^{-1}$\fi}
\newcommand{\ergcmsA}	{\ifmmode {\rm erg\,cm}^{-2}\,{\rm s}^{-1}\,{\rm \AA}^{-1}
\else erg\,cm$^{-2}$\,s$^{-1}$\,\AA$^{-1}$\fi}
\newcommand{  \ergcmsHz  }{\ifmmode{\rm erg\,cm}^{-2}\,{\rm s}^{-1}\,{\rm Hz}^{-1}
                       \else ergs\,cm$^{-2}$\,s$^{-1}$\,Hz$^{-1}$\fi}
\newcommand{\kev}	{\ifmmode {\rm keV} \else keV\fi}

\newcommand{\mic}	{\ifmmode {\rm \mu m} \else $\mu$m\fi}
\newcommand{\vFWHM}	{\ifmmode v_{\mbox{\tiny FWHM}} \else $v_{\mbox{\tiny FWHM}}$\fi}
\newcommand{\vBLR}	{\ifmmode v_{\mbox{\tiny BLR}} \else $v_{\mbox{\tiny BLR}}$\fi}
\newcommand{\sigBLR}	{\ifmmode \sigma_{\mbox{\tiny BLR}} \else $\sigma_{\mbox{\tiny BLR}}$\fi}
\newcommand{\vNLR}	{\ifmmode v_{\mbox{\tiny NLR}} \else $v_{\mbox{\tiny NLR}}$\fi}
\newcommand{\tauBLR}	{\ifmmode \tau_{\mbox{\tiny BLR}} \else $\tau_{\mbox{\tiny BLR}}$\fi}

\newcommand{\Hubble}	{\ifmmode {\rm km\,s}^{-1}\,{\rm Mpc}^{-1} \else km\,s$^{-1}$\,Mpc$^{-1}$\fi}
\newcommand{\NDunit}	{\ifmmode {\rm Mpc}^{-3} \else Mpc$^{-3}$\fi}
\newcommand{\LFunit}	{\ifmmode {\rm Mpc}^{-3}\,{\rm mag}^{-1} \else Mpc$^{-3}$\,mag$^{-1}$\fi}
\newcommand{\MFunit}	{\ifmmode {\rm Mpc}^{-3}\,{\rm dex}^{-1} \else Mpc$^{-3}$\,dex$^{-1}$\fi}

\newcommand{\Msun}{\ifmmode M_{\odot} \else $M_{\odot}$\fi}
\newcommand{\Lsun}{\ifmmode L_{\odot} \else $L_{\odot}$\fi}
\newcommand{\Zsun}{\ifmmode Z_{\odot} \else $Z_{\odot}$\fi}
\newcommand{\mpyr}{\ifmmode \Msun\,{\rm yr}^{-1} \else $\Msun\,{\rm yr}^{-1}$\fi}

\newcommand{\qnote}{\ifmmode q_{0} \else $q_{0}$\fi}
\newcommand{\Hnote}{\ifmmode H_{0} \else $H_{0}$\fi}
\newcommand{\hnote}{\ifmmode h_{0} \else $h_{0}$\fi}
\newcommand{\anote}{\ifmmode a_{0} \else $a_{0}$\fi}
\newcommand{\tnote}{\ifmmode t_{0} \else $t_{0}$\fi}

\def\gsim{\;\rlap{\lower 2.5pt \hbox{$\sim$}}\raise 1.5pt\hbox{$>$}\;}
\def\lsim{\;\rlap{\lower 2.5pt \hbox{$\sim$}}\raise 1.5pt\hbox{$<$}\;}

\newcommand{  \Halpha   }{\ifmmode {\rm H}\alpha \else H$\alpha$\fi}

\newcommand{  \ha       }{\Halpha}
\newcommand{  \Hbeta    }{\ifmmode {\rm H}\beta \else H$\beta$\fi}

\newcommand{  \hb       }{\Hbeta}
\newcommand{  \Hgamma   }{\ifmmode {\rm H}\gamma \else H$\gamma$\fi}
\newcommand{  \Hdelta   }{\ifmmode {\rm H}\delta \else H$\delta$\fi}
\newcommand{  \Lya      }{\ifmmode {\rm Ly}\alpha \else Ly$\alpha$\fi}
\newcommand{  \Lyb      }{\ifmmode {\rm Ly}\beta \else Ly$\beta$\fi}
\newcommand{  \Pa       }{\ifmmode {\rm P}\alpha \else P$\alpha$\fi}
\newcommand{  \Pb       }{\ifmmode {\rm P}\beta \else P$\beta$\fi}
\newcommand{  \Bra      }{\ifmmode {\rm Br}\alpha \else Br$\alpha$\fi}
\newcommand{  \Brg      }{\ifmmode {\rm Br}\gamma \else Br$\gamma$\fi}
\newcommand{  \hii      }{\ifmmode {\rm H}\,\textsc{ii} \else H\,\textsc{ii}\fi}
\newcommand{  \hei      }{\ifmmode {\rm He}\,\textsc{i} \else He\,\textsc{i}\fi}
\newcommand{  \heii     }{\ifmmode {\rm He}\,\textsc{ii} \else He\,\textsc{ii}\fi}
\newcommand{  \HeIIuv   }{\ifmmode {\rm He}\,\textsc{ii}\,\lambda1640 \else He\,\textsc{ii}\,$\lambda1640$\fi}
\newcommand{  \HeIIop   }{\ifmmode {\rm He}\,\textsc{ii}\,\lambda4686 \else He\,\textsc{ii}\,$\lambda4686$\fi}
\newcommand{  \CII	}{\ifmmode \left[{\rm C}\,\textsc{ii}\right]\,\lambda157.74\,\mu{\rm m} \else [C\,{\sc ii}]\ $\lambda157.74\,\mu{\rm m}$\fi}
\newcommand{  \cii	}{\ifmmode \left[{\rm C}\,\textsc{ii}\right] \else [C\,{\sc ii}]\fi}

\newcommand{  \ciii     }{\ifmmode {\rm C}\,\textsc{iii}\right] \else C\,\textsc{iii}]\fi}
\newcommand{  \CIII     }{\ifmmode {\rm C}\,\textsc{iii}\right]\,\lambda1909 \else C\,\textsc{iii}]\,$\lambda1909$\fi}
\newcommand{  \civ      }{\ifmmode {\rm C}\,\textsc{iv}  \else C\,\textsc{iv}\fi}
\newcommand{  \CIV      }{\ifmmode {\rm C}\,\textsc{iv}\,\lambda1549 \else C\,\textsc{iv}\,$\lambda1549$\fi}
\newcommand{  \NIIopt   }{\ifmmode \left[{\rm N}\,\textsc{ii}\right]\,\lambda6584 \else [N\,\textsc{ii}]\,$\lambda6584$\fi}
\newcommand{  \nii      }{\ifmmode \left[{\rm N}\,\textsc{ii}\right]  \else [N\,\textsc{ii}]\fi}
\newcommand{  \niii     }{\ifmmode {\rm N}\,\textsc{iii} \else N\,\textsc{iii}\fi}
\newcommand{  \NIII     }{\ifmmode {\rm N}\,\textsc{iii}\,\lambda4640 \else N\,\textsc{iii}\,$\lambda4640$\fi}
\newcommand{  \niv      }{\ifmmode {\rm N}\,\textsc{iv}  \else N\,\textsc{iv}\fi}
\newcommand{  \NIVuv    }{\ifmmode {\rm N}\,\textsc{iv}\,\lambda1486 \else N\,\textsc{iv}\,$\lambda1486$\fi}
\newcommand{  \nv       }{\ifmmode {\rm N}\,\textsc{v}   \else N\,\textsc{v}\fi}
\newcommand{\oi}{\ifmmode \left[{\rm O}\,\textsc{i}\right] \else [O\,{\sc i}]\fi}
\newcommand{\OI}{\ifmmode \left[{\rm O}\,\textsc{i}\right]\,\lambda6300 \else [O\,{\sc i}]$\,\lambda6300$\fi}
\newcommand{\oii}{\ifmmode \left[{\rm O}\,\textsc{ii}\right] \else [O\,{\sc ii}]\fi}
\newcommand{\OII}{\ifmmode \left[{\rm O}\,\textsc{ii}\right]\,\lambda3727 \else [O\,{\sc ii}]\,$\lambda3727$\fi}
\newcommand{\oiii}{\ifmmode \left[{\rm O}\,\textsc{iii}\right] \else [O\,{\sc iii}]\fi}
\newcommand{\OIII}{\ifmmode \left[{\rm O}\,\textsc{iii}\right]\,\lambda5007 \else [O\,{\sc iii}]\,$\lambda5007$\fi}
\newcommand{  \OIIIbf   }{\ifmmode {\rm O}\,\textsc{iii}\,\lambda3133 \else O\,\textsc{iii}\,$\lambda3133$\fi}
\newcommand{  \OIIIuv   }{\ifmmode {\rm O}\,\textsc{iii}\,\lambda1663 \else O\,\textsc{iii}\,$\lambda1663$\fi}
\newcommand{  \oiv      }{\ifmmode {\rm O}\,\textsc{iv}  \else O\,\textsc{iv}\fi}
\newcommand{  \OIVuv    }{\ifmmode {\rm O}\,\textsc{iv}\,\lambda1402  \else O\,\textsc{iv}\,$\lambda1402$\fi}
\newcommand{  \OIVIR    }{\ifmmode {\rm O}\,\textsc{iv}\,25.9\,\mu {\rm m} \else O\,\textsc{iv}\,$25.9\,\mu$m\fi}
\newcommand{  \ovi      }{\ifmmode {\rm O}\,\textsc{vi}   \else O\,\textsc{vi}\fi}
\newcommand{  \Ovi      }{\ifmmode {\rm O}\,\textsc{vi}\,\lambda1035 \else O\,\textsc{vi}\,$\lambda1035$\fi}
\newcommand{  \nei      }{\ifmmode {\rm Ne}\,\textsc{i}   \else Ne\,\textsc{i}\fi}
\newcommand{  \neii     }{\ifmmode {\rm Ne}\,\textsc{ii}  \else Ne\,\textsc{ii}\fi}
\newcommand{  \NeiiIR   }{\ifmmode {\rm Ne}\,\textsc{ii}\,12.8\,\mu {\rm m} \else Ne\,\textsc{ii}\,$12.8\,\mu$m\fi}
\newcommand{  \neiii    }{\ifmmode {\rm Ne}\,\textsc{iii} \else Ne\,\textsc{iii}\fi}
\newcommand{  \neiv     }{\ifmmode {\rm Ne}\,\textsc{iv}  \else Ne\,\textsc{iv}\fi}
\newcommand{  \nev      }{\ifmmode {\rm Ne}\,\textsc{v}   \else Ne\,\textsc{v}\fi}
\newcommand{  \NevIR    }{\ifmmode {\rm Ne}\,\textsc{v}\,24.3\,\mu {\rm m} \else Ne\,\textsc{v}\,$24.3\,\mu$m\fi}
\newcommand{  \nevi     }{\ifmmode {\rm Ne}\,\textsc{vi}  \else Ne\,\textsc{vi}\fi}
\newcommand{  \mgi      }{\ifmmode {\rm Mg}\,\textsc{i} \else Mg\,\textsc{i}\fi}
\newcommand{  \mgii     }{\ifmmode {\rm Mg}\,\textsc{ii} \else Mg\,\textsc{ii}\fi}
\newcommand{  \MgII     }{\ifmmode {\rm Mg}\,\textsc{ii}\,\lambda2798 \else Mg\,\textsc{ii}\,$\lambda2798$\fi}
\newcommand{  \sii      }{\ifmmode {\rm S}\,\textsc{ii} \else S\,\textsc{ii}\fi}
\newcommand{  \siii     }{\ifmmode {\rm S}\,\textsc{iii} \else S\,\textsc{iii}\fi}
\newcommand{  \siv      }{\ifmmode {\rm S}\,\textsc{iv} \else S\,\textsc{iv}\fi}
\newcommand{  \sili     }{\ifmmode {\rm Si}\,\textsc{i}   \else Si\,\textsc{i}\fi}
\newcommand{  \silii    }{\ifmmode {\rm Si}\,\textsc{ii}  \else Si\,\textsc{ii}\fi}
\newcommand{  \Siliv    }{\ifmmode {\rm Si}\,\textsc{iv}  \else Si\,\textsc{iv}\fi}
\newcommand{  \SilIVuv  }{\ifmmode {\rm Si}\,\textsc{iv}\,\lambda1400  \else Si\,\textsc{iv}\,$\lambda1400$\fi}
\newcommand{  \AlIII   }{\ifmmode {\rm Al}\,\textsc{iii}\,\lambda1857 \else Al\,\textsc{iii}\,$\lambda1857$\fi}
\newcommand{  \Aliii   }{\ifmmode {\rm Al}\,\textsc{iii} \else Al\,\textsc{iii}\fi}
\newcommand{  \caii     }{\ifmmode {\rm Ca}\,\textsc{ii} \else Ca\,\textsc{ii}\fi}
\newcommand{  \feii     }{\ifmmode {\rm Fe}\,\textsc{ii} \else Fe\,\textsc{ii}\fi}
\newcommand{  \feiii    }{\ifmmode {\rm Fe}\,\textsc{iii} \else Fe\,\textsc{iii}\fi}
\newcommand{  \Kalpha   }{\ifmmode {\rm K}\alpha \else K$\alpha$\fi}

\newcommand{ \Lhb   }{\ifmmode L_{\hb} \else $L_{\hb}$\fi}
\newcommand{ \Lha   }{\ifmmode L_{\ha} \else $L_{\ha}$\fi}
\newcommand{ \fwhb  }{\ifmmode {\rm FWHM}\left(\hb\right) \else FWHM(\hb)\fi}
\newcommand{\sighb  }{\ifmmode \sigma\left(\hb\right) \else $\sigma\left(\hb\right)$\fi}
\newcommand{ \ewhb  }{\ifmmode {\rm EW}\left(\hb\right) \else EW(\hb)\fi}
\newcommand{ \fwha  }{\ifmmode {\rm FWHM}\left(\ha\right) \else FWHM(\ha)\fi}
\newcommand{ \ewha  }{\ifmmode {\rm EW}\left(\ha\right) \else EW(\ha)\fi}
\newcommand{ \Lmg   }{\ifmmode L\left(\mgii\right) \else $L\left(\mgii\right)$\fi}
\newcommand{ \fwmg  }{\ifmmode {\rm FWHM}\left(\mgii\right) \else FWHM(\mgii)\fi}
\newcommand{ \Lciv  }{\ifmmode L\left(\civ\right) \else $L\left(\civ\right)$\fi}
\newcommand{ \fwciv }{\ifmmode {\rm FWHM}\left(\civ\right) \else FWHM(\civ)\fi}
\newcommand{ \fwhm  }{\ifmmode {\rm FWHM} \else FWHM\fi} 
\newcommand{ \voff  }{\ifmmode v_{\rm off} \else $v_{\rm off}$\fi} 
\newcommand{ \vmax  }{\ifmmode v_{\rm max} \else $v_{\rm max}$\fi} 

\newcommand{ \mumg  }{\ifmmode \mu\left(\mgii\right) \else $\mu\left(\mgii\right)$\fi}
\newcommand{ \fmg   }{\ifmmode f\left(\mgii\right) \else $f\left(\mgii\right)$\fi}
\newcommand{ \muciv }{\ifmmode \mu\left(\civ\right) \else $\mu\left(\civ\right)$\fi}
\newcommand{ \fciv  }{\ifmmode f\left(\civ\right) \else $f\left(\civ\right)$\fi}
\newcommand{  \auvo     }{\ifmmode \alpha_{\nu,{\rm UVO}} \else $\alpha_{\nu,{\rm UVO}}$\fi}
\newcommand{  \Ledd     }{\ifmmode L_{\rm Edd} \else $L_{\rm Edd}$\fi}
\newcommand{  \lamLlam  }{\ifmmode \lambda L_{\lambda} \else $\lambda L_{\lambda}$\fi}
\newcommand{  \lLl      }{\ifmmode \lambda L_{\lambda} \else $\lambda L_{\lambda}$\fi}
\newcommand{  \nuLnu    }{\ifmmode \nu L_{\nu} \else $\nu L_{\nu}$\fi}
\newcommand{  \nLn      }{\ifmmode \nu L_{\nu} \else $\nu L_{\nu}$\fi}
\newcommand{  \Luv      }{\ifmmode L_{1450} \else $L_{1450}$\fi}
\newcommand{  \Lop      }{\ifmmode L_{5100} \else $L_{5100}$\fi}
\newcommand{  \lLop     }{\ifmmode \log\left(\Lop/\ergs\right) \else $\log\left(\Lop/\ergs\right)$\fi}
\newcommand{  \Lthree   }{\ifmmode L_{3000} \else $L_{3000}$\fi}
\newcommand{  \lLthree  }{\ifmmode \log\left(\Lthree/\ergs\right) \else $\log\left(\Lthree/\ergs\right)$\fi}
\newcommand{  \Lsix      }{\ifmmode L_{6200} \else $L_{6200}$\fi}
\newcommand{  \lLisx     }{\ifmmode \log\left(\Lop/\ergs\right) \else $\log\left(\Lop/\ergs\right)$\fi}
\newcommand{  \Lxray    }{\ifmmode L_{\rm X} \else $L_{\rm X}$\fi}

\newcommand{  \Lhard    }{\ifmmode L_{\rm 2-10} \else $L_{\rm 2-10}$\fi}
\newcommand{  \Lsoft    }{\ifmmode L_{\rm 0.5-2} \else $L_{\rm 0.5-2}$\fi}

\newcommand{\Fthree}{\ifmmode F_{3000} \else $F_{3000}$\fi}
\newcommand{\fuv}{\ifmmode f_{\lambda}\left(1450{\rm \AA}\right) \else $f_{\lambda}\left(1450 {\rm \AA}\right)$\fi}
\newcommand{\fthree}{\ifmmode f_{\lambda}\left(3000{\rm \AA}\right) \else $f_{\lambda}\left(3000{\rm \AA}\right)$\fi}
\newcommand{\fH}{\ifmmode f_{\lambda}\left(1.65\micron\right) \else
$f_{\lambda}\left(1.65\micron\right)$\fi}

\newcommand{\fbol}{\ifmmode f_{\rm bol} \else $f_{\rm bol}$\fi}
\newcommand{\fbolwv}{\ifmmode f_{\rm bol}\left(\lambda\right) \else $f_{\rm bol}\left(\lambda\right)$\fi}
\newcommand{\fbolopt}{\ifmmode f_{\rm bol}\left(5100{\rm \AA}\right) \else $f_{\rm bol}\left(5100{\rm \AA}\right)$\fi}
\newcommand{\fbolthree}{\ifmmode f_{\rm bol}\left(3000{\rm \AA}\right) \else $f_{\rm bol}\left(3000{\rm \AA}\right)$\fi}
\newcommand{\fboluv}{\ifmmode f_{\rm bol}\left(1450{\rm \AA}\right) \else $f_{\rm bol}\left(1450{\rm \AA}\right)$\fi}

\newcommand{\fbolbat}{\ifmmode f_{\rm bol}\left(14-150\,\kev\right) \else $f_{\rm bol}\left(14-150\,\kev\right)$\fi}
\newcommand{\fbolhard}{\ifmmode f_{\rm bol}\left(2-10\,\kev\right) \else $f_{\rm bol}\left(2-10\,\kev\right)$\fi}

\newcommand{\fobs}{\ifmmode f_{\rm obs} \else $f_{\rm obs}$\fi}

\newcommand{  \mbh      }{\ifmmode M_{\rm BH} \else $M_{\rm BH}$\fi}
\newcommand{  \lmbh     }{\ifmmode \log\left(\mbh/\Msun\right) \else $\log\left(\mbh/\Msun\right)$\fi} 
\newcommand{  \lledd    }{\ifmmode L/L_{\rm Edd} \else $L/L_{\rm Edd}$\fi}
\newcommand{  \mmedd    }{\ifmmode \dot{m}/\dot{m}_{\rm \,Edd} \else $\dot{m}/\dot{m}_{\rm \,Edd}$\fi}
\newcommand{  \Lbol     }{\ifmmode L_{\rm bol} \else $L_{\rm bol}$\fi}
\newcommand{  \lbol     }{\ifmmode L_{\rm bol} \else $L_{\rm bol}$\fi}
\newcommand{  \lLbol    }{\ifmmode \log\left(\Lbol/\ergs\right) \else $\log\left(\Lbol/\ergs\right)$\fi} 
\newcommand{  \Lagn     }{\ifmmode L_{\rm AGN} \else $L_{\rm AGN}$\fi}
\newcommand{  \lagn     }{\ifmmode L_{\rm AGN} \else $L_{\rm AGN}$\fi}

\newcommand{  \tgrow     }{\ifmmode t_{\rm growth} \else $t_{\rm growth}$\fi}
\newcommand{  \tAD     }{\ifmmode t_{\rm acc} \else $t_{\rm acc}$\fi}
\newcommand{  \tacc    }{\ifmmode t_{\rm acc} \else $t_{\rm acc}$\fi}
\newcommand{  \tUni      }{\ifmmode t_{\rm Universe} \else $t_{\rm Universe}$\fi}

\newcommand{  \Mdotin	}{\ifmmode \dot{M}_{\rm infall} \else $\dot{M}_{\rm infall}$\fi}
\newcommand{  \Mdotbh	}{\ifmmode \dot{M}_{\rm BH} \else $\dot{M}_{\rm BH}$\fi}
\newcommand{  \Mdotad	}{\ifmmode \dot{M}_{\rm AD} \else $\dot{M}_{\rm AD}$\fi}
\newcommand{  \Mdotacc	}{\ifmmode \dot{M}_{\rm acc} \else $\dot{M}_{\rm acc}$\fi}
\newcommand{  \Mdotthin	}{\ifmmode \dot{M}_{\rm thin} \else $\dot{M}_{\rm thin}$\fi}
\newcommand{  \Mdotdisk	}{\ifmmode \dot{M}_{\rm disk} \else $\dot{M}_{\rm disk}$\fi}

\newcommand{  \Mindot	}{\ifmmode \dot{M}_{\rm infall} \else $\dot{M}_{\rm infall}$\fi}
\newcommand{  \Mbhdot	}{\ifmmode \dot{M}_{\rm BH} \else $\dot{M}_{\rm BH}$\fi}
\newcommand{  \Maddot	}{\ifmmode \dot{M}_{\rm AD} \else $\dot{M}_{\rm AD}$\fi}
\newcommand{  \Maccdot	}{\ifmmode \dot{M}_{\rm acc} \else $\dot{M}_{\rm acc}$\fi}
\newcommand{  \Mthdot	}{\ifmmode \dot{M}_{\rm thin} \else $\dot{M}_{\rm thin}$\fi}
\newcommand{  \Mdsdot	}{\ifmmode \dot{M}_{\rm disk} \else $\dot{M}_{\rm disk}$\fi}

\newcommand{  \as	}{\ifmmode a_{\rm *} \else $a_{\rm *}$\fi}
\newcommand{  \avec	}{\ifmmode \vec{a}_{\rm *} \else $\vec{a}_{\rm *}$\fi}
\newcommand{  \re	}{\ifmmode \eta      	 \else $\eta$\fi}
\newcommand{  \RISCO	}{\ifmmode R_{\rm ISCO}  \else $R_{\rm ISCO}$\fi}

\newcommand{  \mseed    }{\ifmmode M_{\rm seed} \else $M_{\rm seed}$\fi}
\newcommand{  \mbul     }{\ifmmode M_{\rm bulge} \else $M_{\rm bulge}$\fi} 
\newcommand{  \mstar    }{\ifmmode M_{*} \else $M_{*}$\fi} 
\newcommand{  \mgal     }{\ifmmode M_{*} \else $M_{*}$\fi} 
\newcommand{  \mhost    }{\ifmmode M_{\rm host} \else $M_{\rm host}$\fi}
\newcommand{  \mmsmall  }{\ifmmode M_{\rm BH}/M_{*} \else $M_{\rm BH}/M_{*}$\fi}
\newcommand{  \mmlarge  }{\ifmmode M_{*}/M_{\rm BH} \else $M_{*}/M_{\rm BH}$\fi}

\newcommand{  \mmdotlarge}{\ifmmode \dot{M}_*/\Mbhdot \else $\dot{M}_*/\Mbhdot$\fi}
\newcommand{  \mmdotsmall}{\ifmmode \Mbhdot/\dot{M}_* \else $\Mbhdot/\dot{M}_*$\fi}

\newcommand{  \mmwp     }{\ifmmode \left(M_{*}/M_{\rm BH}\right) \else $\left(M_{*}/M_{\rm BH}\right)$\fi}
\newcommand{  \ml       }{\ifmmode M_{*}/L_{*} \else $M_{*}/L_{*}$\fi}
\newcommand{  \mlwp     }{\ifmmode \left(M_{*}/L\right) \else $\left(M_{*}/L\right)$\fi}
\newcommand{  \mlk      }{\ifmmode \left(M_{*}/L_{K}\right) \else $\left(M_{*}/L_{K}\right)$\fi}
\newcommand{  \sigs     }{\ifmmode \sigma_{*} \else $\sigma_{*}$\fi}
\newcommand{  \Reff     }{\ifmmode R_{\rm e} \else $R_{\rm e}$\fi}
\newcommand{  \Rvir     }{\ifmmode R_{\rm vir} \else $R_{\rm vir}$\fi}
\newcommand{  \Rtwo     }{\ifmmode R_{200} \else $R_{200}$\fi}
\newcommand{  \Rfive    }{\ifmmode R_{500} \else $R_{500}$\fi}
\newcommand{  \Rgrp     }{\ifmmode R_{\rm grp} \else $R_{\rm grp}$\fi}
\newcommand{  \nser     }{\ifmmode n_{\rm s} \else $n_{\rm s}$\fi}
\newcommand{  \LSF      }{\ifmmode L_{\rm SF}  \else $L_{\rm SF}$\fi}
\newcommand{  \LFIR     }{\ifmmode L_{\rm FIR} \else $L_{\rm FIR}$\fi}
\newcommand{  \Lfir     }{\ifmmode L_{\rm FIR} \else $L_{\rm FIR}$\fi}
\newcommand{  \LTIR     }{\ifmmode L_{\rm TIR} \else $L_{\rm TIR}$\fi}
\newcommand{  \Ltir     }{\ifmmode L_{\rm TIR} \else $L_{\rm TIR}$\fi}

\newcommand{  \mdyn     }{\ifmmode M_{\rm dyn} \else $M_{\rm dyn}$\fi} 
\newcommand{  \mgas     }{\ifmmode M_{\rm gas} \else $M_{\rm gas}$\fi} 
\newcommand{  \mh       }{\ifmmode M_{\rm h} \else $M_{\rm h}$\fi}
\newcommand{  \mhalo    }{\ifmmode M_{\rm halo} \else $M_{\rm halo}$\fi}
\newcommand{  \sfr      }{\ifmmode {\rm SFR} \else SFR\fi}

\newcommand{ \Lcii     }{\ifmmode L_{\cii} \else $L_{\cii}$\fi}
\newcommand{ \fwcii  }{\ifmmode {\rm FWHM}\cii \else FWHM\cii\fi}

\newcommand{\bj}{\ifmmode b_{\rm J} \else $b_{\rm J}$\fi}

\newcommand{\iab}{\ifmmode i_{\rm AB} \else $i_{\rm AB}$\fi}

\newcommand{\jab}{\ifmmode J_{\rm AB} \else $J_{\rm AB}$\fi}
\newcommand{\hab}{\ifmmode H_{\rm AB} \else $H_{\rm AB}$\fi}
\newcommand{\kab}{\ifmmode K_{\rm AB} \else $K_{\rm AB}$\fi}

\newcommand{\jveg}{\ifmmode J_{\rm Vega} \else $J_{\rm Vega}$\fi}
\newcommand{\hveg}{\ifmmode H_{\rm Vega} \else $H_{\rm Vega}$\fi}
\newcommand{\kveg}{\ifmmode K_{\rm Vega} \else $K_{\rm Vega}$\fi}

\def\arcsec{\hbox{$^{\prime\prime}$}}

\newcommand{  \Chisq    }{\ifmmode \chi^{2} \else $\chi^{2}$}
\newcommand{  \nelec    }{\ifmmode n_{e} \else $n_{e}$\fi}     % electron density
\newcommand{  \nh       }{\ifmmode n_{\rm H} \else $n_{\rm H}$\fi}     % hydrogen density
\newcommand{  \Ncol     }{\ifmmode N_{\rm col} \else $N_{\rm col}$\fi} % column density
\newcommand{  \NH       }{\ifmmode N_{\rm H} \else $N_{\rm H}$\fi}     % column density

\def\arcsec{\hbox{$^{\prime\prime}$}}
\def\ion#1#2{#1$\;${\small\rm\@Roman{#2}}\relax}

\newcommand{\SiX}{\ifmmode \left[{\rm Si}\,\textsc{x}\right]\,\lambda14300 \else [Si\,{\sc x}]\,$\lambda14300$\fi}
\newcommand{\SiVI}{\ifmmode \left[{\rm Si}\,\textsc{vi}\right]\,\lambda19640 \else [Si\,{\sc vi}]\,$\lambda19640$\fi}
\newcommand{\SXI}{\ifmmode \left[{\rm S}\,\textsc{xi}\right]\,\lambda19196 \else [S\,{\sc xi}]\,$\lambda19196$\fi}
\newcommand{\SVIII}{\ifmmode \left[{\rm S}\,\textsc{viii}\right]\,\lambda9915 \else [S\,{\sc viii}]\,$\lambda9915$\fi}
\newcommand{\SIX}{\ifmmode \left[{\rm S}\,\textsc{ix}\right]\,\lambda12520 \else [S\,{\sc ix}]\,$\lambda12520$\fi}
\newcommand{\FeXIII}{\ifmmode \left[{\rm Fe}\,\textsc{xiii}\right]\,\lambda10747 \else [Fe\,{\sc xiii}]\,$\lambda10747$\fi}
\newcommand{\SiXI}{\ifmmode \left[{\rm Si}\,\textsc{xi}\right]\,\lambda19320 \else [Si\,{\sc xi}]\,$\lambda19320$\fi}

\newcommand{\RCO}{\ifmmode 0.64\pm0.09 \else $0.63\pm0.09$\fi}

\title[${\rm ^{12}CO(J=2-1)/(J=1-0)}$ in EMPIRE]{%Narrowing down 
New Constraints on the ${\rm ^{12}CO(2-1)/(1-0)}$ Line Ratio Across Nearby Disc Galaxies}

\author[J. den Brok et al.]{J.~S. den Brok,$^{1}$
D. Chatzigiannakis,$^{1}$
F. Bigiel,$^{1}$
J. Puschnig,$^{1}$
A. T. Barnes,$^{1}$
A. K. Leroy,$^{2}$
\and
M. J. Jim{\'e}nez-Donaire,$^{3}$
A. Usero,$^{3}$
E. Schinnerer,$^{4}$
E. Rosolowsky,$^{5}$
C. M. Faesi,$^{6}$
\and
K. Grasha,$^{7}$
A. Hughes,$^{8}$
J.~M.~D.~Kruijssen,$^{9}$
D. Liu,$^{4}$
L. Neumann,$^{1}$
J. Pety,$^{10,11}$
\and
M. Querejeta,$ ^{3}$
T. Saito,$^{4}$
A. Schruba,$^{12}$
S. Stuber$^{4}$\\% List of institutions
$^{1}$ Argelander-Institut für Astronomie, Universität Bonn, Auf dem Hügel 71, 53121 Bonn, Germany\\
$^{2}$Department of Astronomy, The Ohio State University, 4055 McPherson Laboratory, 140 West 18th Avenue, Columbus, OH 43210, USA\\
$^{3}$Observatorio Astron{\'o}mico Nacional (IGN), C/Alfonso XII 3, E-28014 Madrid, Spain\\
$^{4}$Max Planck Institute for Astronomy, Königstuhl 17, D-69117 Heidelberg, Germany\\
$^{5}$4-183 CCIS, University of Alberta, Edmonton AB T6G 2E1, Alberta, Canada\\
$^{6}$Dept. of Astronomy, University of Massachusetts - Amherst, 710 N. Pleasant Street, Amherst, MA 01003, USA\\
$^{7}$Research School of Astronomy and Astrophysics, Australian National University, Canberra, ACT 2611, Australia\\
$^{8}$Université de Toulouse, UPS-OMP, F-31028 Toulouse, France ; CNRS, IRAP, Av. du Colonel Roche BP 44346, F-31028 Toulouse cedex 4, France\\
$^{9}$Astronomisches Rechen-Institut, Zentrum für Astronomie der Universität Heidelberg, Mönchhofstraße 12-14, 69120 Heidelberg\\
$^{10}$IRAM, 300 rue de la Piscine, F-38406 Saint Martin d’Hères, France\\
$^{11}$Sorbonne Université, Observatoire de Paris, Université PSL, École normale supérieure, CNRS, LERMA, F-75005, Paris, France\\
$^{12}$Max-Planck Institut für Extraterrestrische Physik, Giessenbachstraße 1, 85748 Garching, Germany}%Argelander-Institut für Astronomie, Universität Bonn, Auf dem Hügel 71, 53121 Bonn, Germany}\\

\date{Accepted 2021 March 12. Received 2021 March 09; in original form 2020 May 06 }

\pubyear{2018}

\begin{document}
\label{firstpage}
\pagerange{\pageref{firstpage}--\pageref{lastpage}}
 \maketitle

% Abstract of the paper
\begin{abstract}
Both the CO(2-1) and CO(1-0) lines are used to trace the mass of molecular gas in galaxies. Translating the molecular gas mass estimates between studies using different lines requires a good understanding of the behaviour of the CO(2-1)-to-CO(1-0) ratio, $R_{21}$. %In this study 
We compare new, high quality CO(1-0) data from the IRAM 30-m EMPIRE survey to the latest available \mbox{CO(2-1)} maps from HERACLES, PHANGS-ALMA, and a new IRAM 30-m M51 Large Program. This allows us to measure $R_{21}$ across the full star-forming disc of nine nearby, massive, star-forming spiral galaxies at 27\arcsec (${\sim} 1{-}2$ kpc) resolution. We find an average $R_{21} = \RCO$ %for our sample
when we take the luminosity-weighted mean of all individual galaxies. 
%{\bf \textcolor{red}{The line ratio uncertainty incorporates physical galaxy-to-galaxy variation and uncertainties in absolute flux scale that combined we estimate to be of the order 15\%.}}
%This value broadly agrees with previous work, however remains moderately lower than previously measured. 
This result is consistent with the mean ratio for disc galaxies that we derive from single-pointing measurements in the literature, $R_{\rm 21, lit}~=~0.59^{+0.18}_{-0.09}$.
%In addition, we combined and homogenized unresolved line ratio values from  previous publications finding for disc galaxies a value of $R_{\rm 21, lit.} = 0.59^{+0.18}_{-0.09}$. Our result is in agreement with this value.
The ratio %we find in the EMPIRE sample 
shows weak radial variations compared to the point-to-point scatter in the data. 
%The strongest effect being a ${\sim} 15\%$ central enhancement in six of our nine targets. 
In six out of nine targets the central enhancement in $R_{21}$ with respect to the galaxy-wide mean is of order $\sim 10{-}20\%$.
We estimate an azimuthal scatter of $\sim$20\% in $R_{21}$ at fixed galactocentric radius but this measurement is limited by our comparatively coarse resolution of 1.5\, kpc. %and we expect that we would see more azimuthal variation at higher resolution
 We find mild correlations between $R_{21}$ and CO brightness temperature, IR intensity, 70-to-160\,$\mu$m ratio, and IR-to-CO ratio. All correlations indicate that $R_{21}$ increases with gas surface density, star formation rate surface density, and the interstellar radiation field. %While uncertainties in the data can cause scatter in the ratio for an individual source, the 
%The combination of all the observations has helped us to narrow down the value of the ratio and will help future ISM studies in constraining temperatures, densities and opacities.  
\end{abstract}

% Select between one and six entries from the list of approved keywords.
% Don't make up new ones.

%%%%%%%%%%%%%%%%%%%%%%%%%%%%%%%%%%%%%%%%%%%%%%%%%%

%%%%%%%%%%%%%%%%% BODY OF PAPER %%%%%%%%%%%%%%%%%%

\section{Introduction}
\label{sec:intro}

Carbon monoxide (CO) is the most abundant molecule in the interstellar medium after molecular hydrogen (H$_2$). Unlike H$_2$, CO has a permanent dipole moment and its rotational transitions can be excited at low temperatures. The two lowest rotational transitions of the main CO molecule, \mbox{$^{12}$C$^{16}$O $J=1\rightarrow0$}, hereafter \mbox{CO(1-0)}, and $^{12}$C$^{16}$O $J=2\rightarrow1$, hereafter \mbox{CO(2-1)}, are among the brightest millimeter-wave spectral lines emitted by galaxies. They have critical densities of $\rm n_{\rm crit,\rm 1{-} \rm 0} \sim 2{,}000 ~{\rm cm}^{-3} $ and $\rm n_{\rm crit,\rm 2{-} \rm 1} \sim 10{,}000 ~{\rm cm}^{-3} $ for a fully molecular gas with a temperature of $T~=~10~ \rm K$ and optically thin transitions. Given typical optical depths { for CO(1-0)} of $\tau \sim 5{-}10$, line trapping effects lower the effective critical density even further, to ${\sim} 100{-}1{,}000$~$\rm{ cm}^{-3}$. This is comparable to the mean density of molecular gas in galaxies \citep[for more see reviews by][]{Bolatto2013,Heyer2015,Shirley2015}. As a result of their brightness, low excitation requirement, and locations at favorable frequencies for observations from the ground, both transitions are often used to trace the mass of molecular gas in galaxies.

ALMA, NOEMA and other mm-wave facilities now regularly map both CO(2-1) and \mbox{CO(1-0)} line emission across large areas and large samples of galaxies. It is increasingly important to be able to quantitatively compare results obtained using these different lines. Physically, the CO(2-1)-to-CO(1-0) line ratio, $R_{21}$, should depend on the temperature and density of the gas and on the optical depths of the lines \citep[e.g., see][]{Sakamoto1994,Sakamoto1997,Penaloza2017,Penaloza2018}. Thus, understanding how $R_{21}$ varies in response to the local environment also has the prospect to provide information regarding the physical conditions of the molecular gas. 

The $R_{21}$ ratio has been studied in both the Milky Way \citep[e.g.][]{Hasegawa1997,Hasegawa1997_2,Sakamoto1997,Sawada2001,Yoda2010} and nearby galaxies \citep[e.g.][]{Eckart1990,Casoli1991a,Lundgren2004,Crosthwaite2007,Leroy2009,Koda2012,Leroy2013,Druard2014,Saintonge2017,Law2018, Koda2020, Yajima2020}. Milky Way studies highlight a correlation between the $R_{21}$ ratio and density, with $R_{21}$ dropping with decreasing gas density from the centers to the edges of molecular clouds \citep[e.g.,][]{Hasegawa1997_2}. 

Studies of individual other galaxies often find higher $R_{21}$ in the central kpc compared to the outer parts \citep[e.g.,][]{Braine1992,Leroy2009, Leroy2013, Koda2020, Yajima2020}. This radial behaviour could be explained if the average temperature and/or density of molecular gas drops with galactocentric radius. Independent evidence suggests that both temperature and density are often enhanced in galaxy centers \citep[e.g.,][]{Mangum2013,Gallagher_b2018,Sun2018, Donaire2019}. {Other work has focused on azimuthal variations in well-resolved galaxies with strong spiral arms, especially M51. There studies indicate enhanced excitation in the spiral arms and bar ends compared to the interarm regions} \citep{Koda2012,Vlahakis2013,Law2018,Koda2020}.

However, our quantitative knowledge of how $R_{21}$ varies across galaxies remains limited. Extensive CO(2-1) mapping has only been possible for $\sim 10$ years and there have been only a limited number of mapping surveys that cover both CO(1-0) and CO(2-1) in the same sample of galaxies. As a result, the magnitude of the observed variations in $R_{21}$ remain fairly weak, with the typical range of values found in spiral galaxies spanning from $0.5{-}0.9$ and often much less inside a single galaxy. This is easily within the range where even modest calibration uncertainties and heterogeneous data can obscure real astrophysical signal. Furthermore, much of the extragalactic mapping work has been confined to single-galaxy studies \citep[e.g.,][]{Crosthwaite2007, Koda2012,Vlahakis2013, Law2018, Koda2020}.

In this paper we present the full disc mapping of CO(1-0) and CO(2-1) observations. CO(1-0) is obtained with the IRAM 30-m telescope, from the EMIR Multi-Line Probe of the ISM Regulating Galaxy Evolution (EMPIRE) \citep[]{Bigiel2016,Donaire2019}  and for CO(2-1) we use the  {latest} available data, selecting from the \textit{HERA CO-Line Extragalactic Survey} (HERACLES) \citep[]{Leroy2009}, the \textit{Physics at High Angular resolution in Nearby Galaxies} (PHANGS) survey (A.~K.\ Leroy et al., in prep.) or the IRAM 30-m M51 Large Program (J.~S.\ den Brok et al., in prep.). Thus, each line is covered by a homogeneous, deep, wide-area mapping survey. Together they probe $R_{21}$ across a sample of nine nearby spiral galaxies. Our main goals are to derive robust galaxy-wide mean values of the $R_{21}$ ratio and to investigate how $R_{21}$ varies systematically across the discs of these galaxies. 

In Section\,\ref{sec:observations} we present the data and define the physical quantities we use. Our analysis of the $R_{21}$ ratio is presented in Section\,\ref{analysis}, where we examine the distribution of the ratio, its radial and azimuthal variations, as well as the possible correlations between $R_{21}$ and physical properties such as CO brightness temperature and IR emission. We discuss our results and compare them to results from previous observations in Section\,\ref{discussion}. We summarize our findings in Section\,\ref{Summary}.

\section{Observations} 
\label{sec:observations}

\begin{table}
 \caption{Galaxy Sample}
 \label{tab:empire_gal}
 \begin{tabular}{llcccccc}
  \hline \hline
 Name & RA & DEC & D & $i$ & PA \\
 & (J2000) & (J2000) & (Mpc) & (deg) & (deg) \\
 \hline
 NGC 0628 & 01:36:41.8 & 15:47:00 & 9.0 & 7 & 20 \\
 NGC 2903 & 09:32:10.1 & 21:30:03 & 8.5 & 65 & 204 \\
 NGC 3184 & 10:18:17.0 & 41:25:28 & 13.0 & 16 & 179 \\
 NGC 3627 & 11:20:15.0 & 12:59:30 & 9.4 & 62 & 173 \\
 NGC 4254 & 12:18:50.0 & 14:24:59 & 16.8 & 32 & 55 \\
 NGC 4321 & 12:22:55.0 & 15:49:19 & 15.2 & 30 & 153 \\
 NGC 5055 & 13:15:49.2 & 42:01:45 & 8.9 & 59 & 102 \\
 NGC 5194 & 13:29:52.7 & 47:11:43 & 8.4 & 20 & 172 \\
 NGC 6946 & 20:34:52.2 & 60:09:14 & 7.0 & 33 & 243 \\
 \hline 
 \end{tabular}
\\ \begin{flushleft}
Notes: Adopted from \cite{Donaire2019}.
\end{flushleft}
\end{table}

\subsection{Galaxy Sample}

Our sample consists of the nine nearby star-forming disc galaxies targeted by the EMPIRE survey \citep{Bigiel2016,Donaire2019}. We list their names, orientations, and adopted distances in Table\,\ref{tab:empire_gal}. For a more detailed description of the properties of our sample we refer to \citet[]{Donaire2019}. Summarizing, our targets are all massive, star-forming disc galaxies, with stellar masses of $10 <\log_{10}(M_\star / M_\odot) <10.6$, 
%${\rm log_{10}}(M_{\ast})$, between  $  10.0 {-} 10.6~{\rm log_{10}}(M_{\odot})$, 
metallicities from half--solar to solar, and star formation rate surface densities in the range $ {2.8 {-} 21} \times 10^{-3}{ M_{\odot} \rm yr ^{-1} \rm kpc^{-2}}$.

\subsection{EMPIRE CO(1-0) Data}\label{sec:co10_obs}

EMPIRE mapped the entire optical discs of these galaxies in several 3\,mm emission lines using the EMIR receiver. One main goal of EMPIRE is to understand how the dense gas fraction depends on the environment within and among galaxies. To achieve this goal, EMPIRE obtained deep, extended maps of high critical density lines that trace dense gas, such as HCN (1-0), HCO$^+$ (1-0), and HNC(1-0). In order to measure the dense gas fraction, EMPIRE also required a high quality tracer of the total molecular gas. This was accomplished by mapping the $^{12}$CO(1-0) and $^{13}$CO(1-0) lines. 

We employ the $^{12}$CO(1-0) data from EMPIRE for eight galaxies (PI Jim\'enez-Donaire, projects 061-51 and 059-16, \citet{Donaire2019}; PI Cormier, project D15-12 for NGC 5055, \citet{Cormier2018}). For NGC~5194, we use the $^{12}$CO(1-0) data from the PdBI Arcsecond Whirlpool Survey (PAWS) \citep[]{Pety2013,Schinnerer2013}. This was also obtained by the IRAM~30-m using an almost identical strategy to the EMPIRE project.
%\autoref{tab:empire_gal} gives a basic overview of our sample.

These CO(1-0) maps cover the full disc of each galaxy, with an angular resolution of $27''$ (${\sim} 1{-}2$ kpc)\footnote{{When we quote the beam of single dish maps, we refer to the effective beam size, which combines the primary beam of the telescope and that of the gridding kernel. All observations used short dump times that critically sampled the beam along the scan direction. See \cite{Mangum2007} for general information on gridding kernels and see \citet{Leroy2009}, \citet{Donaire2019}, and \citet{Herrera2020} for information on the HERACLES, EMPIRE, and PHANGS-ALMA maps.}}. They have rms noise between $13{-}24$\,mK in each $4$\,km\,s$^{-1}$ channel. For full details regarding the observing strategy, reduction, and data products see \citet[]{Donaire2019}.

{We estimate the overall calibration uncertainty of EMPIRE to be $\sim5\%$ and we expect this to mostly be a multiplicative factor that scales the whole map. Most EMPIRE maps were made by combining many observing sessions that each covered the whole galaxy. Therefore we expect the maps to be well-calibrated internally. We regularly observed line calibrators as part of the EMPIRE observing strategy. \citet{Cormier2018} and \citet{Donaire2019} showed that the absolute flux calibration of the EMPIRE data showed rms variation of $\sim4{-}8\%$ from session to session.}

Given the overall brightness of the CO(1-0) line, this calibration term represents the dominant source of uncertainty over the inner region of many galaxies.\footnote{{ For CO(1-0), considering all individual lines-of-sight, we find 839/5416 points have S/N>20, compared to 3528/5416 points with S/N>3. For \mbox{CO(2-1)} on the other hand, we have 1130 points with S/N>20 compared to 4134 with S/N>3.}}

\subsection{CO(2-1) Data}\label{sec:co21_obs}

We compare the EMPIRE CO(1-0) data to CO(2-1) maps {from the IRAM 30-m and ALMA. In each case, we picked the highest quality available CO(2-1) map. All of the CO(2-1) data have higher native resolution than the CO(1-0) maps. We convolved them to the resolution of EMPIRE using a Gaussian kernel with width determined by subtracting the current beam from the target beam in quadrature. We then aligned the CO(2-1) data to the EMPIRE astrometric grid and rebinned to $4$~km~s$^{-1}$ channels.}

{For NGC 5194 (M51)  we use observations from a new IRAM 30-m Large Program (PI: Toshiki Saito, project 055-17; den Brok et al., in prep.). The goal of this program is to obtain sensitive observations of 1-mm and 3-mm CO isotopologue transitions and thereby to improve constraints of ISM physical quantities. These observations were carried out using the EMIR instrument at the IRAM 30-m telescope with a total of 172\,h. The program included new CO(2-1) observations, which we use here. At $27''$ resolution and 4~km~s$^{-1}$ channel width, this new CO(2-1) cube has rms noise $5.6$~mK. }

{The observing strategy and instrument for the M51 EMIR maps closely resemble that of EMPIRE. Therefore, we expect that the uncertainty in the amplitude calibration of the M51 EMIR CO(2-1) map is also similar to that of the EMPIRE maps and consider ${\sim} 5\%$ a good estimate.}

{ALMA observed CO(2-1) emission from NGC 0628, NGC 2903, NGC 3627, NGC 4254, and NGC 4321 part of the PHANGS-ALMA survey (A. K. Leroy, E. Schinnerer et al., in preparation). PHANGS-ALMA is using ALMA's 12-m, 7-m and total power antennas to observe CO(2-1) emission from a large sample of nearby. We begin with the cubes made from combining the 12-m, 7-m, and total power observations. Because we convolve the data to $27''$ for our analysis and the ALMA single dishes have a ${\sim} 27''$ beam, the total power data contribute almost all of the information for our analysis. As a result, the details of interferometric imaging are secondary. The PHANGS-ALMA total power pipeline is described by \citet{Herrera2020}. After convolving to $27''$, the PHANGS-ALMA cubes have on average an rms noise of $1{-}2$~mK in each 4~km~s$^{-1}$ channel.}

{The PHANGS-ALMA data are calibrated using observations of solar system objects or Galactic SF regions that are pinned to the ALMA interferometric calibration scheme. The overall uncertainty in the flux calibration should be about $5\%$ at 1mm. The flux of total power observations targeting the same PHANGS-ALMA galaxy on different days scatter from one another by $\pm 3\%$ \citep{PHANGSPipeline}, in good agreement with the aforementioned 5\% \citep[and with][]{Bonato2018}.}

For the remaining three sources, NGC 3184, NGC 5055, and NGC 6946, we take observations from the HERA CO-Line Extragalactic Survey (HERACLES) \citep[]{Leroy2009}. HERACLES surveyed CO(2-1) emission from 48 nearby galaxies. These maps have appeared previously in \citet[][]{Schruba2011,Schruba2012,Leroy2013,Sandstrom2013}. {After gridding,} the HERACLES maps have a native resolution of $13.3\arcsec$. {After matching to the EMPIRE $27''$ beam and velocity grid, the HERACLES cubes have rms noise between $5{-}11$~mK per 4~km~s$^{-1}$ channel.}

{The overall flux scale of HERACLES is uncertain at the ${\sim} 6{-}20\%$ level \citep[see][and Appendix \ref{app:comp_HERA_ALMA}]{Leroy2009}. Also, the HERACLES maps combine information from multiple receiver pixels that can have gain uncertainties relative to one another. We investigate the internal gain variations of the HERA pixels and compare HERACLES, ALMA, and EMIR data for galaxies with multiple maps in Appendix \ref{app:comp_HERA_ALMA}. This analysis yields the gain uncertainty mentioned above and also leads us to prefer ALMA or EMIR maps when available because their calibration should be more stable (i.e. their calibration uncertainties are likely to be well-described by a single gain factor).
%{\bf \textcolor{red}{This analysis of HERA leads us to prefer ALMA data over HERA.}}
}

{As with the CO(1-0) data, the high signal to noise of the CO(2-1) data means that calibration often represents the dominant source of uncertainty. Anywhere that the HERACLES data exceed $\mathrm{S/N}=5{-}10$, and anywhere that the ALMA and EMIR data exceed $\mathrm{S/N}=20$, calibration will dominate our uncertainty on the brightness temperature.
{{Below the S/N threshold, the uncertainty is dominated by the random noise. }}\\
%The quality of the calibration drives our preference for the ALMA and EMIR maps over the HERACLES maps. ToDo: add Ash's suggestion. 
{{We need the best intensity accuracy possible for this study. Comparing ALMA/EMIR to HERACLES, ALMA/EMIR has both lower absolute and relative flux uncertainty across the map. Although all the galaxies have been also observed with HERA, we, therefore, choose the ALMA/EMIR over HERACLES were possible in our analysis. In summary, the associated uncertainties for the CO(2-1) emission line from ALMA are 5\,per cent, EMIR are 5-10\,per cent and HERA are 6-20\,per cent.}}
See Appendix \ref{app:comp_HERA_ALMA} for more details.}

%{\bf In Appendix \ref{app:comp_HERA_ALMA}, we provide a more in--depth analysis of the calibration uncertainty of the instruments.}

%\textbf{The general RMS noise in brightness temperature units is found to be around $\sigma_{\rm RMS} \approx 0.17$\, K  per  $2.5$\,km\,s$^{-1}$ channel. It can vary a little between cubes.}

%In order to compare to the EMPIRE CO(1-0) maps, we convolved the higher resolution CO(2-1) cubes to the coarser 27$\arcsec$ resolution of the EMPIRE data using a Gaussian kernel. We then aligned the data to the EMPIRE astrometric grid and rebinned to $4$~km~s$^{-1}$ channels. After matching the CO(2-1) observations to the EMPIRE resolution, the data have rms noise between $5{-}11$ mK in each $4$~km~s$^{-1}$ channel.
%{\bf Generally, the systematic errors dominate over the statistical errors for the HERACLES observations.}
\begin{figure*}
\includegraphics[width=\textwidth]{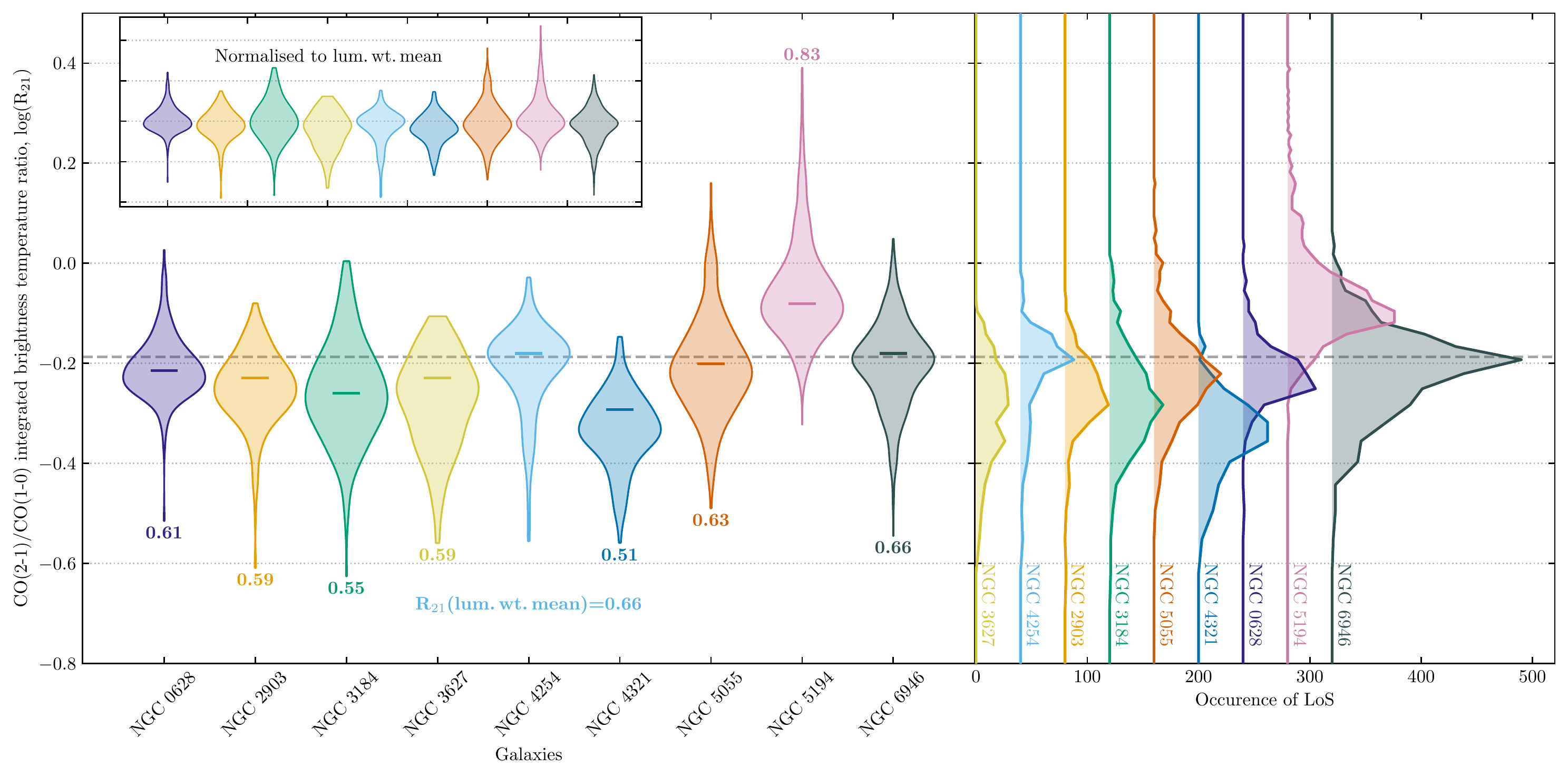}%[scale=0.28]{quartile.png}
\caption{%{\bf[Updated by Ash - main text needs updating?]} 
\textbf{Histograms of the CO(2-1)/CO(1-0) integrated brightness temperature ratio, log($R_{21}$), for nine nearby star-forming spiral galaxies.} The \textit{left panel} shows distributions of log($R_{21}$) as violin histograms. These histograms only show results for the positions that have $\mathrm{S/N}>3$ integrated brightness temperature detections in both the CO(1-0) and CO(2-1) maps. These histograms treat each line of sight equally. The \textit{luminosity-weighted mean} $R_{21}$ value for each galaxy appears as a coloured horizontal line inside each histogram, with the value reported above or below each violin histogram. All data, regardless of S/N, are included within the calculation of the luminosity weighted mean, as shown in Table\,\ref{tab:med_ratio}. {{{We note that the uncertainty incorporates physical galaxy-to-galaxy scatter as well as uncertainties in flux scale calibration.}}} The \textit{inset} within the left panel shows the $R_{21}$ violin histograms for each galaxy normalised to their luminosity weighted mean $R_{21}$ values. The \textit{right panel} shows the histogram of the combined $R_{21}$ distribution for all targets. The dashed grey line extending across all panels shows $\langle R_{21}\rangle = 0.65$, the non-weighted mean of $R_{21}$ if including all lines-of-sight with $\mathrm{S/N}>3$ (see last column of Table \ref{tab:med_ratio}).}
\label{fig:hist_R}
\end{figure*}

\subsection{Far-Infrared Data}\label{sec:IR_obs}

We compare the $R_{21}$ ratio to infrared (IR) maps at wavelengths of 70, 160, and 250\,$\mu$m from the \textit{Herschel} space telescope. These were compiled and processed to match the EMPIRE beam and astrometric grid by \citet[]{Donaire2019}. For seven of our targets, the data come from the KINGFISH survey \citep[][]{Kennicutt2011}. For NGC~5194, the data come from the Very Nearby Galaxies Survey \citep[][]{Parkin2013}. NGC~2903 lacks \textit{Herschel} data. As a result we cannot determine the 70-to-160\,$\mu$m ratio or the TIR luminosity surface brightness in Section\,\ref{correlations} for this galaxy.

\subsection{Measured Quantities}
%\subsection{Physical quantities}
\label{sec:Physical}

We follow a similar analysis path to the $^{13}$CO-focused study of \citet{Cormier2018} and the HCN-focused study of \citet{Donaire2019}. We measure the $R_{21}$ ratio as a function of galactocentric radius, the 70-to-160$\,\mu$m ratio, CO brightness temperature, total IR surface brightness, and the TIR-to-CO ratio.

\textit{CO(2-1)/(1-0) ratio, $R_{21}$}: We define $R_{21}$ as the line-integrated CO(2-1) surface brightness divided by the line-integrated \mbox{CO(1-0)} surface brightness. For both lines, the line-integrated surface brightness has units of K~km~s$^{-1}$. 

{Note that our brightness temperature-based definition of $R_{21}$ differs from the flux density based values often quoted in the high redshift literature \citep[e.g.,][]{ARAVENA10, DADDI10, BOTHWELL13, ARAVENA14, ARAVENA16}. Using the velocity-integrated flux density definition, one would expect thermalized lines to show a ratio of about four. Using the brightness temperature scale, the line ratio for a thermalized line will be about unity \citep[e.g., see][]{Solomon2005}, or slightly lower due to deviations from the Rayleigh-Jeans approximation. }

\textit{$R_{21}$ for individual lines of sight}: We calculate $R_{21}$ for each line of sight. When doing so, we use exactly the same velocity range for the integral over both lines. To define this velocity range, we use for sightlines outside of the 0.23\,$r_{25}$ aperture the velocities covered by \textsc{Hi} 21-cm line emission \citep[mostly from THINGS;][]{Walter2008} as an independent estimate for the velocity range likely to be covered by CO. For lines of sight within the 0.23\,$r_{25}$ aperture, where the ISM is mostly molecular, we use the CO(2-1) emission as a proxy for the velocity range. This way we make sure that the broad, central CO lines are fully included. We chose the CO(2-1) line instead of the CO(1-0) because our CO(2-1) maps have higher S/N than our CO(1-0) maps.

{Note that because our sampling scheme oversamples the beam by a factor of $4$, measurements for $R_{21}$ from adjacent lines of sight are correlated and not independent. We take this into account in our presentations of results.}

\textit{$R_{21}$ from spectral stacking}: {In addition to measuring $R_{21}$ for individual lines of sight,} we employ a spectral stacking method to explore possible correlations between the $R_{21}$ ratio and various physical quantities. In this approach, we bin the data by some other quantity, for example TIR surface brightness. We construct an average CO(2-1) and CO(1-0) spectrum for each bin. We estimate the mean $R_{21}$ in that bin by dividing the integrated brightness temperature calculated from each binned spectrum.

The method is described in detail by \citet{Schruba2011,JDonaire2017} and \citet{Cormier2018}. We regrid each spectrum so that the local mean velocity now corresponds to $v=0$~km~s$^{-1}$. For this application, we use the velocity field derived from the \textsc{Hi} 21-cm data to estimate the local mean velocity. After regridding the spectra, we average together all spectra in each bin. Because the large-scale velocity gradient has been removed, spectra from different parts of the galaxy average coherently. 

{We derive the integrated brightness temperature from each stacked spectrum. We pick the velocity range for this integral by first fitting the spectrum. We use either a single-Gaussian profile or a double-horn profile, whichever fits better. The double-horn profile offers a better description of the broad, flat-topped emission lines found in some of our galaxy centers. We set the velocity range for direct integration of the spectrum to cover everywhere that the fit exceeds 1\% of the peak brightness temperature. Note that the fit is \textit{only} used to set boundaries over which we integrate the spectrum.}  

{We only present stacked measurements of $R_{21}$ when both lines have an integrated emission above $3\sigma$ of the rms noise. In practice, our stacks almost always achieve much higher signal to noise than this. In Table \ref{tab:SNR_stacks} we summarize the signal to noise ratio for our stacks as a function of galactocentric radius. Inside $r_{\rm gal} < 9$~kpc $>95\%$ of the stacked $R_{21}$ measurements have signal to noise $>11$ for both emission line brightness temperature measurements, comparable to the very high{, pixel based} signal to noise threshold { values} used by \citet{Koda2012} and \citet{Koda2020}.}

{\textit{Uncertainties on $R_{21}$}: We compute the uncertainties on the integrated brightness temperature, $\sigma_{\rm Int}$, using the following formula,}
\begin{center}
%\begin{equation}
%   \sigma_{\rm Int.} =\sigma_{\rm rms} \times \Delta \nu_{\rm chan} \times \sqrt{\frac{FWHM_{\rm line}}{\Delta \nu_{\rm chan}}}
%\end{equation}
\begin{equation}
    \label{eq:unc}
    \sigma_{\rm Int} =\sigma_{\rm rms} \times \Delta \nu_{\rm chan} \times \sqrt{n_{\rm chan}}
\end{equation}
\end{center}
where $\sigma_{\rm rms}$ is the 1$\sigma$ rms value of the noise in K measured from the signal-free part of the spectrum, $\Delta \nu_{\rm chan}$ is the channel width in\,$\rm km\, \rm s^{-1}$, and 
%$\rm FWHM_{\rm line}$ the full width half maximum of the fit derived line, in  $\rm km\,\rm s^{-1}$. 
$n_{\rm chan}$ is the number of channels that are integrated together. 

{When we apply Equation~\eqref{eq:unc} to the stacked spectra, we measure the noise, $\sigma_{\rm rms}$, from the signal-free region of the stacked spectrum itself. As a result, this approach properly accounts for the fact that our original pixels oversample the beam.}

%For the stacked line emission, the number of channels is determined by selecting only channels with an intensity above 1\% of the fitted Gaussian peak.

{After estimating $\sigma_{\rm Int.}$ for each line, we estimate the uncertainties on $R_{21}$ by propagating the errors of the CO(1-0) and CO(2-1) integrated brightness temperature.}

%The errors for each integrated brightness temperature measurement are estimated by measuring the rms in individual channels from the line-free part of the spectrum and then accounting for error propagation across the line velocity range. 

%It is also worth noting that high redshift surveys do find a slightly higher value of the $R_{21}$ ratio, compared to the typical values of the ratio reported for the local universe. 

%Given the small observed dynamic range for $R_{21}$, the relative calibration of the maps represents a significant source of systematic uncertainty. 

{Because of the high signal to noise in our CO observations, the systematic uncertainty due to flux calibration often dominates the overall uncertainty in $R_{21}$. For example, in Table \ref{tab:SNR_stacks} we report the median signal to noise ratio for stacks within $9$~kpc is $>30$. This $\sim 3\%$ uncertainty is lower than the systematic uncertainty due to calibration. Our EMPIRE-ALMA or EMPIRE-EMIR galaxies have $R_{21}$ calibration uncertainties $\sim 7\%$. For our three EMPIRE-HERACLES targets, this may be as high as $\sim 20\%$, and at least $10{-}15\%$. At least in the EMIR-ALMA targets, we expect this calibration uncertainty to act as a single multiplicative factor for the map. Thus it will affect the mean value, but not the internal distribution in each galaxy. For the EMPIRE-HERACLES cases, we expect the primary uncertainty to be an overall scaling, but there may be second-order local variations due to the differences in the pixel gains discussed in Appendix \ref{app:comp_HERA_ALMA}.}

%\sout{Note that the small dynamic range in $R_{21}$ compounds the importance of careful calibration. For example for a case with $15\%$ calibration uncertainty on each line, a typical $R_{21} = 0.6$ will have a $1\sigma$ systematic uncertainty range of $0.5{-}0.7$ based only on the uncertainty in the calibration scale. This is already comparable to the dynamic range in our sample.}

%ToDo: replace with 1-2 sentences summary.
%The bandwidth of the HERA receivers did not allow simultaneously line calibrator observations, and the survey used an array receiver with some uncertainty in the relative gains of the pixels.

\textit{70-to-160$\mu$m ratio}: We compare $R_{21}$ to the 70-to-160$\mu$m ratio. This ratio traces the temperature of interstellar dust \citep[e.g.,][]{Draine2007}. Because most of the large grains in a galaxy are in thermal equilibrium with the local interstellar radiation field \citep[see textbook by ][]{Drain2011}, this ratio also acts as a tracer of the interstellar radiation field (ISRF). {Note that at the average densities and temperatures of molecular clouds traced by CO emission, we do not expect the dust and gas to collisionally couple and share the same temperature \citep{Drain2011}, so we do expect the IR color to directly trace the ISRF by not the gas temperature.} We measure the 70-to-160\,$\mu$m ratio after convolving the \textit{Herschel} $70\,\mu$m and $160\,\mu$m maps to match the EMPIRE resolution.

\textit{Total IR surface brightness}: We compare $R_{21}$ to the total IR (TIR) luminosity  per unit area. We use the TIR surface brightness as an observational proxy for the amount of embedded recent star formation. This tracer has the advantage compared to other SFR tracers, as it traced the embedded SFR, which means the recent SFR might affect the state of the molecular ISM. We follow the same approach as our previous work \citep[e.g.,][]{Usero_2015,Bigiel2016,JDonaire2017,Cormier2018}. 

We combine \textit{Herschel} 70, 160, and 250$\mu$m data in order to estimate the TIR surface brightness. First, we convolve each band to our common resolution of 27$\arcsec$ and place them onto the EMPIRE astrometric grid. Then we combine the bands, following \citet{2013MNRAS.431.1956G},
\begin{center}
\begin{equation}
    S_{\rm TIR} = \sum c_i S_i 
\end{equation}
\end{center}
where $S_{\rm TIR}$ refers to the TIR surface brightness, $S_i$ to the brightness in the given \textit{Herschel}   band $i$, and $c_i$ to the calibration coefficient from combined brightness. We use the specific calibration coefficients provided for each galaxy, with the exception of NGC~5194 where we use the generic calibration provided by \citet{2013MNRAS.431.1956G}, since this galaxy was not explicitly studied.

We focus on TIR surface brightness because it represents a simple, reproducible quantity that is closely related to the local surface density of recent star formation. We do not implement any specific conversions or consider second order effects like IR cirrus. Our analysis also does not hinge on any numerical conversion of TIR surface brightness to a star formation rate. For a detailed discussion of the use of TIR as a star formation rate proxy and a quantitative comparison to other star formation tracers \citep[e.g. ][]{Cormier2018,Gallagher2018,Donaire2019}.

\section{Results} \label{analysis}

\subsection{Overall Distribution of \texorpdfstring{$R_{21}$}{Lg}} 
\label{overall_dist}

\begingroup
\renewcommand{\arraystretch}{1.5} % Default value: 1
\begin{table*}
 \begin{center}
 \caption{The CO(2-1)/CO(1-0) integrated brightness temperature ratio, $R_{21}$, for nine nearby star-forming spiral galaxies. The upper half of the table shows luminosity-weighted statistics ($R_{21}$). The lower half of the table shows number statistics, which treat each pixel equally ($R_{21}^\mathrm{num}$). We tabulate the galaxy name, values of the mean, and 5th, 16th, 50th (i.e. median), 84th, and  95th percentile ranges of $R_{21}$ for our target galaxies. The second to last column gives the mean value and associated standard deviation of $R_{21}$ across the whole sample. The last column lists the mean and percentiles when considering all significant ($\mathrm{S/N}>3$) lines-of-sight across the whole sample. Note that all values within this table have been calculated for positions with a significant detection in both the CO(1-0) and CO(2-1) integrated brightness temperature maps (i.e. $\mathrm{S/N}>3$). { The bottom rows indicate the number of lines of sight (l.o.s.) for the individual galaxies. The number of all l.o.s. as well as those l.o.s which have both the CO(1-0) and CO(2-1) above the threshold of $3\sigma$ and $10\sigma$.}}
 \label{tab:med_ratio}
 %\begin{tabular}{l c c c c c c}
 %\hline \hline
 %Galaxy & $R_{21}^\mathrm{mean}$ & $R_{21}^\mathrm{16\%}$ & $R_{21}^\mathrm{84\%}$ & $R_{21}^\mathrm{5\%}$ & $R_{21}^\mathrm{95\%}$ \\
%\hline
 %\dots &  \dots &  \dots & \dots &  \dots \\
 %\hline
  %& $R_{21}^\mathrm{num,mean}$ & $R_{21}^\mathrm{num,16\%}$ & $R_{21}^\mathrm{num,84\%}$ & $R_{21}^\mathrm{num,5\%}$ & $R_{21}^\mathrm{num,95\%}$ \\
%\hline
 %\dots &  \dots &  \dots & \dots &  \dots \\
 %\hline \hline
 %\end{tabular}
 \begin{tabular}{l c c c c c c c c c c c}
 \hline \hline
  NGC & 0628 & 2903 & 3184 & 3627 & 4254 & 4321 & 5055 & 5194 & 6946&Galaxy Average& All Sightlines ($>3\sigma$)\\ \hline
 \begin{filecontents*}{Tables/table_weighted_mean.csv}
galaxy,ngcone,ngctwo,ngcthre,ngcfour,ngcfive,ngcsix,ngcseven,ngceight,ngcnine,average,los
R$_{21}^\mathrm{mean}$,0.61,0.59,0.55,0.59,0.66,0.51,0.63,0.83,0.66,0.63$\pm$0.09,0.66
R$_{21}^\mathrm{5\%}$,0.50,0.48,0.41,0.41,0.49,0.38,0.48,0.68,0.51,,0.49
R$_{21}^\mathrm{16\%}$,0.55,0.52,0.46,0.48,0.61,0.44,0.54,0.74,0.58,,0.56
R$_{21}^\mathrm{50\%}$,0.60,0.58,0.54,0.59,0.66,0.50,0.62,0.81,0.65,,0.71
R$_{21}^\mathrm{84\%}$,0.67,0.70,0.64,0.71,0.71,0.60,0.72,0.91,0.73,,0.85
R$_{21}^\mathrm{95\%}$,0.75,0.72,0.78,0.76,0.74,0.68,0.79,1.02,0.83,,0.94
\end{filecontents*}
\csvreader[head to column names, late after line=\\]{Tables/table_weighted_mean.csv}{}
 {\galaxy&\ngcone &\ngctwo&\ngcthre&\ngcfour&\ngcfive&\ngcsix & \ngcseven & \ngceight & \ngcnine &\average&\los}
 \hline
 \begin{filecontents*}{Tables/table_norm_mean.csv}
galaxy,ngcone,ngctwo,ngcthre,ngcfour,ngcfive,ngcsix,ngcseven,ngceight,ngcnine,average,los
R$_{21}^\mathrm{num. mean}$,0.62,0.57,0.57,0.54,0.63,0.47,0.64,0.90,0.65,0.62$\pm$0.12,0.65
R$_{21}^\mathrm{num. 5\%}$,0.52,0.44,0.40,0.37,0.42,0.34,0.44,0.66,0.47,,0.42
R$_{21}^\mathrm{num. 16\%}$,0.55,0.49,0.46,0.43,0.53,0.40,0.51,0.73,0.54,,0.49
R$_{21}^\mathrm{num. 50\%}$,0.61,0.57,0.55,0.54,0.65,0.47,0.61,0.84,0.64,,0.63
R$_{21}^\mathrm{num. 84\%}$,0.70,0.65,0.68,0.66,0.71,0.54,0.75,1.06,0.74,,0.82
R$_{21}^\mathrm{num. 95\%}$,0.79,0.72,0.82,0.73,0.77,0.59,0.97,1.37,0.85,,1.01
\end{filecontents*}
\csvreader[head to column names, late after line=\\]{Tables/table_norm_mean.csv}{}
 {\galaxy&\ngcone &\ngctwo&\ngcthre&\ngcfour&\ngcfive&\ngcsix & \ngcseven & \ngceight & \ngcnine &\average&\los}
 \hline
 \begin{filecontents*}{Tables/los_table.csv}
galaxy,ngcone,ngctwo,ngcthre,ngcfour,ngcfive,ngcsix,ngcseven,ngceight,ngcnine,all
N$_{\rm l.o.s}^{\rm all}$,328,319,741,239,288,318,410,947,1824,5414
N$_{\rm l.o.s}^{>3\sigma}$,292,188,302,200,207,304,370,705,884,3452
N$_{\rm l.o.s}^{>10\sigma}$,153,101,65,136,95,162,201,475,321,1709
\end{filecontents*}
\csvreader[head to column names, late after line=\\]{Tables/los_table.csv}{}
 {\galaxy&\ngcone &\ngctwo&\ngcthre&\ngcfour&\ngcfive&\ngcsix & \ngcseven & \ngceight & \ngcnine &\all&}
 \hline
 \end{tabular}
 \end{center}
\end{table*}
\endgroup
\begin{figure*}
	\begin{center}
	   \includegraphics[width=\textwidth]{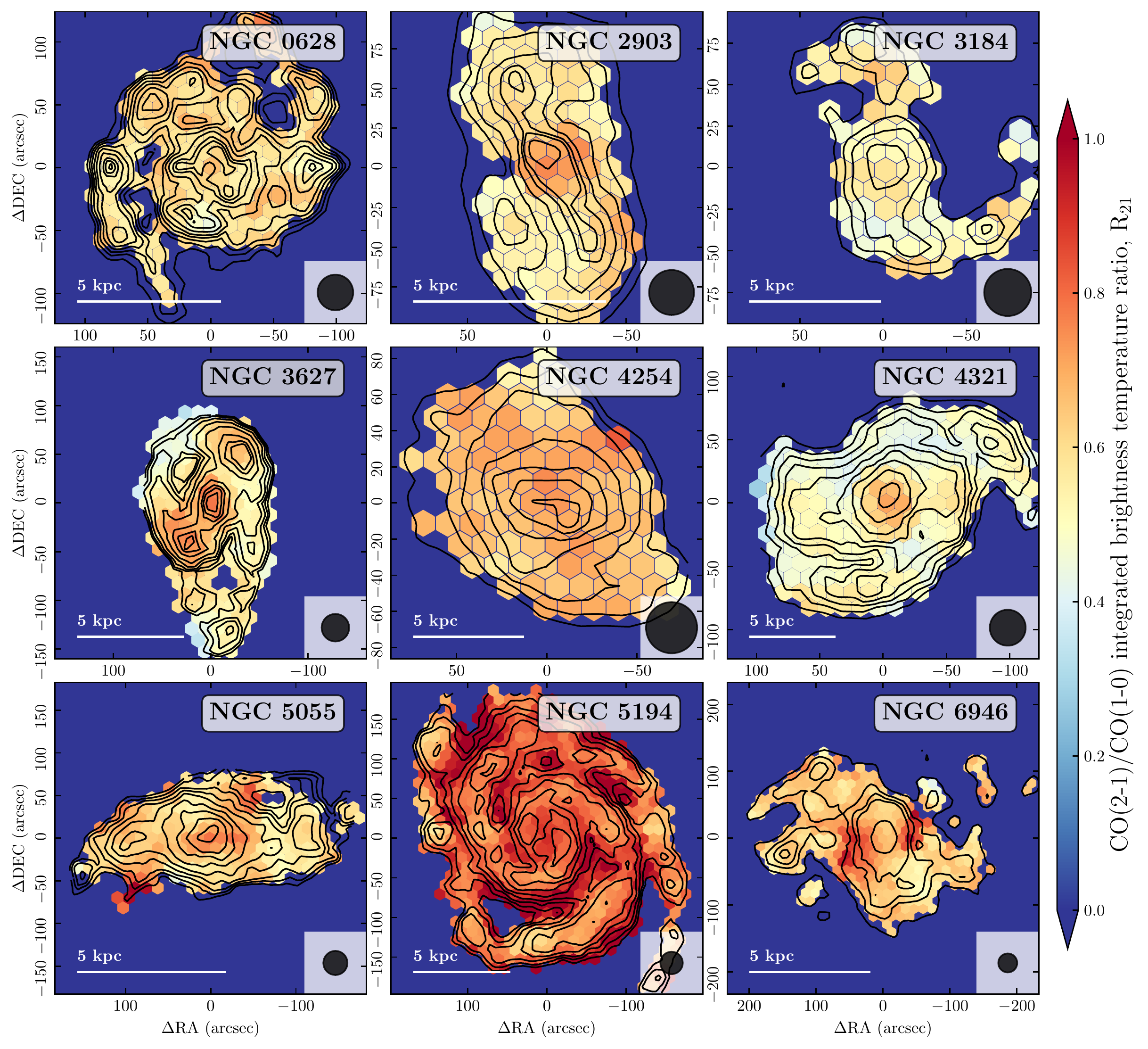}
	\end{center}
\caption{%{\bf[Updated by Ash - main text needs updating?]} 
\textbf{Maps of the CO(2-1)/CO(1-0) integrated brightness temperature ratio, log($R_{21}$), for nine nearby star-forming spiral galaxies.} These maps show $R_{21}$ for all positions that have $\mathrm{S/N} >10$ integrated brightness temperature in both the CO(2-1) and CO(1-0) maps. The overlaid contours show the CO(1-0) integrated brightness temperature, with levels showing 20, 30, 40, 50, 60, 70, 80, 90, 95, 97.5, and 99.5\% of the peak value in the map (see Figure\,\ref{fig:all_maps1_appendix}). The white scale-bar in the bottom left corner of each panel shows a linear scale of 5\,kpc, without accounting for inclination, at the distance of each source (see Table\,\ref{tab:empire_gal}). { The black circle in the bottom right corner indicates the beam size of 27$\arcsec$. Note that the hexagonal grid shows points critically sampling the beam, i.e., adjacent points are spaced by one-half the beam size.} }
\label{fig:R21_maps}
\end{figure*}

We estimate the line ratio, $R_{21}$, for each line of sight that has a measured brightness temperature for both CO(2-1) and CO(1-0) lines. In total, this yields { 5,416} measurements across nine galaxies at~ \mbox{$27''~(\sim 1-2$~kpc)} resolution. 

Figure\,\ref{fig:hist_R} shows the $R_{21}$ distribution of all individual lines of sight for each galaxy, as well as a histogram of the combined $R_{21}$ distribution for the entire sample. These histograms visualize results only for lines of sight with $\mathrm{S/N}>3$ in both lines. We do not find many cases where only one of the two lines is detected{, highlighting that the two lines follow similar distributions and the data sets are well-matched.} { The individual sightlines are arranged in a hexagonal grid (see Figure \ref{fig:R21_maps}), where the points have a half beam separation distance. Consequently, the beam size is oversampled by 4 hexagonal grid pixels.}

Table\,\ref{tab:med_ratio} reports the luminosity-weighted mean value for each galaxy, as well as the 16th/84th, and 5th/95th percentiles. Here ``luminosity-weighted'' means { averaging over the individual $R_{21}$ values} weighted by the { corresponding} CO~(1-0) brightness temperature. We prefer to use these intensity-weighted values for our quantitative results because they map straightforwardly to the results expected from galaxy-integrated measurements.

For individual galaxies, we find luminosity-weighted mean $R_{21}$ ratios ranging between ${\sim} 0.51$ and $0.87$.  In our view, the best characteristic sample-wide value for $R_{21}$ is the mean of the luminosity-weighted mean ratios for the individual galaxies. This is $\langle R_{21}^{\rm mean}\rangle=\RCO$\ with $0.10$ rms scatter from galaxy to galaxy. The uncertainty is the standard deviation between the galaxies. The value of the ratio agrees well with previous measurements of a wider population of galaxies, which tend to lie in the range $0.5{-}0.8$ (Section \ref{sec:intro} and \ref{discussion}).
%The mean of the luminosity-weighted mean line ratio of the individual galaxies. 
We verified that no significant effects are found when different weighting schemes are used. 

%Our slightly lower value, however, may reflect that previous measurements tended to focus on brighter, high excitation galaxies or the inner parts of galaxies. 

In principle our choice of method could affect our derived mean $R_{21}$ if, e.g., a few very bright regions show different $R_{21}$ than the rest of the galaxy or there is a large diffuse component with different $R_{21}$. The small differences among different approaches in Figure \ref{fig:hist_R} and Table \ref{tab:med_ratio} show that this is mostly not the case for our sample. The galaxy-wide mean and intensity-weighted mean show good agreement for most galaxies. Moreover, we find an average ratio of $\langle R_{21}^{\rm mean, norm}\rangle=0.62$ and a standard deviation of $0.12$ when weighting all lines of sight equally, compared to $\langle R_{21}^{\rm mean}\rangle=\RCO$\ when weighting by the luminosity-weighted mean of each galaxy.

Figure \ref{fig:R21_maps} shows the maps of the distribution of the CO line brightness temperature ratio $R_{21}$ across the individual galaxies. We do find evidence for both radial and azimuthal variations. We explore the systematic variation of the ratio within and between individual galaxies in the following sections.

\subsection{Radial variations of \texorpdfstring{$R_{21}$}{Lg}}

\begin{figure*}
\includegraphics[width=1\textwidth, height=18cm]{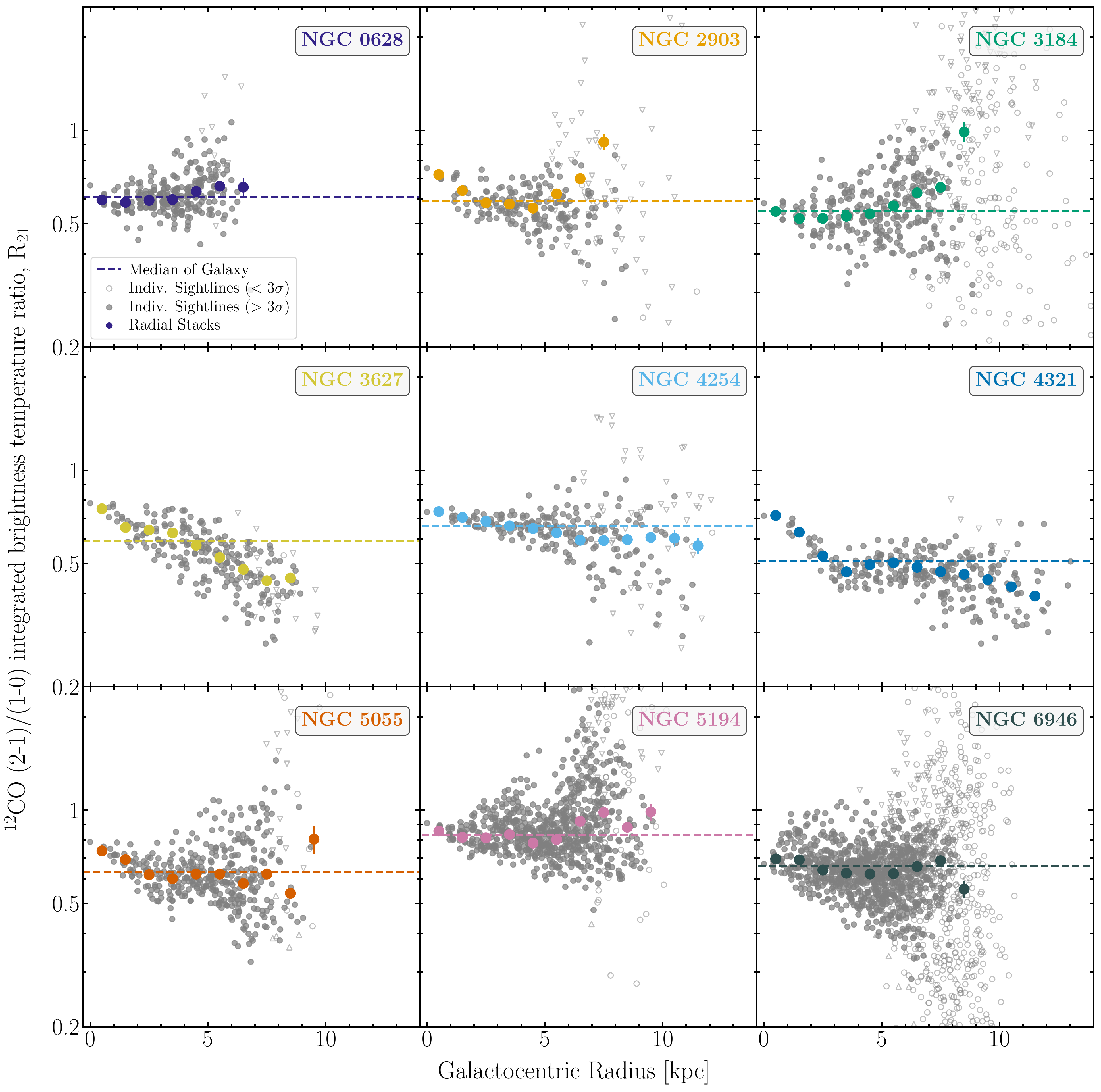}
\caption{\textbf{Radial profiles of the $R_{21}$ ratio.} \textbf{The gray points correspond to individual sightlines with adjacent lines of sight spaced by one half the beam width.} { The line ratio is plotted on a logarithmic scale.} Filled circles indicate data points where both lines have a signal to noise ratio above 3. Data with lower signal to noise appear either as upper or lower limits. Open triangles show where one line is below S/N$=$3. Open circles show points where both lines are below the S/N threshold of 3. {Coloured circles present the stacked values of the line ratio calculated in 27$\arcsec$-wide radial bins following the method described in in Section \ref{sec:Physical}.} The galaxy-wide median value for each individual galaxy appears as a colored dashed line. Six of our nine targets show clear central enhancements in $R_{21}$, but otherwise the stacked profiles show relatively small deviation from the galaxy-wide median.
}
\label{fig:radial}
\end{figure*}

Many quantities, including the star formation rate, molecular gas fraction, and gas density vary as a function of galactocentric radius. In Figure\,\ref{fig:radial}, we present radial profiles of $R_{21}$ for our sample. We plot all individual lines of sight at $27\arcsec$ (${\sim} 1-2$ kpc) {resolution. Recall that for these data, adjacent points are spaced by one-half beam so that the points are not independent}. Filled points show measurements with a signal to noise ratio, $\mathrm{S/N} >3$, on the line ratio, propagated. Open symbols indicate measurements with $\mathrm{S/N} < 3$.

\begin{figure*}
	%\begin{center}
    %		\includegraphics[width=0.5\textwidth, height=9cm]{radial_stc_all.eps}%
	%	\includegraphics[width=0.5\textwidth, height=9cm]{radial_stc_all_norm.eps}
	%\end{center}
	\includegraphics[width=0.75\textwidth]{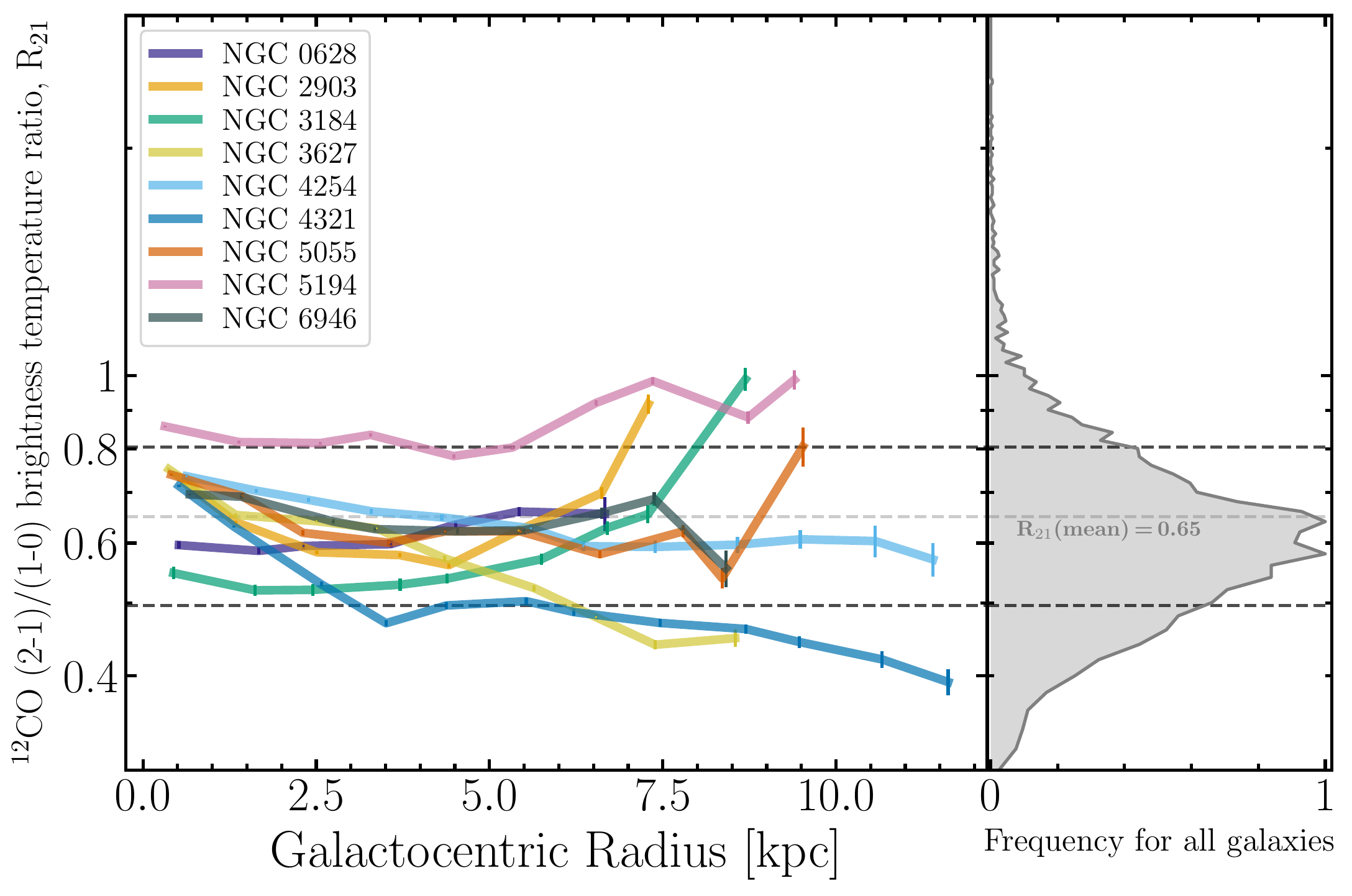}
\caption{\textbf{Stacked profiles of the CO line brightness temperature ratio, $R_{21}$, as a function of galactocentric radius, plotted using a $y$-axis logarithmic scale.} Stacked radial profiles use a bin width of 1.5 kpc. We only plot stacked bins where both CO(1-0) and CO(2-1) integrated, stacked lines are detected above a $\mathrm{S/N}>3$, but otherwise include all data. Error bars indicate the propagated uncertainty of the integrated brightness temperature of the stacked spectra. The histogram, reproduced from Figure \ref{fig:hist_R}, indicates the distribution of all lines of sight were both the CO(1-0) and CO(2-1) integrated line emissions are significantly ($\mathrm{S/N}>3$) detected. The light grey, dashed line indicates the sample-wide mean ratio when considering all lines-of-sight with S/N>3. The darker, dashed lines indicate the 1$\sigma$ scatter over all sightlines. 
The figure illustrates the galaxy to galaxy scatter that accounts for a large fraction of the overall variation of $R_{21}$ in the sample.}
\label{fig:radial_all}
\end{figure*}

Coloured points in Figure\,\ref{fig:radial} show radial profiles of azimuthal-averaged ratios, with error bars indicating the uncertainty on this mean $R_{21}$. 
%This criterion is used only for plotting purposes, since in our analysis we stack all lines of sight regardless. 
For these stacked profiles, we use all of the data, regardless of $\mathrm{S/N}$.
We plot all of the stacked profiles together in Figure\,\ref{fig:radial_all}. 

%We find a general agreement when we compare simple galaxy-wide median vs. a stacked value of the ratio per radial bin.

In Figures \ref{fig:radial} and \ref{fig:radial_all}, our azimuthally averaged measurements of $R_{21}$ show only a small dynamical range across individual galaxy discs for some cases while for other sources the ratio tends to show a negative or positive gradient. In Table \ref{tab:spearman_all}, we report the power-law fit  relating $R_{21}$ to galactocentric radius in each galaxy. We also report the $p$-value (of a linear relation in logarithmic space), which allows us to gauge the significance ($p$-value) of the radial  gradient.

%Given the very high signal to noise in our stacked measurements, these subtle radial trends are all highly significant (see Table \ref{tab:spearman_all}) but their magnitude is not large.

Six out of the nine galaxies show radial variations. Four show stronger radial variations. NGC~2903 shows an initial radial decline then a rise in $R_{21}$ with increasing radius. NGC~3184 shows increasing $R_{21}$ with increasing radius. And NGC 3627 and NGC 4321 show a strong while NGC 4254 shows a moderate decreasing trend as a function of the radius.

Figure\,\ref{fig:radial_all} also shows that in 6 of our 9 targets, $R_{21}$ appears higher in the central kpc than at intermediate radii, $\sim 1{-}6$~kpc. This central enhancement is most prominent in NGC 2903, NGC 3627, NGC 4321 and NGC 5055. Other galaxies, for example NGC 4254, show little or no central enhancement. 
%In NGC 0628 and NGC 3627, we observe a decrease rather than an upturn in the central bin. 
On average, the central $R_{21}$ bin ($0{-}1.5$ kpc) for our targets is 16$\%$ higher (median is 15\%) compared to the luminosity-weighted average of the rest of the galaxy. %This is smaller than the galaxy-to-galaxy offsets, so that in Figure \ref{fig:radial_all}, the main visible effect is galaxy-to-galaxy scatter, not the radial profiles.

Outside a galactocentric radius of about $6$~kpc, we find highly variable behaviour among our sample, with some galaxies showing increasing $R_{21}$, some showing decreasing $R_{21}$, and some being flat. Figure \ref{fig:radial} shows that these breaks in the profile often coincide with the emergence of a large amount of low signal to noise data. Using different techniques to bin the data and estimate the binned ratio yield large discrepancies. We are, therefore, hesitant to over-interpret them. Sensitive multi-line observations of outer discs will help illuminate whether CO excitation does change dramatically in the outer parts of disc galaxies.

% I moved commented text to here:

%Red points in Figure\,\ref{fig:radial} show radial profiles constructed using the spectral stacking method described by \citet{2011AJ....142...37S} and \citet{2017ApJ...836L..29J,Cormier2018}. We regrid each spectrum so that the local mean velocity now corresponds to $v=0$~km~s$^{-1}$. For this application, we use the velocity field derived from the CO(2-1) data to derive the reference velocity. After regridding the spectra, we average together all spectra in each $27\arcsec$-thick radial bin, corrected for inclination. Because the effects of the velocity field have been removed, the spectra average coherently. 

%The right panel of that figure shows the same profiles, but now normalizing each profile by its galaxy-wide median value, as stated in Table\,\ref{tab:med_ratio}. That is, the right panel removes the overall galaxy-to-galaxy variations in $R_{21}$ in order to highlight the radial trends.

%After removing galaxy-to-galaxy scatter in the right panel, we often see a mild increase in $R_{21}$ towards galaxy centers. 

%, larger that $10 \%$, including a sharp decrease outside 8 kpc. In contrast, NGC 0628, NGC 2903 and NGC 3184 each show a larger than $10 \%$ increase outside 5 kpc.

\subsection{Azimuthal Variations of \texorpdfstring{$R_{21}$}{Lg}} 
\label{variations}

Figure\,\ref{fig:R21_maps} also shows variation in $R_{21}$ at fixed galactocentric radius. The star formation rate surface density and gas column density also vary azimuthally, with the most striking features due to the influence of spiral arms and bars. {NGC 3627, 5194, NGC6946, and to some extent NGC 2903, 4321 and 5055 show clear spatial variations in the CO line ratio. NGC 5194 shows higher $R_{21}$ ratios in the interarm. We note, that this stands in contrast to previous findings. \citet[]{Koda2012} found the a higher line ratio in the arm region as opposed to the interarm region in this galaxy. For NGC 3627, we find a higher line ratio in the center and at the bar ends. NGC 6946 shows regions with enhanced line ratio towards the east and west of the center. NGC 2903, 4321 and 5055 all show an increase of the CO line ratio in the central $\sim 1-2$\,kpc region. The other three sources do not show any clear spatial variations.}

At $27''$ resolution, our ability to distinguish arm and interarm regions is limited, especially in the inner parts of galaxies where most of the molecular gas resides. The most straightforward imprint of azimuthal $R_{21}$ variations on our data is to increase the observed scatter in $R_{21}$ at fixed radius, e.g., as suggested in Figure \ref{fig:radial}.

\begin{figure}
    \centering
	\includegraphics[width=\columnwidth]{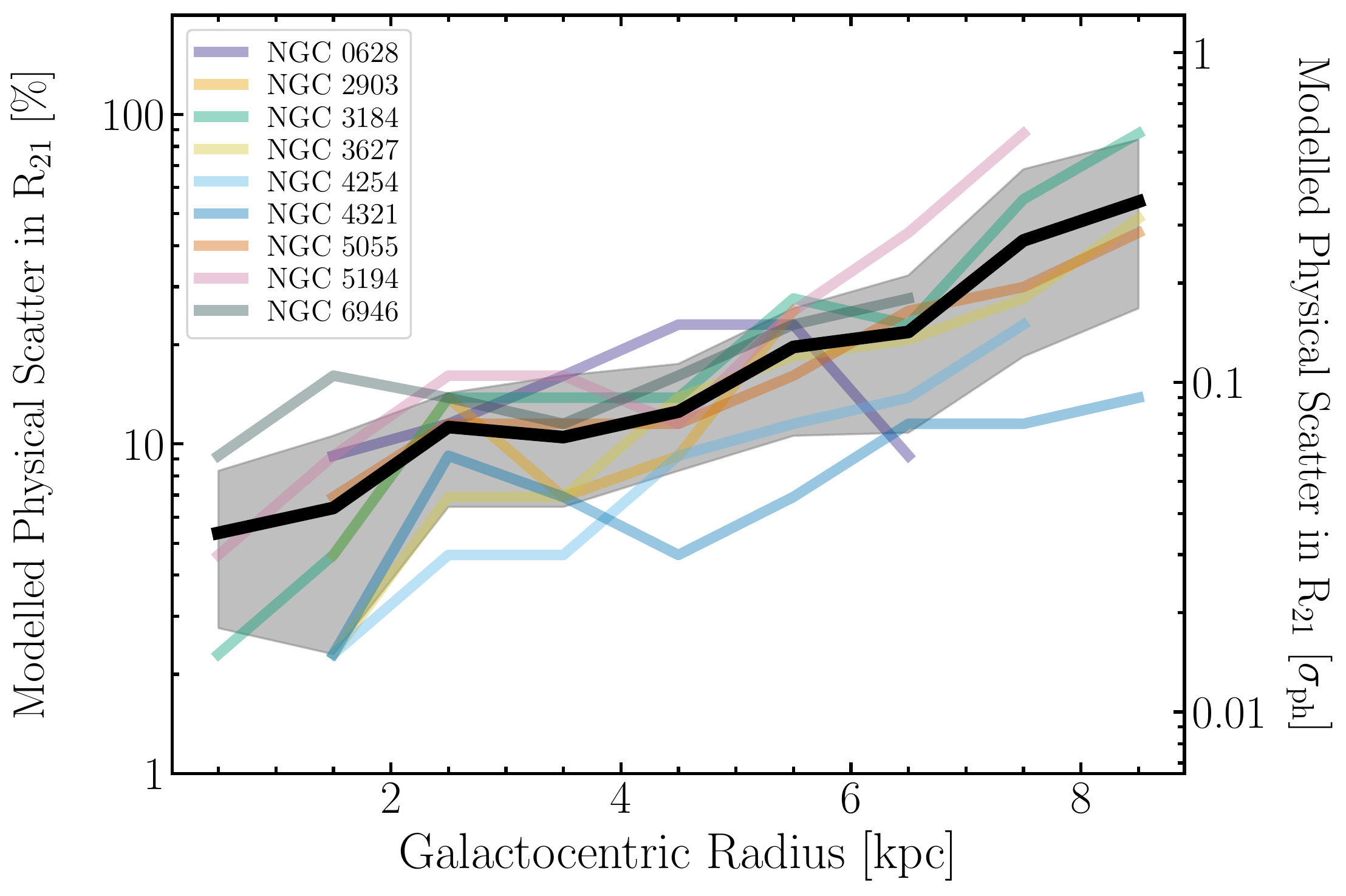}
    \caption{\textbf{Inferred physical scatter in the $R_{21}$ ratio for each radial bin.} The plot shows the physical scatter in each radial bin inferred from our modelling. We have already removed the effects of observational noise from the plotted scatter via a forward-modeling Monte Carlo calculation. The left $y$-axis indicates the scatter in percentage of the line ratio, while the right $y$-axis describes the scatter's actual value (converting using a fixed $R_{21}=0.65$). Colored lines show individual galaxies. The black line and gray region show the median and $\pm1\sigma$ range combining all galaxies. We observe increasing scatter in $R_{21}$ towards large galactocentric radii, with scatter ${\gtrsim}$ 20\% typical at radii $>6$~kpc. This increase in scatter may reflect a large variation in physical conditions at large radius or the inability to separate physically distinct regions at the $27''$ ($\sim1-2$\, kpc) resolution of our measurements.}
    \label{fig:az_model}
\end{figure}

%\begin{figure}
%    \centering
%	\includegraphics[width=\columnwidth]{var_model.eps}
%    \caption{Scatter of the $R_{21}$ ratio per radial bin, that best %describes our data, provided from our modelling, after 1000 %iterations. The value of scatter presented is disentangled from %the observational noise, that is calculated and accounted for, for %every radial bin. We observe an increase in the value of scatter %towards the edges of the galaxies, that seems to peak at ${\sim}$ %0.1 dex.}
%    \label{fig:az_model}
%\end{figure}

To quantify these azimuthal variations in $R_{21}$, we measure the scatter in the ratio at a fixed galactocentric radius. We disentangle the physical variation from scatter due to observational noise using a forward modelling process, which we describe in Appendix \ref{ap:forward_model}. Briefly, we use a Monte Carlo approach and the known observational errors to determine how much physical variation must be present in each radial bin to match the observed scatter. We plot the results of this calculation in Figure \ref{fig:az_model}, { where we repeat the modelling routine for every individual galaxy}. We show best-fit physical variation in $R_{21}$ as a function of radius for each galaxy. 

On average, the intrinsic scatter rises from $\lesssim 10\%$ in the inner bins to $\gtrsim 20\%$ outside a galactocentric radius of $6$~kpc. This has roughly the same magnitude as the observed galaxy-to-galaxy scatter. Note, however, that we do not expect calibration uncertainties to play as large a role in the scatter observed \textit{within} a galaxy. Taking this into account the physical scatter within galaxies may be larger than the physical scatter among galaxies. Also note that our azimuthal scatter calculations consider each pixel equally. A luminosity-weighted calculation would suppress faint regions and lower the magnitude of the measured scatter. A reasonable overall conclusion from this is that in our data, point-to-point scatter has a magnitude greater than or equal to galaxy-to-galaxy variations, and appears stronger than radial variations.

We also applied the same Monte Carlo based analysis on the complete data of all lines of sights as a whole. This way we can estimate the overall physical scatter. The physical scatter estimated after accounting for different calibration uncertainties for the individual instruments is about 8\%. 

%, with the exception of $ {\sim} 2{-}3$ galaxies where large systematic deviations are found at large galactic radii.

%Generally, the azimuthal scatter in $R_{21}$ approaches or remains slightly below the galaxy-to-galaxy scatter. Both have an rms magnitude of about 20\%. 

%We also want to point out that we do not find any meaningful correlations between the physical scatter derived by our modeling and the distance of the galaxy. We believe that the lack of such a correlation is due to the fact that our modeling takes the observational noise of each galaxy into consideration.

%In \autoref{fig:az_model} the intrinsic scatter tends to increase with increasing galactocentric radius, typically peaking at ${\sim} 0.07{-}0.1$ dex at ${\sim} 5-10$ kpc. 

{ As stated before, for NGC 5194 (M51)} we measure spatial variations {that have the opposite sense of} those reported by \citet[]{Koda2012}. { 
%We find values in the arm region that mostly agree with those in \citet{Koda2012}. However, while they reported lower ratio values in the interarm region (often $0.4 {-} 0.6$), we find significantly larger values \mbox{($\sim 0.9-1$)}.
In the arm region, we find a $R_{21}$ value ${\sim} 0.8$, which is in agreement with the value found by \cite{Koda2012} in the arm region. However, in the interarm region, we find larger values ($R_{21}{\sim}0.9{-}1$), while they find lower values ($R_{21}{\sim}0.4{-}0.6$).
We examine in detail and discuss possible causes for the difference in Appendix \ref{app:comp_NRO_EMIR}, and find the disagreement to stem from differences in the CO~(1-0) maps used for the analysis. {{In particular, the NRO map used by \citet{Koda2011} shows more emission in the interarm region than the IRAM 30m PAWS CO(1-0) map.}}} Beyond a galactocentric radius of 2~kpc, NGC~5194 shows the highest scatter in our sample and Figure \ref{fig:R21_maps} does show the strongest arm-interarm contrast. At our resolution, this contrast manifests as rms physical scatter of $15{-}20\%$ between $r_{\rm gal} = 2$ and 6~kpc. \citet{Koda2012} report a standard deviation of ${\sim} 0.1{-}0.15$ and mean $R_{21} \approx 0.7$, so our numbers for both the mean line ratio and scatter appear { to be overall slightly larger than theirs, which is mostly driven by the differing values within the interarm region.}

%Our relatively poor $27\arcsec$ ${\sim} 1.5$ kpc spatial resolution, compared to the $19.7 \arcsec$ one from \citet[]{Koda2012}, also likely plays a role in averaging out point-to-point scatter. 

As Figure \ref{fig:all_maps1_appendix} shows, our $27''$ resolution only coarsely resolves the dynamical features in our targets. Spiral arms and bars are visible at large radii in many targets. However we cannot distinguish the upstream and downstream sides of these features and they can be almost entirely suppressed in the inner galaxies. Physical conditions can vary dramatically across a spiral arm \citep[e.g.,][]{Schinnerer2010, Schinnerer2017}. Thus, we expect our large beam to blur together regions with a wide range of temperature and densities, especially in the inner parts of galaxies. This effect is expected to be even stronger, when a bright arm region lies next to a faint interarm region, the wider spacing between, e.g., arms and other discrete regions in the outer parts of galaxies may partially explain the increased scatter at large radii. Future work at higher physical scales offers the prospect to give much more insight on local variations of $R_{21}$.

\begin{table*}
    \caption{\textbf{Fits and correlation coefficients for individual galaxies.} Results from fitting a power law of form $R_{21,{\rm norm}}^{\rm fit} = C\cdot x^{m}$ to the stacked ratios in Figures \ref{fig:correlations_all} and \ref{fig:sfe}. The fit is performed as a linear fit in logarithmic space. We normalized the line ratio by the luminosity weighted, galaxy wide mean (see Table \ref{tab:med_ratio}). The Pearson $p$-value indicates the significance of the linear correlation in logarithmic space.
    We only performed the fit, if the $p$ value is below 0.05. Furthermore, the Spearman's rank correlation coefficient, $r_s$, is given.  For CO(1-0) the fitting range was set to exclude points below $I_{\rm CO(1-0)}<10$ K~km~s$^{-1}$. A minus indicates that no fit could be made. For NGC 0628 \& NGC 3184, we do not have $I_{\rm CO(1-0)}>10$\,K\,km\,s$^{-1}$ data and for NGC 2903 we do not have IR data from \textit{Herschel}.}
    \label{tab:spearman_all}
    \begin{center}
    \begin{tabular}{c c c c c c c c c c c} \hline \hline
    && NGC 0628 & NGC 2903 & NGC 3184 & NGC 3627 & NGC 4254 & NGC 4321 & NGC 5055 & NGC 5194 & NGC 6946 \\ \hline
    \multirow{ 4}{*}{$\begin{matrix} \text{CO(1-0)} \\ \text{[K km/s]} \end{matrix}$}& m & -- & 0.17&-- &0.23 &0.73 &0.18 &0.16 & -- & -- \\
    & C & -- & 0.61 & -- &0.46 &0.82 & 0.62 & 0.62 & -- & -- \\
    & p & -- & 0.048 & -- & 0.030 & 0.035& $6.4\times 10^{-4}$ &0.026 & 0.26& 0.24\\
    & $r_s$ & -- & 1.0 & -- & 1.0 & 1.0 & 1.0 &1.0 & 0.60& 0.80\\\hline
    \multirow{ 4}{*}{$\begin{matrix} \text{CO(2-1)} \\ \text{[K km/s]} \end{matrix}$}& m & -0.0027& 0.14&0.13 &0.24 &0.14 &0.17 &0.076 & -- & -- \\
    & C &1.0 & 0.70 &0.86 &0.56 &0.74 &0.69 &0.82 & -- & -- \\
    & p &0.0 &0.0050 &0.0 &0.0016 &0.034 &0.0053 &0.029 & 0.08 &0.06 \\
    & $r_s$ &-1.0 &1.0 &1.0 &1.0 & 1.0 &1.0 &0.9 & 0.68 &0.77 \\\hline
    \multirow{ 4}{*}{PACS 70/160}& m & 0.15 &-- &0.19 &0.43 &-- &0.40 &0.21 &0.20 &0.20 \\
    & C &1.2 &-- &1.3 &1.5 &-- &1.6 &1.3 &1.2 &1.2 \\
    & p &0.039 &-- &0.047 &$2.8\times10^{-6}$&0.07 &0.00033 & 0.0083 &0.00046 &0.03 \\
    & $r_s$ &0.9 &-- &1.0 &1.0&1.0 &1.0 & 0.82 &1.0 &0.83 \\\hline
    \multirow{ 4}{*}{$\begin{matrix} \Sigma_{\rm TIR} \\ [{\rm W\,kpc^{-1}}] \end{matrix}$}& m &  --  &-- & -- &0.20 &0.080 &0.20 &0.082 &0.020 &0.1 \\
    & C & -- & --&  -- & $6.7\times 10^{-8}$&0.0016 &$1.1\times10^{-7}$ &0.0014 &0.20 &0.00025 \\
    & p &0.40 &--&0.27 &0.00082 &0.034 &0.0023 &0.0073 &0.037 &0.026  \\
    & $r_s$ &0.4 &--&0.8 &1.0 &1.0 &1.0 &0.94 &0.86 &0.94  \\\hline
    \multirow{ 4}{*}{$\begin{matrix} \Sigma_{\rm TIR}/{\rm CO(1-0)} \\ [({\rm W\,kpc^{-1}})/({\rm K\,km/s})] \end{matrix}$}& m & 0.24&-- &0.2 & -- &0.11 & -- & -- &0.17 & --   \\
    & C & -- & -- & $9.8\times10^{-8}$ & -- & 0.00021 & -- & -- &$1.3\times10^{-6}$ & -- \\
    & p &0.15 &--&0.0 &0.70 &0.0 &0.20 &0.13 &0.011 &0.088  \\
    & $r_s$ &1.0 &--&1.0 &1.0 &1.0 &1.0 &1.0 &1.0 &0.8  \\\hline
    \multirow{ 4}{*}{$\begin{matrix} \Sigma_{\rm TIR}/{\rm CO(2-1)} \\ [({\rm W\,kpc^{-1}})/({\rm K\,km/s})] \end{matrix}$}& m & -- &-- &-0.063 &0.12 &-0.053& -- &-- & -- & -- \\
    & C & -- & -- &$1.4\times10^2$ &$7.4\times10^{-5}$ &64 &-- &-- &-- &--   \\
    & p & 0.25&--&0.0 &0.0 &0.0 &0.60 &0.54 &0.24 &0.18  \\
    & $r_s$ & 1.0&--&-1.0 &1.0 &-1.0 &-0.50 &0.5 &1.0 &-1.0  \\\hline
    
    \end{tabular}
    \end{center}
\end{table*}

%\begin{table*}
%\caption{---}
%\label{tab:av_powerlaw_all}
 %\begin{center}
 %\begin{tabular}{lccccc}
 % \hline \hline
 % Galaxy & Radius & CO(1-0) Intensity & CO(2-1) Intensity & 70-to-160 $\mu$m ratio & TIR surface brightness\\
 % \hline

%    \hline \hline
% \end{tabular}
% \end{center}
%\end{table*}

\begin{figure*}
	\begin{center}
		\includegraphics[width=0.75\textwidth]{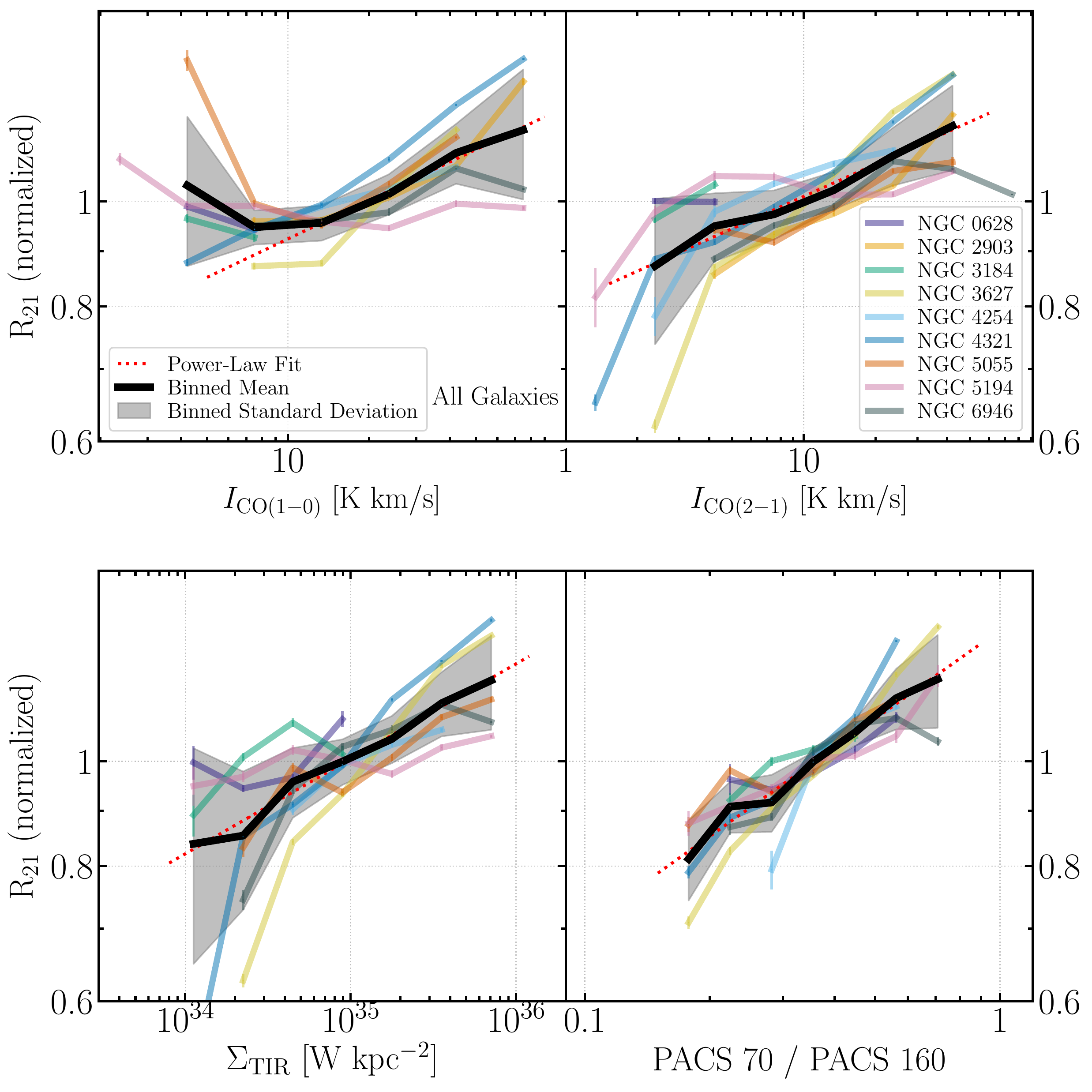}
	\end{center}
\caption{\textbf{Stacked measurements of $R_{21}$ on a logarithmic scale in bins of CO(1-0), CO(2-1) brightness temperature, TIR surface brightness, and IR color.} In each panel, we bin each galaxy by the quantity on the $x$-axis. Then, we measure stacked line ratios in each bin for each galaxy. For stacking {in bins of CO(1-0), CO(2-1) brightness temperature, TIR surface brightness, and IR color}, we only included lines of sight with a $\mathrm{S/N}>10$ for both CO(1-0) and CO(2-1) in order to reduce noise effects and make the trend more robust. The individual lines are normalized by the median stacked CO ratio value of the corresponding galaxy. The black line is the binned mean combining all galaxies. Only bins with at least a stacking result from three galaxies are included in the mean line. We observe positive correlations between $R_{21}$ and CO brightness temperature, IR surface brightness, and IR color. These have the sense that $R_{21}$ increases along with gas surface density, star formation activity, and dust temperature. The red dotted line indicates the power law fit. The fitting range and results are listed in Table \ref{tab:fitting}. We also report rank correlation coefficients in Table \ref{tab:spearman_all}.}
\label{fig:correlations_all}
\end{figure*}

\begin{figure*}
    \centering
    \includegraphics[width = \textwidth]{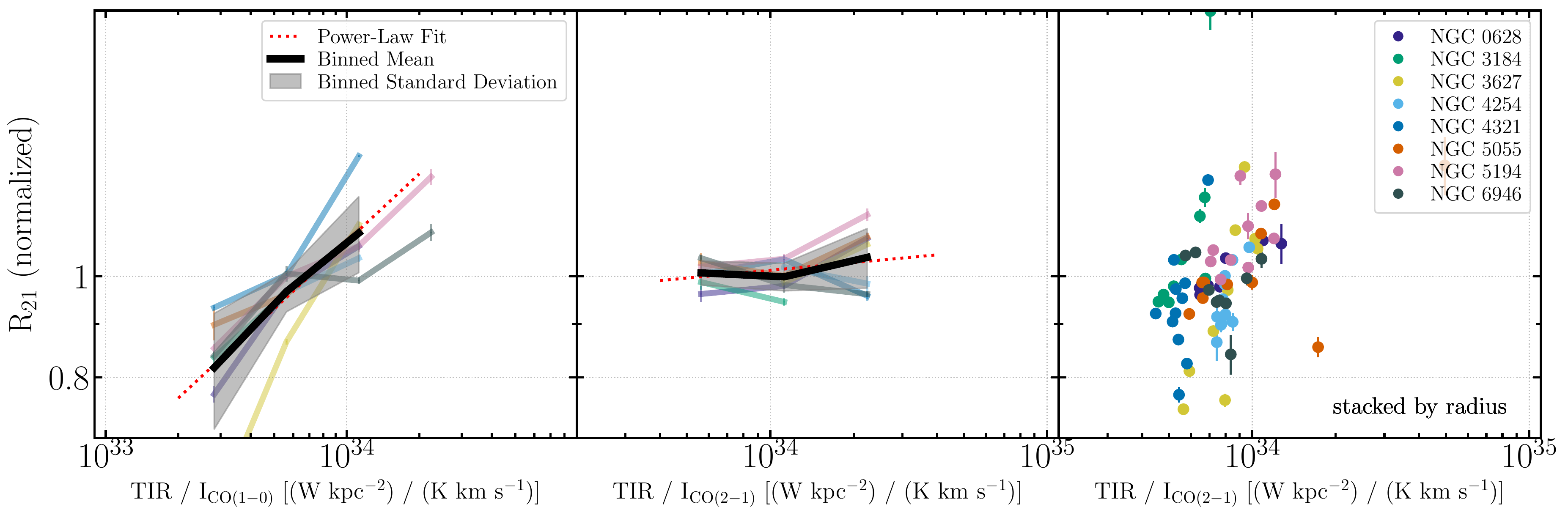}
    \caption{\textbf{Stacked measurements of the $R_{21}$ ratio as a function of TIR-to-CO ratios.} The TIR-to-CO ratio is closely related to the star formation rate per unit gas mass, a key figure for many studies using both lines. The figure shows three plots of stacked $R_{21}$ as a function of TIR-to-CO(1-0) and TIR-to-CO(2-1). In the left panel, we show $R_{21}$ as a function of TIR-to-CO(1-0), stacking by TIR-to-CO(1-0). In the middle panel, we show $R_{21}$ as a function of TIR-to-CO(2-1), stacking by TIR-to-CO(2-1). In both of these stacks the quantity being stacked correlates with the quantity used for stacking. However, the correlation with the CO(2-1) intgrated brightness temperature is much weaker. In the right panel, we show $R_{21}$ as a function of TIR-to-CO(2-1), but now stacked by radius, an independent quantity, to remove this bias in the stacks. The underlying trend appears to be a moderate positive correlation between $R_{21}$ and TIR-to-CO(2-1), consistent with the results in Figure \ref{fig:correlations_all}. The inner, high density parts of galaxies both show higher star formation per unit gas and higher excitation.}
    \label{fig:sfe}
\end{figure*}

%\begin{table}
% \caption{Spearman's rank correlation coefficients
%  with significance (quoted as a $p$-value in parenthesis) relating the stacked line ratio $R_{21}$ to stacked TIR-to-CO ratios. In the case that we only have two datapoints, we did not calculate the significance.}
% \label{tab:sfe}
% \begin{center}
 %\begin{tabular}{lccc}
 % \hline \hline
 % Galaxy & TIR/I\textsubscript{CO(1-0)} & TIR/I\textsubscript{CO(2-1)} & TIR/I\textsubscript{CO(1-0)} (Radial stacks)\\
 % \hline
 %   NGC 0628 & 1.00 (0.00) & -0.80 (0.20) & 0.93 (0.00)\\[2pt]
 %   NGC 3184 & 1.00 (0.00) & -0.50 (0.67) & 1.00 (0.00)\\[2pt]
 %   NGC 3627 & 1.00 (0.00) & -0.50 (0.67) & 0.57 (0.18)\\[2pt]
 %   NGC 4254 & 1.00 (0.00) & -1.00 (0.00) & -0.43 (0.40)\\[2pt]
 %   NGC 4321 & 1.00 (0.00) & -1.00 (0.00) & 0.00 (1.00)\\[2pt]
 %   NGC 5055 & 1.00 (0.00) & -0.50 (0.67) & 0.70 (0.04)\\[2pt]
 %   NGC 5194 & 1.00 (0.00) & -1.00 (0.00) & 0.34 (0.28)\\[2pt]
 %   NGC 6946 & 1.00 (0.00) & 0.37 (0.47) & -0.24 (0.46)\\[2pt] 
 %   \hline \hline
 %\end{tabular}
% \begin{tabular}{l c c c} \hline \hline
%         Galaxy & TIR/I$_{\rm CO(1-0)}$  & TIR/I$_{\rm CO(2-1)}$ & TIR/I$_{\rm CO(2-1)}$ (Rad. Stacks) \\ \hline
%         \csvreader[head to column names, late after line=\\]{Tables/spearman.csv}{}
%        { \galaxy & \TIRcoone & \TIRcotwo & \TIRovercoone}
%        \hline
%    \end{tabular}
% \end{center}
%\end{table}

\begin{table}
 \caption{\textbf{Fitting results for stacked, normalized profiles combining all galaxies.} Results from fitting a power law of form $R_{21,{\rm norm}}^{\rm fit} = C\cdot x^{m}$ to the mean of the stacked, normalized quantities in Figures \ref{fig:correlations_all} and \ref{fig:sfe} (black line in figure). The fitting range indicates the range along the $x$-axis over which the fit is performed.}
 \label{tab:fitting}
 \begin{center}
 \begin{tabular}{l c c c c} \hline \hline
         Parameter & Fitting range &C & m &$r_s$ \\
         ($x$-axis) & (units of param.)  & &  \\\hline
         CO(1-0) & $10-75$&0.70 & 0.12&1.0\\
         CO(2-1) &$2-43$& 0.81 & 0.099&1.0\\
         PACS 70/ PACS 160 &$0.2-0.7$& 1.3 & 0.27&1.0\\
          $\Sigma_{\rm TIR}$&$10^{34}-10^{35.85}$& 8.5$\times 10^{-4}$ & 0.088& 1.0\\
         $\Sigma_{\rm TIR}$/CO(1-0) &$10^{33.45}-10^{34.05}$&5.2 $\times10^{-8}$ & 0.20 &1.0\\
         $\Sigma_{\rm TIR}$/CO(2-1) &$10^{33.75}-10^{34.35}$& 0.15 & 0.025&0.5\\ 
        \hline
    \end{tabular}
 \end{center}
\end{table}

\subsection{Correlations with CO brightness temperature and IR emission} \label{correlations}

We also compare $R_{21}$ to the local intensities of \mbox{CO(1-0)} and \mbox{CO(2-1)} emission, the local TIR surface brightness, and the local 70-to-160\,$\mu$m ratio. These observed quantities { indirectly trace physical conditions that should affect CO excitation}, so that this analysis can highlight the physical drivers of the $R_{21}$ variations observed in the previous two sections. 

{ We compare to these specific quantities because they are directly observable and also indirectly related to conditions which we expect to affect excitation. Though we observe at coarse physical resolution, we expect that CO(1-0) and CO(2-1) emission trace the molecular gas surface density and more indirectly trace gas volume density. High gas densities will be associated with thermalization and a higher $R_{21}$. The IR color traces the dust temperature, which in turn is set by strength of the interstellar radiation field \citep{Drain2011}. The radiation field also illuminates photon-dominated regions and should play a key role in heating the gas. All other things equal, we expect warmer gas to be more nearly thermalized. Along similar lines, the TIR surface brightness indicates the level of star formation activity. We expect that this indirectly relates to both the heating of the gas and the gas density, with denser gas forming more stars, on average.}

%While the exact interplay between the different physical quantities is complex, providing comparisons between the line ratio $R_{21}$ and a single physical parameter will potentially reveal correlations to first order.} These {\bf physical quantities} are also observable quantities, and these comparisons could help to predict $R_{21}$ variations in other surveys.

Figure\,\ref{fig:correlations_all} shows the mean normalized $R_{21}$ (normalized with respect to the galaxy-internal luminosity-weighted mean; see Table \ref{tab:med_ratio}) calculated from spectral stacks as a function of each quantity of interest. We show results for each galaxy separately and show results stacking the data in bins of \mbox{CO(1-0)} brightness temperature, \mbox{CO(2-1)} brightness temperature, TIR surface density, and 70-to-160\,$\mu$m ratio. Table~\ref{tab:spearman_all} reports the results form fitting a power law relating $R_{21}$ to each quantity for all individual points. In addition, the $p$-value is indicated for the correlation in log-space, describing the tightness of the correlation. Finally, we also report the Spearman's rank correlation coefficient, $r_s$. For the stacking, we only included sightlines with $\mathrm{S/N}>10$ in both the CO(1-0) and CO(2-1) data to make sure that the trends in the ratio are not noise dominated. The black line indicates the binned mean line ratio and the grey band describes the binned standard deviation. Table \ref{tab:fitting} lists the results from fitting a power law to the binned mean line ratio indicating the strength of the trend.

%We also report the average power slope, derived by fitting lines to all individual lines of sight, in the log-log space, in order to better highlight the general trends between the $R_{21}$ ratio and the physical quantities, we test it against.

%for the stacked data. \autoref{tab:spearman_all} reports the rank correlation relating $R_{21}$ to the same quantities but now using each individual data point.

%Galaxy-to-galaxy variations still appear prominent in Figure\,\ref{fig:correlations_all}. In order to check the degree to which variations in $R_{21}$ internal to each galaxy may be driven by these quantities, we normalize $R_{21}$ by the galaxy-wide median for each galaxy. However the scatter at fixed value is not significantly smaller than when looking at all $R_{21}$ values together, with no meaningful differences in the observed trends.

%In \autoref{fig:correlations_all_norm} we normalize $R_{21}$ by the galaxy-wide median. This highlights the degree to which variations in $R_{21}$ internal to each galaxy may be driven by these quantities.

\textbf{CO(1-0) brightness temperature:} The top left panel of Figure\,\ref{fig:correlations_all} shows the stacked, normalized $R_{21}$ as a function of CO(1-0) brightness temperature. The bin width of the stacks is set to 0.25 in logarithmic space. %Though the stacks have high signal to noise, we use the individual data to sort and bin the data. This, and the correlated axes of the plot, can lead to unreliable results at low intensity.
At higher intensities, we observe a tendency to find higher $R_{21}$ in the highest brightness temperature bins. The highest CO(1-0) brightness temperature almost always appears in the galaxy center, so this reflects the same central enhancements noted in the radial profiles. Overall, Figure \ref{fig:correlations_all} reveals a positive relationship between $R_{21}$ and CO(1-0) brightness temperature.

Because of the correlated axes, low signal to noise CO(1-0) measurements will lead to an artificial upturn at the low brightness temperature end driven by sorting predominantly noise measurements, as it can be seen in the top left panel. Supporting this view, no such feature appears in the radial, CO(2-1), or IR intensity plots. 
%We therefore bin by CO(1-0) intensity only above 1\,K\,km/s in these two panels, where we expect the bins to be dominated by significant measurements.

%This upturn at low intensities in all galaxies is expected since these lower intensity bins become dominated by noise. We stack by CO(1-0) intensity and measure $R_{21} \propto 1/$CO(1-0). The stacking exercise does not remove the effect of correlated axes unless one stacks by a third, independent variable. Our best estimate, then is that the lowest bins in the top left part of Figures \ref{fig:correlations_all} and \ref{fig:correlations_all_norm} reflects signal to noise effects. 

\textbf{CO(2-1) brightness temperature:} The top right panel of Figure \ref{fig:correlations_all} shows stacked $R_{21}$ as a function of CO(2-1) brightness temperature. We also chose a bin width of 0.25 in logarithmic space. As with CO(1-0), we observe a positive correlation between $R_{21}$ and CO(2-1).  In general, we tend to find moderately higher $R_{21}$ in high brightness temperature regions. %NGC 628 represents a clear outlier from this trend, showing a declining $R_{21}$ with increasing CO(2-1) intensity.

%Similar to CO(1-0), we might expect a bias --- but this time towards a positive slope --- at the lowest intensities and therefore we choose to focus less on the lowest bin. 

\textbf{Total IR luminosity surface brightness:} The bottom left panel of Figure \ref{tab:spearman_all} shows stacked $R_{21}$ as a function of TIR surface brightness. The bin width for the total infrared surface brightness is 0.3 in logarithmic space. TIR surface brightness traces embedded star formation activity and scales with molecular gas surface density, so we expect similar results to stacking by CO brightness temperature. Again, we observe a positive correlation where the IR-bright parts of our sample show moderately higher $R_{21}$. Because TIR tends to be measured at high signal to noise and represents an independent quantity from CO(2-1) and CO(1-0) this correlation spans a larger dynamic range than the CO(1-0) and CO(2-1)-based stacks and should be less subject to systematics. As in the previous panels, we observe a positive correlation between TIR surface brightness and $R_{21}$. In regions with more star formation per unit area, $R_{21}$ tends to be higher.

\textbf{70-to-160\,$\mu$m ratio:} In the bottom right panel of Figures \ref{fig:correlations_all}, we plot $R_{21}$ stacked as a function of IR colour. The stacks have a bin with of 0.1 in logarithmic space. IR colour traces dust temperature and the interstellar radiation field. The axes are not correlated, though the lowest bin may again suffer from some sampling and signal to noise concerns.

As above, we find a positive correlation between 70-to-160 $\mu$m ratio and $R_{21}$. 
%The tendency is that stronger interstellar radiation fields correspond to moderately higher excitation for the low-$J$ CO. 
$R_{21}$ tends to be higher with stronger interstellar radiation field.
This fits with an overall pattern that systems with more intense star formation activity also tend to have higher dust temperatures, denser gas, and more nearly thermal excitation in their CO lines. A higher dust temperature does correspond to a higher $R_{21}$. %However, Figure\,\ref{fig:correlations_all} shows that dust temperature does not explain the offsets among our galaxies. Rather the 70-to-160 $\mu$m ratio appears to be a good predictor of the internal variations in $R_{21}$

%Notably, none of these quantities ``sorts'' our data in a way that explains the substantial galaxy-to-galaxy scatter in $R_{21}$. Instead, these trends can explain some of the internal variations observed as a function of radius and azimuth.

Taken together, the CO-bright, IR-bright, high 70-to-160\,$\mu$m ratio regions of our targets show moderately higher $R_{21}$ than the cooler, fainter regions. These trends appear significant, with most galaxies showing a trend with these external parameters. The overall magnitude of the trends is a $ {\sim} 20{-}30 {\%}$ change in the ratio across the sample. 

Perhaps surprisingly, these trends appear \textit{internal} to galaxies. They do not appear to explain the observed galaxy-to-galaxy offset in $R_{21}$. They can explain some of the internal radial and azimuthal variations observed. The residual galaxy-to-galaxy offsets must either be driven by different physics or be due to flux calibration uncertainties.

{ We also note, that while $R_{21}$ in NGC 5194 shows discrepancies in spatial variation with previous findings \citep{Koda2012}, the trends discussed in this subsection are actually in agreement. The discrepancy is mostly due to differences in the fainter, interarm region, thus an agreement in the trends with environmental parameters spanning the entire galactic radial range is not unexpected.}

\subsection{TIR-to-CO ratio and \texorpdfstring{$R_{21}$}{Lg}}

Many CO surveys over the last two decades have focused on measuring the gas depletion time, star formation scaling relations, or related quantities. In these studies, the figure of merit is often the star formation rate per unit molecular gas mass. Both CO(1-0) and CO(2-1) line emissions are commonly used to estimate the molecular gas mass. We use a simple observational proxy, the TIR-to-CO ratio, to explore how $R_{21}$ depends on the SFR per unit molecular gas. For the bin width, we chose 0.3 in logarithmic space.

In Figure\,\ref{fig:sfe}, we plot $R_{21}$ stacked by the TIR-to-CO(1-0) ratio, the TIR-to-CO(2-1) ratio, and galactocentric radius. We explore all three stacks because of the correlated nature of the axes. We might expect an artificial correlation between $R_{21}$ and TIR-to-CO(1-0) when stacking by TIR-to-CO(1-0). In noisy or scattered data, low CO(1-0) data points will scatter to both high $R_{21}$ and high TIR-to-CO(1-0) values, potentially creating an artificial correlation. A similar effect could introduce an artificial anti-correlation comparing $R_{21}$ to TIR-to-CO(2-1). Because the stacking approach uses values for individual data points to assign them to bins, it will not necessarily reduce this effect via averaging.

In the left panel of Figure\,\ref{fig:sfe}, we plot $R_{21}$ stacked by the TIR-to-CO(1-0) ratio. The profiles show a clear positive correlation between $R_{21}$ and TIR-to-CO(1-0) for all galaxies. Correlated axes could however be tilting the trend in this direction. 

In the central panel of Figure\,\ref{fig:sfe}, we instead plot $R_{21}$ as a function of TIR-to-CO(2-1), binned using the TIR-to-CO(2-1) ratio. That is, we change the line used for the stack. Again, the correlated axes potentially affect the stack, this time producing a mild correlation between $R_{21}$ and TIR-to-CO(2-1) for most galaxies.

With this in mind, the right panel of Figure\,\ref{fig:sfe} where $R_{21}$ is plotted as a function of the TIR-to-CO(2-1) ratio, i.e., the same axes as in the left panel, but now stacked by galactocentric radius. Radius represents an independent variable that should minimize bias in the stacks. This figure shows more scatter and a somewhat smaller dynamic range compared to the previous two stacks. When stacking by radius, there is an overall tendency for TIR-to-CO(2-1) and $R_{21}$ to be positively correlated. The behaviour is less universal than we saw when stacking by TIR-to-CO(2-1). %When stacking by radius, the galaxies show strong positive correlations, one a modest positive correlation, and four show insignificant correlations.

Overall, this result appears consistent with the results in the previous section. We tend to find high $R_{21}$, TIR-to-CO(1-0), 70-to-160\,$\mu$m ratio, and CO brightness temperature in the inner parts of galaxies.

%driven mostly by low $R_{21}$ and TIR-to-CO(1-0) values in NGC~3627 and high values in the inner part of NGC~5194 (M51). About half of our sample shows a weak internal correlation between TIR-to-CO(1-0) and $R_{21}$ when stacking by radius. 

%In the stacks, this amounts to a perfect rank correlation coefficient within each individual galaxy (see \autoref{tab:sfe})
%, showing a clear positive correlation between the two variables). 

%We might expect an artificial correlation between $R_{21}$ and TIR-to-CO(1-0) when stacking by TIR-to-CO(1-0). In noisy or scattered data, low CO(1-0) data points will scatter to both high $R_{21}$ and high TIR-to-CO(1-0).

%After accounting for correlated axes, \autoref{fig:sfe} suggests that $R_{21}$ does correlate with the local star formation per unit molecular gas. This correlation tends to be weak in magnitude inside of a galaxy, and has the same sense as the correlations with 70-to-160$\mu$m color, CO intensity, and IR surface brightness seen above. The inner regions of galaxies tend to show higher efficiencies (shorter gas depletion times) and also tend to show increased CO excitation.

\section{Discussion}
\label{discussion}
%Overall, we find $R_{21} \approx 0.66$ with almost not values above $1$. This implies that some fraction is not optically thick gas and shows some degree of subthermal excitation \citep[compare the simulations from, e.g.,][]{2017MNRAS.465.2277P}.

\subsection{Comparison to \textbf{Single Pointing} Literature Measurements}\label{sec:lit_research}

%Our targets are local, star-forming disc galaxies. They show a mean $R_{21} \approx 0.65$. 

\begin{figure}
		\includegraphics[width=1\columnwidth]{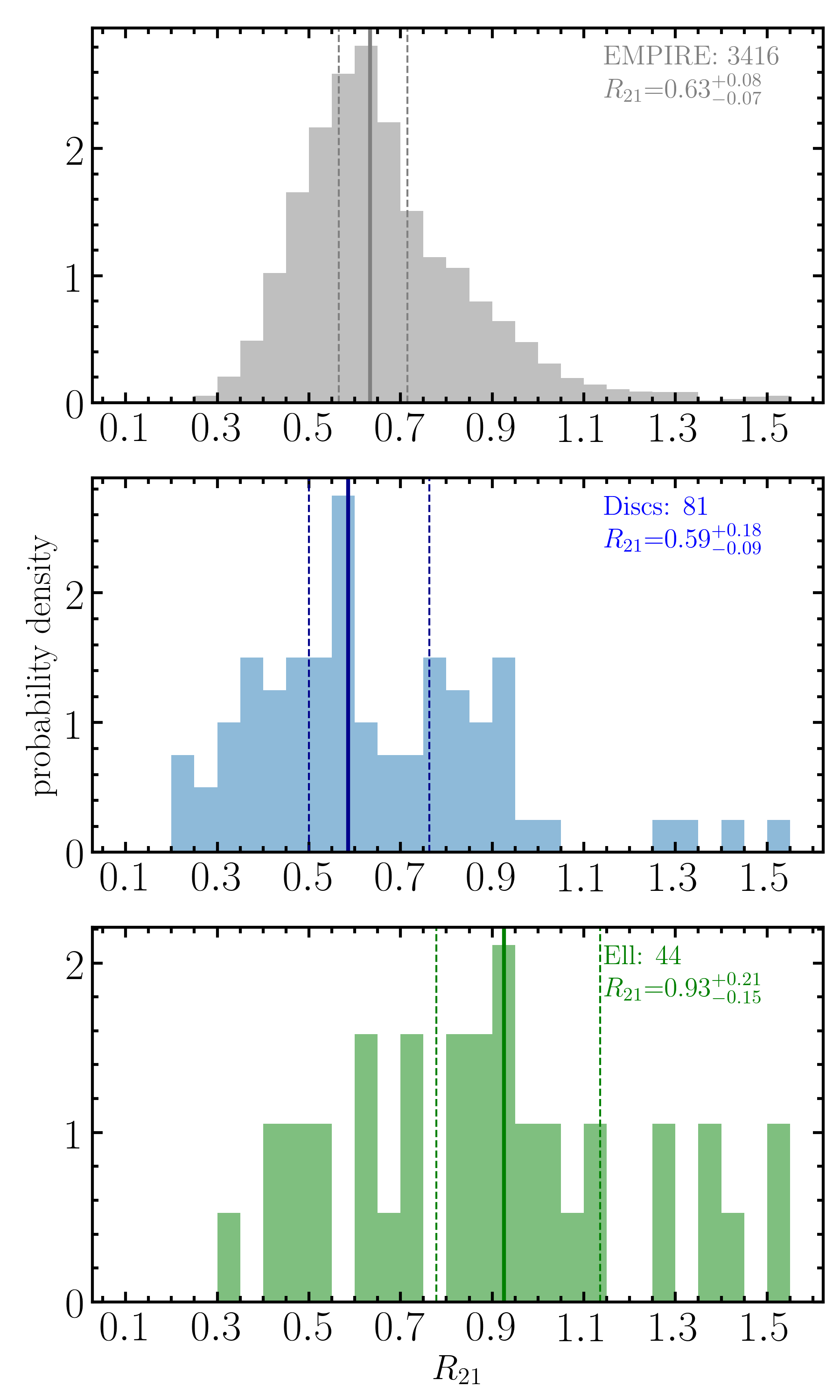}
\caption{\textbf{Distributions of $R_{21}$ from literature studies.} The distribution of $R_{21}$ for individual lines of sight in EMPIRE (top panel) and a compilation of $R_{21}$ estimates from the literature broken into late-type disc (middle panel) and early-type elliptical (bottom) galaxies
\citep[Data from][]{Albrecht2004,Albrecht2007,ATLAS3D,Boeker2003,Boselli1994,Chini1996,Combes2007,Curran2000,Evans2005,Lavezzi1999,Leon1998,Ocana-Flaquer2010,Strong2004,Wiklind1995}.}
\label{fig:co_archive}
\end{figure}

In order to compare our results to literature values, we compiled and homogenized {single-pointing} CO observations from a number of publications \citep{Boselli1994,Wiklind1995,Chini1996,Leon1998,Lavezzi1999,Curran2000,Boeker2003,Albrecht2004,Strong2004,Evans2005,Albrecht2007,Combes2007,Ocana-Flaquer2010,ATLAS3D}. To create a set of reference measurements:

\begin{enumerate}
\item We first tabulate CO(1-0) and CO(2-1) line brightness temperatures, errors, and beam sizes for 659 galaxies drawn from the references above. When necessary, we converted from antenna temperature scale to main beam temperature scale using the efficiencies provided in the respective publication. Note that we further limited our compilation of measurements using the criteria below. Our final number of reference galaxies in Figure\,\ref{fig:co_archive} is 125.

\item For each target, we obtained optical blue band 25\,mag\,arcsec$^{-2}$ isophotal diameters ($D_{25}$), axis-ratios from RC3 \citep{RC3}, and morphological types through the \textit{NASA Extragalactic Database} (NED). 

\item We filtered the data to only include significant detections ($\mathrm{S/N}>3$) where the beam covered an appreciable part of the galaxy. Some values in the literature are upper limits only, and we do not consider these here. We further require that the FWHM of the smaller beam, typically CO(2-1), covers at least 40\% of a CO scale length (see next point). This typically amounts to requiring that the CO beam covers at least $0.1~D_{25}$.

\item From $D_{25}$, we estimate a CO scale length ($r_{\rm CO}$), assuming: $r_{\rm CO}$=0.23\,D$_{25}$/2 \citep{Young1995,Leroy2008,Leroy2009,AMIGA,Davis2013a}. From this scale length and the known beam sizes, we assume an exponential disc and performed an aperture correction to estimate the full luminosity of the galaxy in each line (e.g., see \citealt{Puschnig2020}). Note that because we focus on the line ratio, the accuracy of the extrapolation to the full galaxy is not important. Only matching the effective area covered by the beams matter.

\item We calculate $R_{21}$ as the ratio of the estimated full-galaxy CO(2-1) luminosity to the full-galaxy CO(1-0) luminosity. After the cuts based on signal to noise and extent, this leaves us with $125$ measurements, $81$ for late-type (``disc'') galaxies and $44$ for early-type (``elliptical'') galaxies.

\end{enumerate}

Figure\,\ref{fig:co_archive} shows histograms of these literature $R_{21}$ measurements. In the figure, we divide the literature sample into disc-like and elliptical galaxies. For disc galaxies we find \mbox{$R_{21}=0.59^{+0.18}_{-0.09}$}. This agrees well with our results for the EMPIRE sample, though the literature distribution appears much broader. Some of this additional scatter likely reflects uncertainty in calibration. Anorther part of the scatter reflects the comparatively lower signal to noise of these data compared to EMPIRE. A full meta-analysis disentangling the sources of physical and observational scatter for the litterature measurements is beyond the scope of this paper. Here we emphasize the general good agreement between our smaller set of high quality, resolved measurements and the literature.

The lower panel in Figure \ref{fig:co_archive} shows that elliptical galaxies exhibit higher excitation. This might be expected if the deeper potential well leads to wider line profiles and thus lower optical depth. Regardless of the explanation, literature observations of early-type galaxies indicate higher apparent $R_{21}$ but also enormous scatter.

Recently, \citet{Saintonge2017} studied $R_{21}$ in $28$ galaxies that are part of the xCOLDGASS galaxy sample. Combining IRAM and APEX observations they found a mean ratio of $R_{21}=0.79\pm0.03$ with scatter of $0.15{-}0.23$. xCOLDGASS includes both disc-like and elliptical galaxies, so the relevant comparison is to our full compilation. For all literature data, we find a mean $R_{21}=0.72$ with a scatter of $\pm 0.15$. This agrees reasonably well with the xCOLDGASS results, especially given the heterogeneous nature of the literature data. The EMPIRE results have lower mean $R_{21}$ compared to the xCOLDGASS IRAM--APEX overlap sample. As discussed by \citet{Saintonge2017}, this may somewhat reflect the  central focus of the xCOLDGASS pointings. Or it may reflect a greater contribution of early-type galaxies to their sample. Future larger mapping surveys will be needed to help synthesize our knowledge of resolved and galaxy-integrated $R_{21}$.

%
%Note that these values are more uncertain compared to those derived from matched-beam,
%large FOV maps, like in our case.

%They combined APEX CO(2-1) observations with a beam size of 27" with IRAM 30m measurements of CO(1-0), which have a beam-size of 22". Given the slightly smaller beam-size of the IRAM data, they also adopt an aperture correction. However, they use for the correction the scale length of the star-forming disc rather than D$_{25}$. 

\subsection{Comparison to Previous Mapping Results}

Combining HERACLES with previously existing \mbox{CO(1-0)} data,  \citep{Leroy2009} initially found a mean $R_{21} \approx 0.8$ with evidence for central enhancements. Subsequently, improved main beam efficiencies for the IRAM 30-m telescope became available and \citet{Usero_2015} carried out pointed spectroscopy using the IRAM 30-m that obtained improved CO(1-0) comparison data. Based on comparing HERACLES to the \citet{Usero_2015} data and a collection of earlier CO(1-0) measurements,  \citet{Leroy2013} found a median $R_{21} \approx 0.67$ with a large scatter of $0.16$~dex or ${\sim} 40\%$ among individual measurements. Our mean $R_{21}$ has an almost identical value to that in \citet{Leroy2013}, but the measured scatter using EMPIRE CO(1-0) is smaller. This likely reflects the much better calibration using EMIR compared to the archival CO(1-0) data, though the smaller sample size may also play a role. Also using HERACLES, but now attempting to homogenize literature mapping data, E. Rosolowsky et al. (private communication) found that a ratio of ${\sim} 0.7$ was typical of the inner parts of galaxies, while $0.5$ was more common in outer discs \citep[see also][]{Rosolowsky2015}. Our results yield slight higher $R_{21}$ at large galactocentric radius (see Figure \ref{fig:radial_all}). Again, we expect the EMPIRE CO(1-0) data to be of higher quality than the HERACLES maps, but the EMPIRE sample size is small.

Our results also agree reasonably well with previous mapping-based results for other nearby galaxies. For example, \citet[]{Crosthwaite2007} found $R_{21} {\approx} 0.8 $ in the central $1'\times1'$ of NGC~6946, while we find a value of $R_{21} = 0.7$ for the same galaxy center. Our measured value of $0.7$ for NGC 5194 (M51) agrees with the typical ratio found by \citet[]{Koda2012} while studying the arm/interarm contrast of the ratio. Other resolved mapping results include $R_{21} \approx 0.8$ for M33 \citep[]{Druard2014} with no obvious radial trends. \citet[]{Lundgren2004} and \citet[]{Koda2020} found $R_{21} \approx 0.77$ for M83 which closely resembles the EMPIRE targets in morphology and stellar mass, showing a similar decreasing trend in $R_{21}$ with galactocentric radius. Again, our EMPIRE results tend towards the low side of the literature value, but well within the previously measured range.

\subsection{\texorpdfstring{$R_{21}$}{Lg} Variations in EMPIRE}

We measure the characteristic value, scatter, and dependence of $R_{21}$ on environment. We do find scatter in $R_{21}$ from galaxy-to-galaxy and within galaxies. { In certain galaxies, }we identify significant correlations with galactocentric radius and other observable quantities. 

One recurring theme in our analysis is that the magnitude of these variations is weak. Put simply, the dynamic range in $R_{21}$ across our sample remains small compared to many of the uncertainties associated with measuring the ratio. The small magnitude of these variations somewhat diminish the utility of $R_{21}$ as a diagnostic of the physical conditions in the gas. Of course, CO(2-1) and CO(1-0) both still represent the most widely used tracers of molecular gas at low redshift. Detailed knowledge of how $R_{21}$ behaves is crucial to our knowledge of molecular gas in galaxies.

%\sout{Other ratios with larger dynamic range, e.g., CO(3-2)/CO(2-1) or CO(3-2)/CO(1-0), may represent better probes of CO excitation }\citep[e.g., see][]{Wilson2009}. 

With that caveat in mind, we discuss the major sources of $R_{21}$ variation in EMPIRE:

\textbf{\textbf{Observed} galaxy-to-galaxy scatter ---} In each of our analyses, galaxy-to-galaxy scatter in $R_{21}$ appears to play a role. The galaxy being considered appears to matter independent of radial gradients, correlation with {local conditions}, or azimuthal variations. 

We checked for correlations between {global} galaxy properties and $R_{21}$ that might explain the galaxy-to-galaxy scatter, including comparing to stellar mass, star formation rate, metallicity, inclination, distance, and morphological type. We found no significant correlation that could explain the observed galaxy-to-galaxy scatter. 
%\sout{stronger than} 
%This appears consistent with the analysis by \citet{Saintonge2017}. 
We emphasize, however, that EMPIRE represents an extremely small sample with a limited range of stellar mass, star-formation rate, and metallicity values. EMPIRE is fundamentally a mapping project, not a representative survey of the local galaxy population.

{One plausible explanation for much of the observed galaxy-to-galaxy is uncertainty in the flux calibration, which we discuss in Section \ref{sec:co21_obs} and Appendix \ref{app:comp_HERA_ALMA}. Given the estimated uncertainties in the amplitude calibration of each data set, ${\sim} 6\%$ for EMPIRE, $\approx 5\%$ for PHANGS-ALMA and M51 Large Program, and ${\sim} 20\%$ for HERACLES, we expect $\sim 7{-}20\%$ scatter in $R_{21}$ based on calibration uncertainty alone. We measure rms scatter of $10{-}15\%$ from galaxy-to-galaxy, so it seems highly likely that much of the galaxy-to-galaxy scatter that we observe is caused by flux calibration uncertainties.}

Building a quantitative understanding of galaxy-to-galaxy variations in $R_{21}$ places strong requirements on the data. Given the small dynamic range in the ratio, one needs high signal to noise and high precision absolute flux calibration. To avoid uncertain aperture corrections, one needs to observe and cover the same area in both lines. Although EMPIRE and the IRAM--APEX subset from xCOLDGASS \citep{Saintonge2017} represent good first steps, obtaining such carefully calibrated, high signal to noise data sets still represents a future goal.

%We have not exhausted the full range of possible drivers. Our best hypothesis is that some combination of gas density, variable heating (perhaps including cosmic rays), the recent star formation history, and geometry combine with modest calibration uncertainties to yield the observed galaxy-to-galaxy scatter. A larger sample and more detailed local estimates of physical conditions will shed more light on this matter. 

\textbf{Trends within galaxies ---} Within galaxies, we find a clear, but weak systematic variation of $R_{21}$ as a function of environment. We examined correlations with CO brightness temperature, TIR surface brightness, IR color, and TIR-to-CO ratio. After accounting for biases and disregarding low signal to noise regions, these all show the same trend, i.e. higher $R_{21}$ values in regions with higher gas surface density, hotter dust, and more star formation. 

\textit{Galaxy centres.} One major driver for these trends is that we observe a higher $R_{21}$ in the centre of galaxies compared to the discs. The average enhancement is $15 {\%}$ compared to the galaxy-wide luminosity-weighted mean, but several individual cases show much stronger nuclear enhancements. NGC~2903, NGC327, NGC~4321, and NGC~5055 all show strong (${\sim}$50\%) central enhancements in $R_{21}$. 

%In the Appendix, we used improved new CO(2-1) maps from ALMA, and these further reinforce that in three strongly barred targets (NGC~2903, NGC~3627, and NGC~4321) we observe significant central enhancements in $R_{21}$. 

Though not uniquely associated with bars, these central enhancements do seem strongest in the strongly barred members of our sample. NGC~2903, NGC~3627, and NGC~4321 all have prominent bars that visibly interact with the molecular gas. In these cases our coarse resolution likely causes us to underestimate the strength of the $R_{21}$ enhancement, because the nuclear star-forming regions, where we expect the $R_{21}$ enhancements to be strongest, are compact (often $\sim 0.5$~kpc) compared to our $27''$ ($\sim 1-2$~kpc) beam.

In unbarred galaxies, we often observe flatter $R_{21}$ profiles, e.g., in NGC~628, NGC~3184, or NGC~4254. NGC~5194 (M51) remains an ambiguous case, with our newer EMIR mapping data showing evidence for a flatter $R_{21}$ profile than the HERA maps (see Appendix~\ref{app:comp_HERA_ALMA}).

A similar drop in $R_{21}$ from the centre of the galaxies towards their discs has also been found by many previous studies. Milky Way studies show values close to unity in the central kpc of the Galaxy, dropping to 0.75 at 4 kpc and to ${\sim} 0.6$ at 8 kpc from the Galaxy centre \citep[]{Sakamoto1997,Sawada2001}. Similarly, \citet{Casoli1991} report a value of $\sim$ 1 in the nuclei of nearby spirals (at $\sim 500$~pc scales) compared to $0.5 {-} 0.7$ in their discs. Studying the nearby spiral IC~342, \citet{Eckart1990} found a drop from $R_{21}~\approx~1.1$ to $R_{21}~\approx~0.7{-}0.8$ around 500 pc from the starburst nucleus. A high average ratio of ${\sim}0.9$ is also found by \citet{Braine1992} in the central kpc of 36 nearby galaxies. Similar radial trends have been found by \citet{Saito2017} when studying the spatially-resolved $R_{21}$ ratio { in NGC 1614}. Using RADEX modelling they find a radial kinetic temperature gradient that mirrors the observed $R_{21}$ trend. Using HERACLES with lower quality CO(1-0) data but a larger sample, \citet{Leroy2009,Leroy2013} noted a similar trend in resolved maps of nearby galaxies.

Furthermore, variation in the CO line ratio could be driven to some parts by the presence of an active galactic nuclei (AGN) within the galaxy. Four of the galaxies in our sample, NGC 3627, 4321, 5055, and 5194, are know to host an AGN. 

\textit{Correlation with physical conditions.} We observe positive correlations of $R_{21}$ with CO brightness temperature, TIR surface brightness, and IR color. All of these quantities tend to decrease with increasing galactocentric radius, so these trends likely express the same underlying physics as the radial gradients. 

{Physically, the IR color reflects the interstellar radiation field heating the dust. At the typical densities associated with molecular clouds, gas and dust have different temperatures and are not collisionally coupled \citep[e.g.,][]{Drain2011}. However, the radiation field traced by the dust temperature should also relate to the radiation field illuminating molecular clouds and so indirectly relate to excitation of the gas. We would expect more intensely illuminated clouds to have high temperatures and be more nearly thermalized.}

{Similarly, the TIR surface density traces the heating of the ISM because it indicates the amount of reprocessed, mostly ultraviolet emission. High TIR surface density may also indirectly trace gas density, because high gas densities tend to be associated with high star formation rates \citep[e.g., see more discussion in][]{Donaire2019}. We would expect denser, higher temperature gas to be more nearly thermalized and show a higher $R_{21}$.}

%Physically, increased heating due to the higher radiation field and increased density are both likely to drive {\bf indirectly} the gas to a thermalized state. 

As mentioned above, optical depth effects may also play a role. The line ratio of optically thin gas exceeds unity and a component of diffuse, optically thin gas will drive $R_{21}$ to higher values.

A systematic dependence of $R_{21}$ on, e.g., $\Sigma_{\rm TIR}$ has implications for the slope of derived scaling relations. For example, the scaling relation between $\Sigma_{\rm TIR}$ and $I_{\rm CO}$ corresponds to the molecular version of the Kennicutt--Schmidt law. If $R_{21}$ varies systematically with $\Sigma_{\rm TIR}$ then one expects to derive different slopes if using CO(1-0) or \mbox{CO(2-1)}.

Our results show that this is the case, but also that the effect is modest. For example, in Figure \ref{fig:correlations_all}, $R_{21}$ changes by $\sim 40\%$ as $\Sigma_{\rm TIR}$ changes by a factor of $100$. This would translate to a difference in slope of ${\sim} 0.07$ for a power law relating the two quantities. It seems reasonable to infer that using CO(2-1) instead of CO(1-0) will change the slope of the $I_{\rm CO}$--$\Sigma_{\rm TIR}$ relation by $0.05{-}0.1$.

We caution that the implications of $R_{21}$ for the underlying physical scaling relation, e.g., between $\Sigma_{\rm mol}$ and $\Sigma_{\rm SFR}$, are less clear. The sensitivity of $R_{21}$ to these local physical conditions implies that physical conditions in the molecular gas are changing. Variations in density, temperature, and optical depth will imply changes in the CO-to-H$_2$ conversion factor, $\alpha_{\rm CO}$, for both CO(1-0) and \mbox{CO(2-1)}. Unfortunately, on its own $R_{21}$ does not heavily constrain $\alpha_{\rm CO}$. Future work using a large set of lines and independent constraints on $\alpha_{\rm CO}$ will help map $R_{21}$ variations to $\alpha_{\rm CO}$ variations for both lines.

%This is, because star formation laws (or other scaling relations) are prone to the uncertainty of $\alpha_{CO}$, the conversion factor between CO(1-0) luminosity and molecular gas mass. 

%, does not imply that usage of the CO(1-0) line as a molecular gas tracer is beneficial compared to CO(2-1), even in cases where only one of the lines is available. 

%The latter is many factors larger than the dynamic range of $R_{21}$ seen in Figure \ref{fig:correlations_all}; compare e.g. \cite[Figure 2 therein]{Bolatto2013}.

\textit{Scatter at fixed radius and resolved patterns.} Density and radiation field also vary at fixed galactocentric radius, e.g., due to the effects of spiral density waves and stellar bars. The arm-interarm contrast and small-scale structure of $R_{21}$ have been the focus of several recent papers \citep{Koda2012,Law2018}. Though our ${\sim} 1{-}2$~kpc resolution limits our ability to isolate small scale variations in $R_{21}$, we attempt to quantify the scatter in $R_{21}$ at fixed galactocentric radius in our sample using a forward modeling technique. 

%\sout{We find rms scatter of $\sim 10-20$\% after accounting for the effects of noise. NGC~5194 (M51) shows the largest intrinsic scatter of any target, presumably due to its well-defined grand-design-structure. The magnitude of scatter and existence of strong arm-interarm contrast for M51 appear to agree well with that found by} \citet{Koda2012} {and \citet{Vlahakis2013}. \sout{However, we find an opposite trend to those studies. Our measurements indicate higher $R_{21}$ in the interarm region compared to the arm region, while} \citet{Koda2012} and \citet{Vlahakis2013} \sout{find lower $R_{21}$ in the interarm compared to the arm regions. Studies of the Milky Way} \citep{Sakamoto1997} and M83 \citep{Koda2020} \sout{ have also found higher $R_{21}$ in the arms compared to the interarms.}}

{NGC~5194 (M51) shows the largest intrinsic scatter of any target, presumably due to its well-defined grand-design-structure. Past studies have already highlighted a strong arm-interarm contrast in the CO line ratio in M51 \citep{Koda2012, Vlahakis2013}. The contrast is also strongly visible in our analysis, but we find an opposite trend (we find a high interarm and low arm $R_{21}$ ratio; see Section \ref{variations} and Appendix \ref{app:comp_NRO_EMIR}). Among the literature there is disagreement between the relative and quantitative trend of $R_{21}$ between arm vs interarm. 
Interestingly, when studying NGC6946, \cite{Crosthwaite2007} found $R_{21}>1$ in the interarm as opposed to smaller values in the molecular arm regions. While we cannot confirm such large average absolute values with our observations, the regions which show enhancement in our data overlaps with theirs (we find an average $R_{21}\sim0.9$ in the interarm, with 35\% of the points in that region with S/N$>3$ showing $R_{21}>1$, and only 8\% have $R_{21}>1.2$). Furthermore, \cite{Crosthwaite2002} and \cite{Lundgren2004} have investigated M83 and both report higher line ratio values in the interarm region as well. {{However, the validity of this trend within M83 has recently been disputed by \citet{Koda2020}, who studied the source with ALMA observations.}}
Differences of $R_{21}$ between arm and interarm have been found in several galaxies, however, different studies have presented opposing trends. We believe that this is caused at least partially by coarse spatial resolution and insufficient data quality, and should be investigated in more detail in the future.
}
%{\bf Higher line ratios in the interarm regions have been observed before. Studying NGC6946, \cite{Crosthwaite2007} found $R_{21}>1$ in the interarm as opposed to smaller values in the molecular arm regions. While we cannot confirm such large average absolute values with our observations, the regions which show enhancement in our data overlaps with theirs (we find an average $R_{21}\sim0.9$ in the interarm, with 35\% of the points in that region with S/N$>3$ showing $R_{21}>1$, and only 8\% have $R_{21}>1.2$). Furthermore, \cite{Crosthwaite2002} and \cite{Lundgren2004} have investigated M83 and both find higher line ratio values in the interarm region as well. }

{From point of view of the data, our study differs from \citet{Koda2012} in that we use both new CO(1-0) and CO(2-1) maps obtained using the IRAM 30-m EMIR receiver. The primary difference appears to come from the use of the new CO(1-0) map. We defer a detailed comparison among M51 data sets to the presentation of the new IRAM LP (J.~S.\ den Brok et al. in prep.) and new Submillimeter Array observations (M. Jimenez Donaire et al. in prep.). As emphasized in Section \ref{sec:co10_obs} { and Appendix \ref{app:comp_NRO_EMIR}}, we use what we consider the best available map.}

{{{{Interpreting in terms of ISM physics and molecular cloud conditions},} our finding of enhanced $R_{21}$ in interarm regions as compared to arm regions implies} more excited, perhaps more diffuse{, warm} and optically thin gas in the interarm regions. {In the interarm region, the heating is most likely more efficient due to different cloud composition. Furthermore, \cite{Lundgren2004} suggest that photon-dominated regions (PDRs) around cool stars could be responsible for bright CO emission. This is, because the PDR radiation field is softer, thus the CO can be heated photoelectrically at lower $A_V$ \citep{Spaans1994}. Another possibility would be  ``CO-loud'' gas \citep{Lundgren2004}. Small amounts of optically thin gas could already cause strong emisivity in CO(2-1) \citep{Wiklind1990}. These explanations} might be consistent with the extended diffuse component identified in M51 by \citet{Pety2013}. {They find that $\sim 50$\,\% of the total CO emission originates from larger spatial scales ($>1.3$ kpc), which would be consistent with emission from a diffuse disk of gas at a scale height of $\sim 200$\,pc. A similar fining was made by \cite{Caldu2016} studying M31. The large scatter found} may also reflect the influence of M51's ongoing interaction with the companion galaxy NGC 5195.}

%\sout{Assuming variations take place temporally during the cloud lifecycle, then this is effectively dictated by the uncertainty principle} \citep[as described in][]{Kruijssen2019,Chevance2020}.
%\sout{They find region separation lengths of 100-250 pc. This means that resolving the scale on which $R_{21}$ varies requires arcsecond resolution.}

\textit{Uncertain behavior at large radii.} Despite our use of spectral stacking, many of our $R_{21}$ estimates remain uncertain at low brightness temperature and large radius. In the stacked radial profiles, we see suggestions of large deviations to both low and high $R_{21}$ in some of our targets. Similarly, in the lowest brightness temperature bins of IR color and TIR surface brightness we see hints of significant deviations. It could well be that molecular gas in the outer parts of galaxies is either optically thin, leading to high $R_{21}$, or cool, leading to low $R_{21}$. More sensitive observations of both lines will be required to ascertain the behavior of the ratio in the faint CO emission from the outer discs of galaxies. 

\subsection{Comparison to Radiative Transfer Models}

Following up work of \cite{Leroy2017}, J.~Puschnig et al. (in preparation) have established a set of molecular radiative transfer models,
i.e. the \texttt{Dense Gas Toolbox} \citep{DGT}, which predicts line ratios for a medium with an underlying
density distribution (e.g., a log-normal distribution rather than from a single density). Using CO line optical depths as
previously published by \cite{Cormier2018} for EMPIRE galaxies {(they find $\tau^{12}\approx6$ for 12CO(1-0))}, we now examine the impact of
three physical quantities on $R_{21}$: temperature, mean density, and width of the log-normal density
distribution. The interplay between these quantities can be studied through an interactive
tool\footnote{\url{http://www.densegastoolbox.com/explorer/}}. The models show that $R_{21}$ is most sensitive
to regimes with mean densities lower than $\sim$10$^3$~cm$^{-3}$. Below that value all three quantities are degenerate.
However, above that density the line ratio may only be driven further up by higher temperatures,
regardless of the width of the density distribution (that is proportional to line width or Mach number).
We also recognize that values of $R_{21}>0.8$ are only predicted for temperatures above 35~K, regardless of the mean gas density.
For NGC5194, we may thus conclude that the mean gas density and temperature must be $n_\mathrm{H_2} > 10^3$~cm$^{-3}$ and $T_\mathrm{kin}>$35~K,
throughout the whole disc.

\section{Summary} \label{Summary}

We measure the $^{12}$CO(2-1)/$^{12}$CO(1-0) ~ brightness temperature ratio, $R_{21}$, across the star-forming discs of nine nearby galaxies. We measure CO(1-0) emission from maps obtained by the IRAM 30-m telescope in the context of the EMPIRE survey \citep{Bigiel2016,Donaire2019} and CO(2-1) emission from a mixture of ALMA and IRAM 30-m data (J.~S.~den Brok et al., in prep). We use IRAM 30-m CO~(2-1) maps obtained as part of HERACLES \citep[]{Leroy2009} and a new IRAM Large Program targeting M51. We use ALMA maps obtained as part of the PHANGS-ALMA survey (A.~K.~Leroy et al, in prep.). We measure the distributions and mean values of $R_{21}$ across individual lines of sight, integrated over galaxies, stacking by radius, and stacking as a function of other local conditions. Our main results are:

\begin{enumerate}

\item The luminosity-weighted mean $R_{21}$ for individual galaxies ranges from $0.48{-}0.73$. Within individual galaxies, we observe a typical range of $\pm 0.1$. Over the whole sample, treating galaxies equally we find a mean $R_{21}$ of \RCO . 

\item We compiled and homogenized a set of CO observations from the literature. For 81 disc galaxies, these literature measurements yield \mbox{$R_{21}=0.59^{+0.18}_{-0.09}$}, in good agreement with our mean value.

\smallskip

\item Seven of our nine targets show a central enhancement in $R_{21}$ compared to the disc-averaged value (median enhancment $\sim 15\%$). The magnitude of the deviation varies from galaxy-to-galaxy, but variation at larger radii can be much larger than the ones found towards the centre. Both central enhancements and radial gradients in $R_{21}$ are in agreement with previous work.

\smallskip

\item We find significant correlations between $R_{21}$, CO brightness temperature, TIR surface density, and 70-to-160 $\mu$m ratio. All of these have the expected trend of an increasing ratio when the gas density and radiation field increase.

\smallskip

\item $R_{21}$ also shows azimuthal variation. Using a forward modelling approach, we estimate the intrinsic scatter in $R_{21}$ at fixed galactocentric radius to be $\sim$ 20\% at our ${\sim} 1{-}2$ ~ kpc resolution. 

%These variations appear weaker than some measurements in the literature. This may partially reflect our low resolution.

\smallskip

\item 
These physical trends are not sufficient to explain the majority of the galaxy-to-galaxy variations observed. Given the scale of our calibration uncertainties, we cannot completely rule them out as one of the dominant drivers for these trends.
%None of these physical trends can explain the majority of the galaxy-to-galaxy variations that we observe. 
Instead, the magnitude of this scatter appears consistent with being driven by absolute flux calibration uncertainties.

\smallskip

%This broadly agrees with, but is moderately lower than, previously measured values of the ratio for nearby spiral galaxies.

%Some of the scatter may reflect absolute flux calibration uncertainties. The rest must indicate some combination of local conditions and geometry that our observations do not yet resolve at our ${\sim} 1.5$ ~ kpc resolution.

\end{enumerate}

\section*{Acknowledgements}

The IRAM 30m large program EMPIRE was carried out under project number 206-14 (PI Bigiel), the $^{12}$CO(1-0) observations under projects 061-15 and 059-16 (PI Jim\'enez-Donaire) and D15-12 (PI Cormier). IRAM is supported by INSU/CNRS (France), MPG (Germany) and IGN (Spain). 
MJJD acknowledges support from the Smithsonian Institution as a Submillimeter Array (SMA) Fellow. 
FB, JP, AB and JdB acknowledge funding from the European Union's Horizon 2020 research and innovation programme (grant agreement No 726384/Empire). 
AU acknowledges support from the Spanish funding grants AYA2016-79006-P  (MINECO/FEDER), PGC2018-094671-B-I00 (MCIU/AEI/FEDER), and  PID2019-108765GB-I00 (MICINN).
The work of AKL and MJG is partially supported by the National Science Foundation under Grants No.~1615105, 1615109, and 1653300. AKL also acknowledges partial support from NASA ADAP grants NNX16AF48G and NNX17AF39G. 
ES, DL, and ST acknowledges funding from the European Research Council (ERC) under the European Union’s Horizon 2020 research and innovation programme (grant agreement No. 694343). CMF acknowledges support from the National Science Foundation under Award No. 1903946.
JMDK gratefully acknowledges funding from the Deutsche Forschungsgemeinschaft (DFG) through an Emmy Noether Research Group (grant number KR4801/1-1) and the DFG Sachbeihilfe (grant number KR4801/2-1), as well as from the European Research Council (ERC) under the European Union's Horizon 2020 research and innovation programme via the ERC Starting Grant MUSTANG (grant agreement number 714907). { We thank J.\, Koda  for making available the NRO 45-m CO(1-0) data of NGC 5194.} 

\section*{Data availability}

{The HERACLES and EMPIRE survey data used in this article are publicly available in IRAM repository, at \url{https://www.iram.fr/ILPA/LP001/} and \url{https://www.iram.fr/ILPA/LP015/}, respectively. The PHANGS-ALMA CO maps will be available as part of the first public data release in the first half of 2021 and will be available from the ALMA archive and \url{https://www.phangs.org}. The M51 IRAM Large Program will be publicly available during the first half of 2021 via the IRAM Large Program Archiva. The remaining data underlying this article will be shared on reasonable request to the corresponding author.}

%%%%%%%%%%%%%%%%%%%%%%%%%%%%%%%%%%%%%%%%%%%%%%%%%%

%%%%%%%%%%%%%%%%%%%% REFERENCES %%%%%%%%%%%%%%%%%%

% The best way to enter references is to use BibTeX:

\bibliographystyle{mnras}
\bibliography{bibliography} % if your bibtex file is called example.bib

% Alternatively you could enter them by hand, like this:
% This method is tedious and prone to error if you have lots of references
%\begin{thebibliography}{99}
%\bibitem[\protect\citeauthoryear{Author}{2012}]{Author2012}
%Author A.~N., 2013, Journal of Improbable Astronomy, 1, 1
%\bibitem[\protect\citeauthoryear{Others}{2013}]{Others2013}
%Others S., 2012, Journal of Interesting Stuff, 17, 198
%\end{thebibliography}

%%%%%%%%%%%%%%%%%%%%%%%%%%%%%%%%%%%%%%%%%%%%%%%%%%

%%%%%%%%%%%%%%%%% APPENDICES %%%%%%%%%%%%%%%%%%%%%

%\appendix If you want to present additional material which would interrupt the flow of the main paper,
%it can be placed in an Appendix which appears after the list of references.

\appendix

\section{Overview of Maps}

Figure\,\ref{fig:all_maps1_appendix} displays maps of our nine target galaxies. The left-most column shows $\log_{10} {R}_{21}$, the CO(2-1)/(1-0) integrated brightness temperature ratio (see Section\,\ref{overall_dist}). Columns two and three show the integrated CO(1-0) and CO(2-1) brightness temperature maps (see Sections\,\ref{sec:co10_obs} and \ref{sec:co21_obs}). Column four and five show the \textit{Herschel} 70$\mu$m and 160$\mu$m intensity maps (see Section\,\ref{sec:IR_obs}). The last column shows the total infrared surface brightness (see Section\,\ref{sec:Physical})\footnote{NGC 2903 lacks \textit{Herschel} data.}. 

All of the maps have already been convolved to share the same, 27\arcsec\ angular resolution. They have all been projected onto a hexagonal grid with a grid spacing equal to half the beamsize ($13.5$\arcsec). The maps in Figure\,\ref{fig:all_maps1_appendix} only show sightlines that have significant ($S/N>3$) integrated brightness temperature detections in both \mbox{CO(2-1)} and \mbox{CO(1-0)}.

\begin{figure*}
	\begin{center}
	   \includegraphics[width=\textwidth, trim=0 11mm 0 0mm, clip]{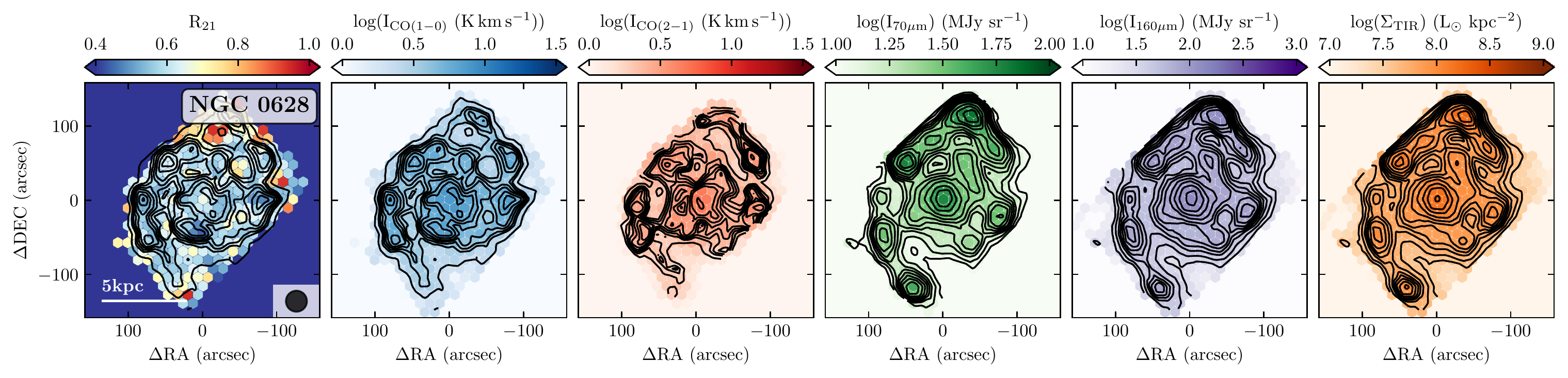}
	   \includegraphics[width=\textwidth, trim=0 11mm 0 7mm, clip]{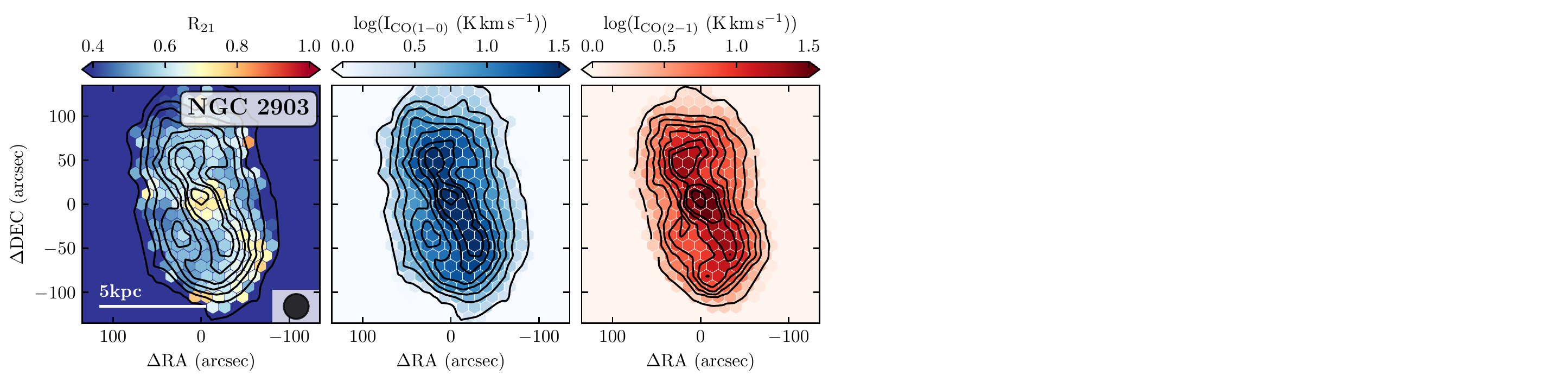}
	   \includegraphics[width=\textwidth, trim=0 11mm 0 7mm, clip]{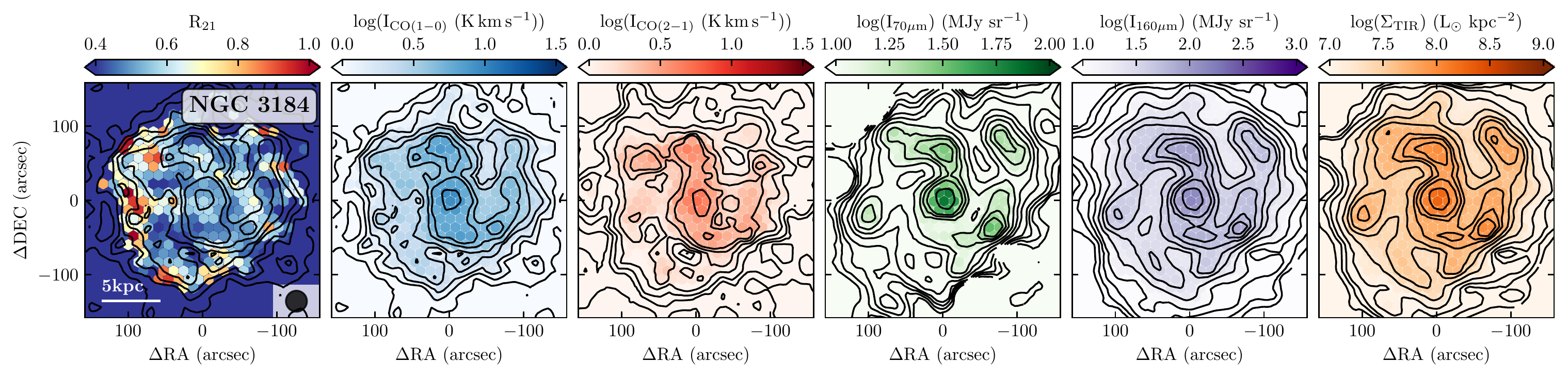}
	   \includegraphics[width=\textwidth, trim=0 11mm 0 7mm, clip]{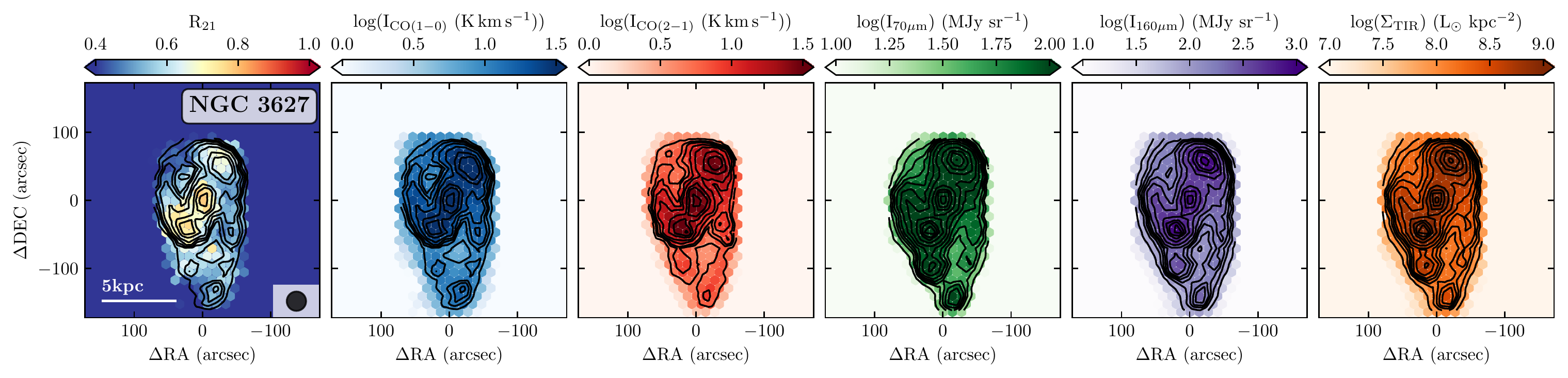}
	   \includegraphics[width=\textwidth, trim=0 11mm 0 7mm, clip]{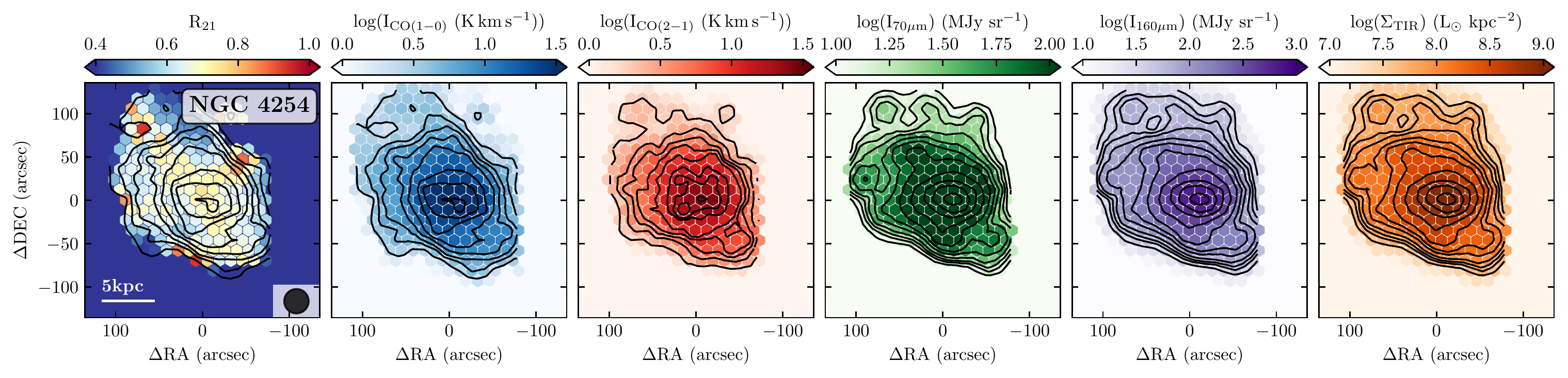}
	   \includegraphics[width=\textwidth, trim=0 0mm 0 7mm, clip]{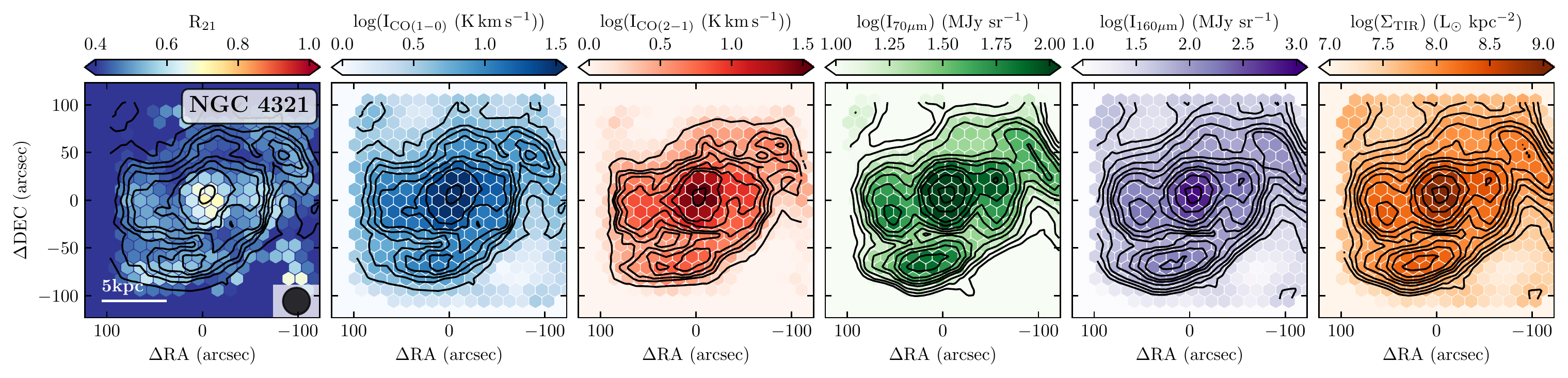}
	\end{center}
\caption{\textbf{Multiwavelength maps used in this paper.} From left to right, each row shows maps of the CO(2-1)/CO(1-0) integrated brightness temperature ratio, log($R_{21}$), CO(1-0) and CO(2-1) integrated brightness temperature, {\it Herschel} 70\,\micron\ and 160\,\micron\ fluxes, and the total infrared surface brightness. There are no \textit{Herschel} data available for NGC 2903. We plot values for positions that have significant ($S/N > 3$) integrated brightness temperature detections within both the CO(2-1) and CO(1-0) maps. For reference, we overlay CO(1-0) integrated brightness temperature contours on the $R_{21}$ map. The remaining panels have contours of their respective colour-scale, in levels of 20, 30, 40, 50, 60, 70, 80, 90, 95, 97.5, and 99.5 of the peak value.}
\label{fig:all_maps1_appendix}
\end{figure*}

\begin{figure*}
	\begin{center}
	   \includegraphics[width=\textwidth, trim=0 11mm 0 0mm, clip]{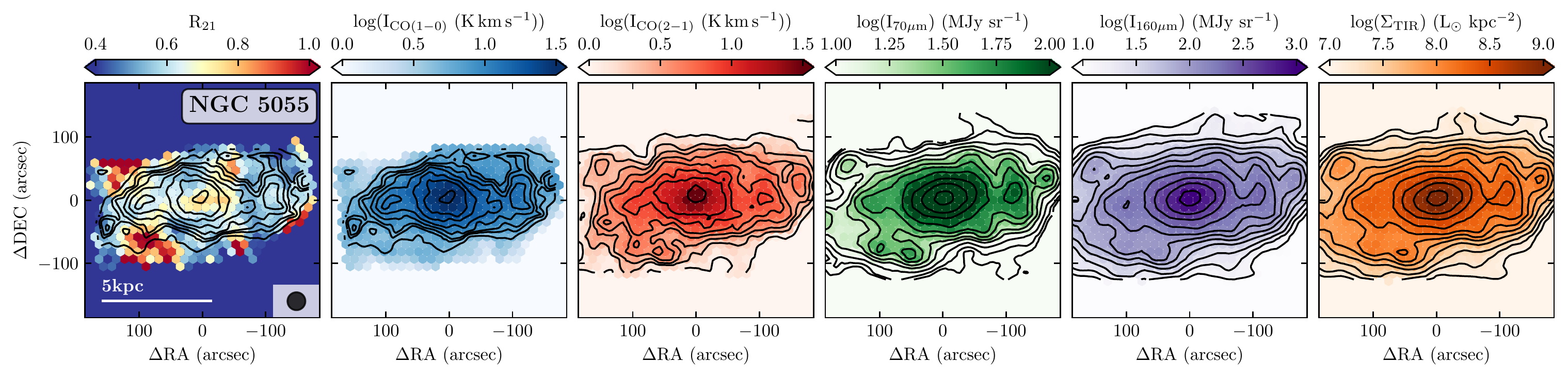}
	   \includegraphics[width=\textwidth, trim=0 11mm 0 7mm, clip]{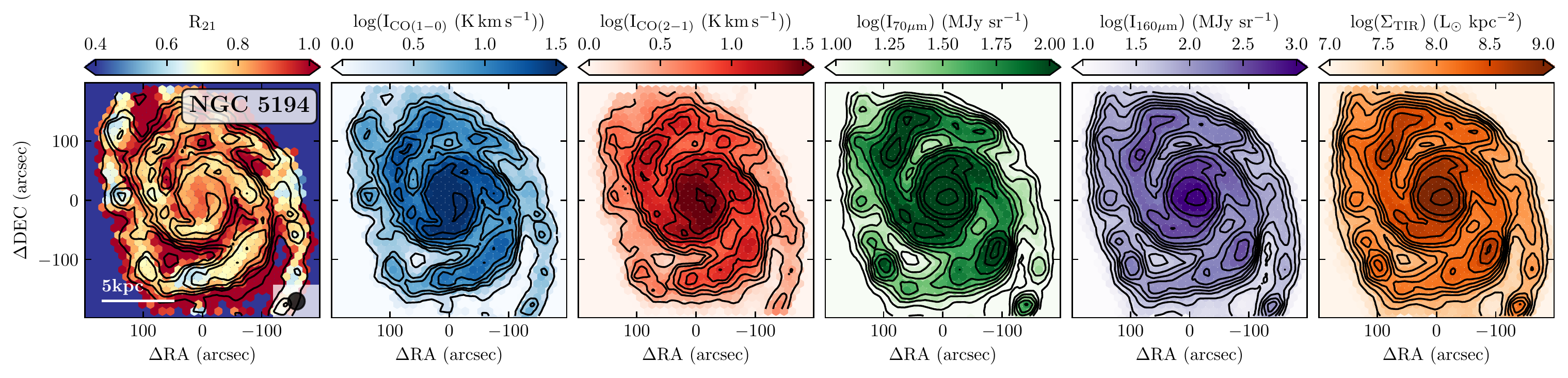}
	   \includegraphics[width=\textwidth, trim=0 0mm 0 7mm, clip]{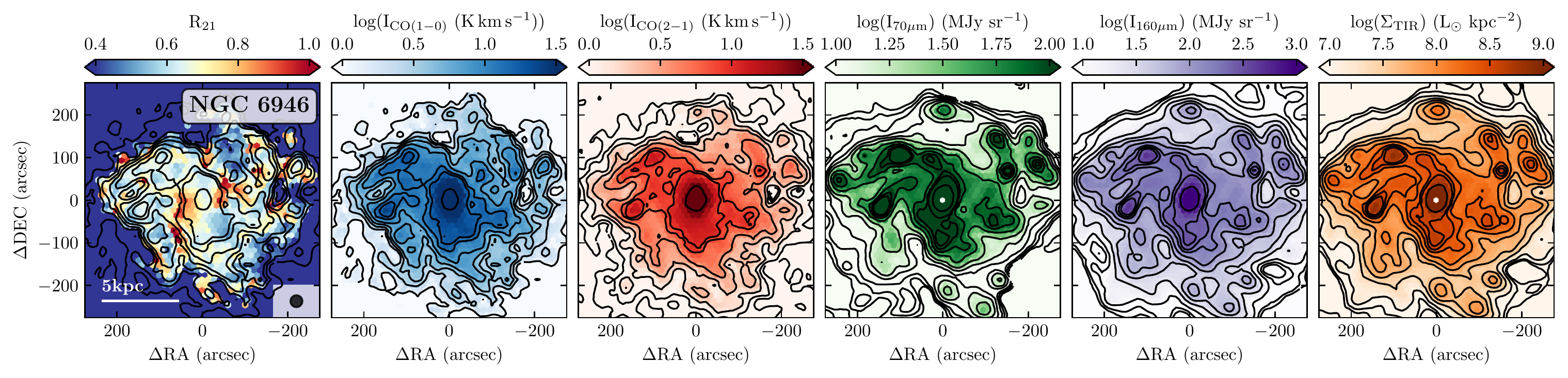}
	\end{center}
\contcaption{}
\label{fig:all_maps2_appendix}
\end{figure*}

\section{Forward-Modelling to Infer Physical Scatter in \texorpdfstring{$R_{21}$}{Lg}}
\label{ap:forward_model}

This section presents our modelling used to estimate the scatter of the $R_{21}$ ratio at a fixed galactocentric radius. In our modelling, we choose to disentangle the intrinsic scatter per radial bin from the observational noise, and model the value of the scatter that best describes our data separately. 

The observed scatter in $R_{21}$ reflects a combination of statistical and physical scatter. Fortunately, we have accurate estimates of the statistical uncertainties. To estimate the physical scatter, we carry out a forward modeling analysis that leverages this knowledge.

To do this, we assume that the true physical distribution of $R_{21}$ is log-normal. This appears to be a reasonable assumption based on the observed distributions, e.g., in Figure \ref{fig:hist_R}.{ The log-normal distribution has strictly speaking no physical meaning, but gives a simple representation of the scatter.} Then, our modeling process proceeds as follows:

\begin{enumerate}
\item We normalize all measured $R_{21}$ ratios to the median value of the ratio in the distribution. 

\item We simulate a set of new data. Each data set has a known physical scatter between 0 and 2~dex. We add Gaussian noise to each new data set based on the known observational uncertainties for the measurement in the data set. 

\item We compare the observed distribution to the simulated distribution and select the simulated data set that best matches our observational data. To selected the best match, we use the Kolmogorov-Smirnov statistics.

\item We adopt the physical scatter in the best-match model distribution as our best estimate of the true physical scatter.
\end{enumerate}

This estimate accounts for the known scatter due to statistical uncertainties, which can be substantial.

%   §In order to ensure a good statistical basis for our modelling, we choose to exclude any radial bins where the number of points is lower than 30. Additionally, we reject any bins where we cannot uniquely assign a value for the astrophysical variation, which is typically at large radii where the fraction of low signal to noise data points is large.

\section{Calibration Uncertainties in HERACLES}
\label{app:comp_HERA_ALMA}

In the main text, we emphasize the importance of knowing the calibration uncertainties for accurate $R_{21}$ estimation. Here we revisit topics related to the calibration of the HERACLES CO~(2-1) maps.

HERACLES was obtained using the HERA receiver array on the IRAM 30-m telescope \citep{Schuster2007}. HERA consists of two nine-receiver arrays, one for each polarization, for a total of 18 pixels. HERACLES was calibrated using the standard IRAM 30-m chopper wheel calibration and converted from antenna temperature to main beam temperature using best estimates for the IRAM forward and main beam efficiencies. The bandwidth of HERA does not allow observations of a Galactic line calibrator. The overhead to observe a flux calibrator with every pixel during each few-hour observing block was prohibitive.

\textbf{Measured Gain Variations:} \citet{Leroy2009} assessed the uncertainty in the HERACLES calibration by building maps from different polarizations and observing sessions. By comparing the intensity of bright point sources, they estimated an overall calibration uncertainty of 20\%.

After that, a more rigorous check was added to the HERACLES pipeline to assess the relative flux calibration of the individual receiver pixels. We took the final cube created from all pixels. Then, we took the location of each observation for each individual receiver pixel. In this way, we simulated the spectrum that we would expect to observe with that pixel. We compared this expected spectrum to the real observed spectrum for that pixel. Based on this comparison, we calculate the best-fit multiplicative factor, the ``pixel gain,'' to match that pixel to the overall cube. The accuracy of the gain measurements is accessible via the $\chi^2$ values obtained from the comparison between simulated and observed spectrum. We measured a gain for each array pixel and observing session, labeling the observing session by the day of the observations.

The measured pixel gain represents the offset in calibration between that pixel and the overall array. This factor does not capture absolute variations in the calibration, it measures how internally well-calibrated the pixels are relative to one another.

\begin{figure*}
	\begin{center}
	   \includegraphics[width=\textwidth]{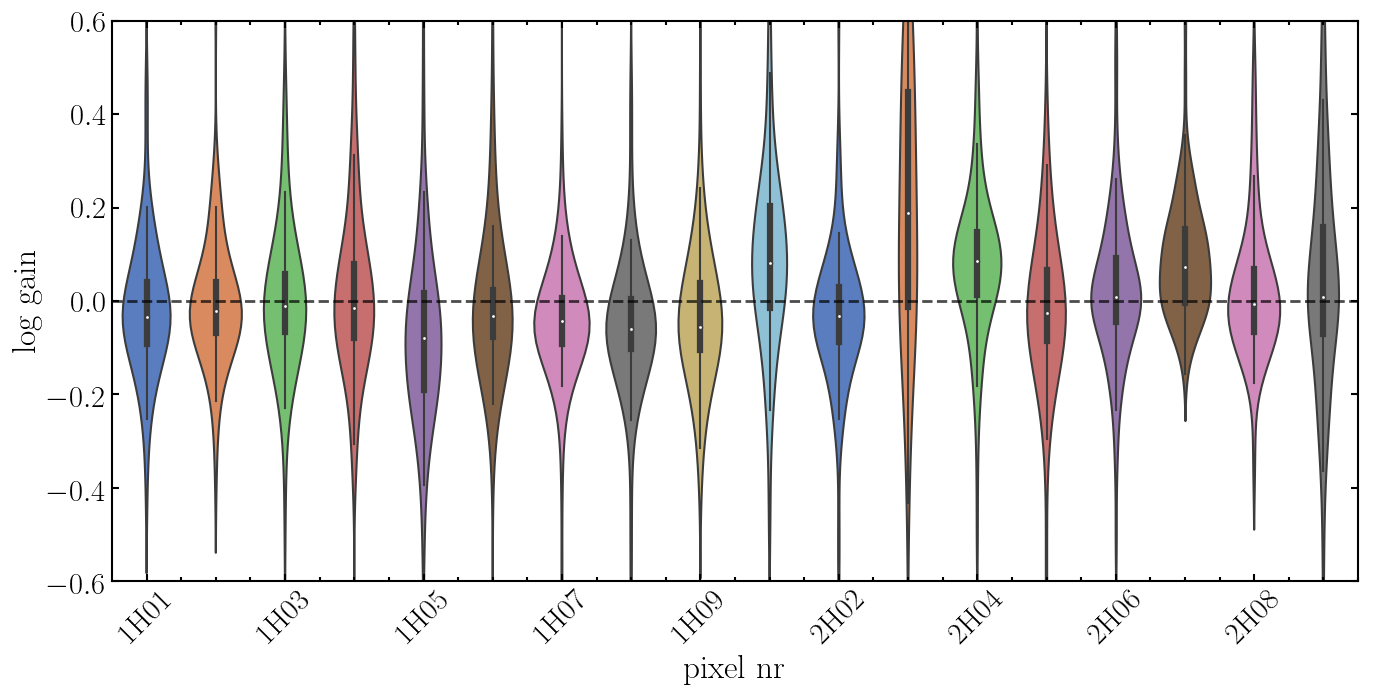}
	\end{center}
\caption{Histograms of log gains per HERA pixel and polarization.}
\label{fig:gain_pix}
\end{figure*}

Figure \ref{fig:gain_pix} shows histograms of the pixel gain for each pixel. We only plot pixel gains with high accuracy, i.e. their $\chi^2$ values lie within $\pm 1\sigma$ of the Gaussian log-$\chi^2$ distribution. Lower signal to noise cases typically represent observations of faint galaxies or empty sky and do not contain the signal needed to fit for the pixel gain.

The figure displays that the gain shows typical rms variation of $\pm 0.10$~dex. Some pixels are less stable than others, with the second polarization (HERA2, labelled ``2H'') showing more scatter than the first polarization.

If the pixel gains were uncorrelated, random, and the coverage of each pixel were spread evenly across the maps, then we expect that the pixel gain uncertainties should average and the calibration uncertainty associated with individual receiver variations would be $\sqrt{18} \approx 4.2$ times lower than the mean individual pixel gain. This represents a lower limit to the calibration uncertainty, which we estimate at $\pm 0.024$~dex or $\pm 6$\%.

In fact, the gains do show some correlation, so that there do not appear to be $18$ truly independent realizations. As mentioned, the two polarizations often appear offset from one another, with the typical offset on any given day of 0.08~dex.

% JP: will do during review
%We also measure how much all pixels on an individual day show offsets from the average, and find that the average of all $18$ gains varies by $\pm XXX$ from day-to-day for the same object. Finally, we check whether individual pixels show correlations as a function of day for the same object, and find that NNN.

Based on this, we find an upper and lower limit uncertainty of $\pm 0.10$~dex and $\pm 0.024$~dex, respectively, corresponding to a flux calibration uncertainty between $\pm 6$\% and $\pm 25$\%. This will not include any additional terms that are covariant among all pixels, like correction for the atmosphere and beam efficiency effects. 

Note that although the HERACLES observing strategy attempted to maximize the number of different pixels observing each part of the sky, local variations in the calibration will be worse due to the fact that not all pixels see all locations.

As an aside, note that we already used these calculated pixel gains to identify and flag the worst receiver-day combinations before producing the maps made publicly available and used in \citet{Schruba2011}, \citet{Schruba2012}, \citet{Leroy2013}, and \citet{Sandstrom2013}.

\begin{figure*}
	\includegraphics[width=0.9\textwidth]{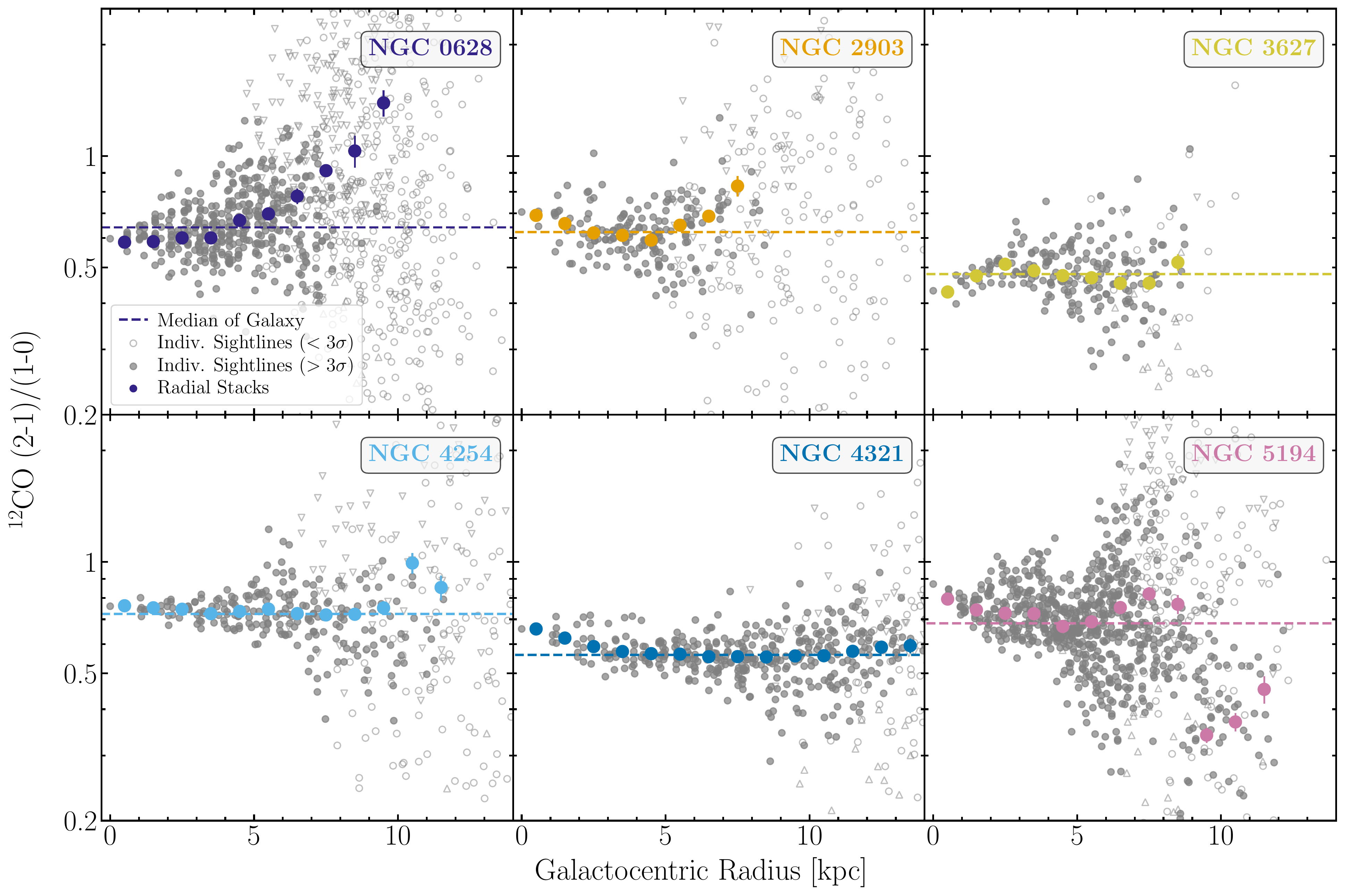}
\caption{Radial profiles of the CO line brightness temperature ratio, $R_{21}$, as a function of galactocentric radius for all the galaxies in the EMPIRE sample, similar to Figure\,\ref{fig:radial}. Here we substituted the CO(2-1) data from HERACLES { for ALMA or the new M51 Large Program data}. Upper and lower limits of individual sightlines are indicated by upwards and downwards triangles. We present the stacked values of the ratio per 27\arcsec radial bin. The dashed line gives the mean line ratio within the galaxy.}
\label{fig:radial_ALMA_cases}
\end{figure*}

\begin{figure*}
	\includegraphics[width=0.75\textwidth]{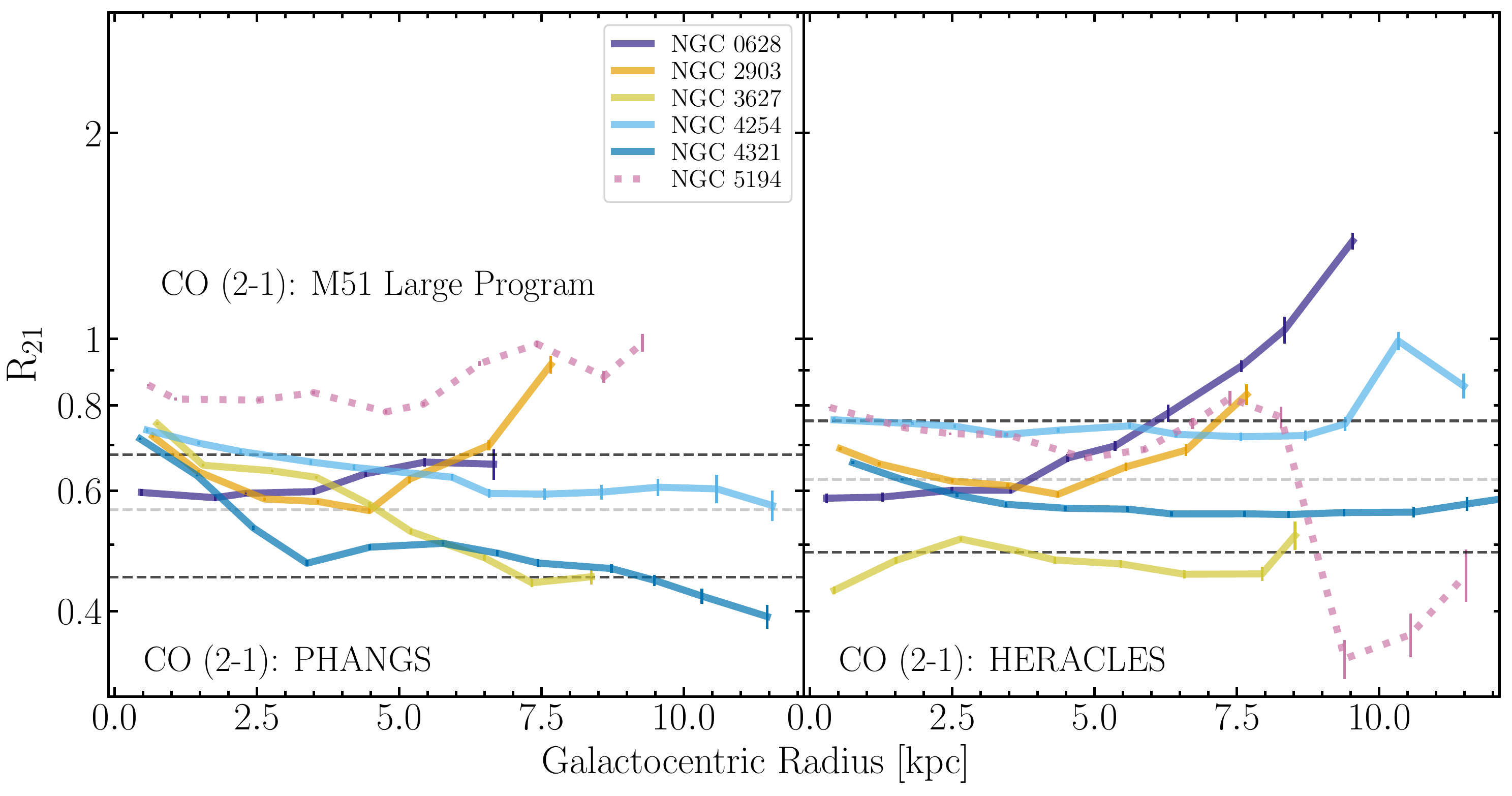}
\caption{Radial profiles of the CO ratio, $R_{21}$, as a function of galactocentric radius for all the galaxies in the EMPIRE sample, similar to Figure\,\ref{fig:radial_all}. The gray, dashed line gives the mean line ratio across all galaxies plotted, while the black, dashed lines indicate the 1$\sigma$ deviation. For the CO(2-1) data set, HERA data was substituted by ALMA and in the case of NGC 5194 with EMIR 1\,mm observations.}
\label{fig:radial_all_ALMA}
\end{figure*}

\textbf{Comparisons in Galaxy With Two Maps:} As a more direct, alternative check, we took the overlap between PHANGS-ALMA, HERACLES, and the new IRAM 30-m map of M51 in our sample and directly calculated how these CO(2-1) maps compared to one another {(see Figures \ref{fig:radial_ALMA_cases} and \ref{fig:radial_all_ALMA})}.

On average, we find consistent results for the mean \mbox{CO(2-1)/CO(1-0)} line ratio when we change the CO(2-1) data set used in the overlapping data set (HERA: $\langle R_{21}\rangle = 0.62\pm0.14$; ALMA: $\langle R_{21}\rangle = 0.56\pm0.11$). That is, there does not seem to be strong evidence that the overall amplitude scale of HERA is biased relative to ALMA or the new 30-m observations obtained with EMIR.

\begin{figure*}
    \centering
    \includegraphics[width = 0.9\textwidth]{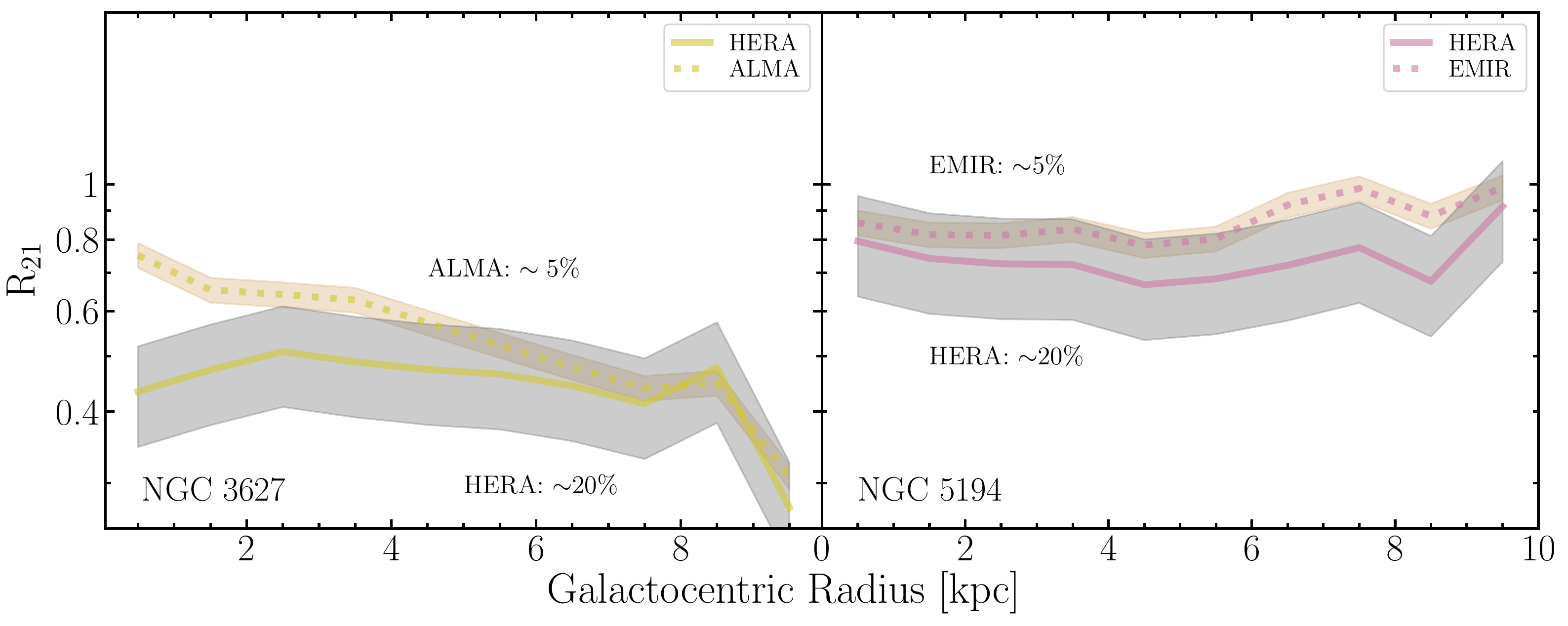}
    \caption{Side-by-side comparison when substituting HERA CO(2-1) with ALMA or EMIR data in the case of NGC~3627 and NGC~5194. The calibrational uncertainties for HERA are estimated to be around 20\% \citep{Leroy2009}, while for EMIR \citep{Carter2012} and ALMA \citep{Bonato2018} it is around 5\%. {We note that the HERA maps in particular of NGC 5194 are less reliable, as this constitutes a commission observation under difficult observing conditions. }The two cases show the strongest discrepancies when substituting the CO(2-1) data sets.}
    \label{fig:comp_Hera_EMIR_ALMA}
\end{figure*}

We do find relatively strong discrepancies in the maps for two galaxies: NGC~3627 and NGC~5194 { (see Figure \ref{fig:comp_Hera_EMIR_ALMA})}. Compared to the new EMIR map by J. den Brok in preparation, the HERACLES map of NGC~5194 is low by a factor of $0.89$. Meanwhile the NGC~3627 shows an offset of $0.73$ from the ALMA map on average, but also a qualitatively different radial structure. 

These were the two earliest galaxies observed with HERA. NGC~5194 was observed as part of commissioning \citep{Schuster2007} and NGC~3627 as part of a pilot program that explored the viability of HERACLES. As a result, they did not yet adopt the rotation, cross-scanning, and offset that became part of the later HERACLES observing strategy. Our recommendation is that the new EMIR and ALMA maps supersede the HERA data for these targets, and we have adopted this approach in this paper.

\begin{figure*}
    \centering
    \includegraphics[width = \textwidth]{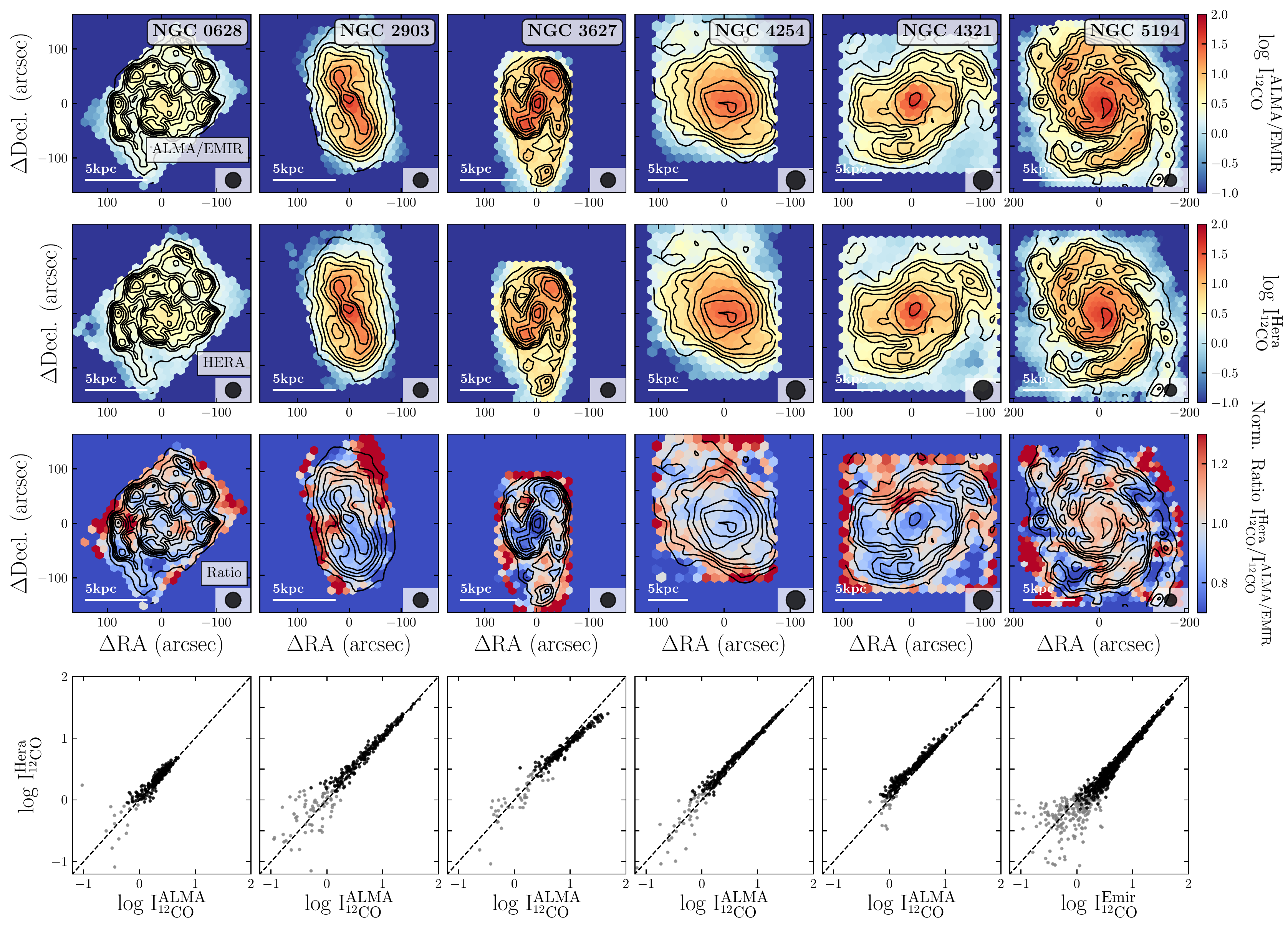}
    \caption{{ In--depth analysis and comparison of the the sources for which we have two CO(2-1) data sets.} The top row shows the {latest} available CO(2-1) data sets (from ALMA or EMIR in the case of NGC 5194). The second row shows the HERACLES CO(2-1) 2D maps. The third row shows the relative spatial variation of the ratio of the two CO(2-1) observations. In the bottom row, the integrated brightness temperatures of individual sightlines are plotted against each other. The dotted line indicates the 1:1 relation. Black points indicate data points for which both CO(2-1) data are detected with S/N$>3$.}
    \label{fig:comp_indepth}
\end{figure*}

For the remaining galaxies with two maps, NGC 628, NGC 2903, NGC 4254, and NGC 4321, we find better agreement. { A more detailed comparison is shown in Figure \ref{fig:comp_indepth}, where the spatial variation in 2D of the ratio of the two different CO(2-1) observations is shown. Overall pairs of CO(2-1) maps mostly show similar morphologies. The global offset in calibration for NGC 3627 and NGC 5194 discussed above is striking. We also see some second-order variations in morphology between the maps, e.g., in the center of NGC 3627 and NGC 4321. Our best estimate is that these reflect pixel gain variations in HERA, which are inducing second-order local calibration uncertainties.} %Statement

\begin{table}
    \caption{Overview of adopted CO(2-1) single dish data sets that we use as complimentary to EMPIRE CO(1-0) for the CO line brightness temperature ratio.}
    \label{tab:summary_obs}
    \centering
    \begin{tabular}{c c c c} \hline \hline
         \textbf{Galaxy}& \textbf{HERA}$^{\rm a}$ & \textbf{ALMA}$^{\rm b}$ & \textbf{EMIR}$^{\rm c}$  \\ \hline
         NGC 0628 & \checkmark & \checkmark  &\\
         NGC 2903 & \checkmark & \checkmark &\\
         NGC 3184 & \checkmark & &\\
         NGC 3627 & \checkmark & \checkmark  &\\
         NGC 4254 & \checkmark & \checkmark  &\\
         NGC 4321 & \checkmark & \checkmark  &\\
         NGC 5055 & \checkmark & &\\
         NGC 5194 & \checkmark & & \checkmark \\
         NGC 6946 & \checkmark & & \\ \hline
    \end{tabular}\\
    \footnotesize
    {\raggedright 
    a) Part of HERACLES \citep{Leroy2009},\\
    b) Part of PHANGS-ALMA-survey (A.~K.\ Leroy et al., in prep.),\\
    c) Part of M51 IRAM 30-m Large Program (J.~S.~den Brok et al., in prep.)
    \par}
\end{table}

\section{Stacked CO Line Measurements}

We apply a stacking technique to improve the S/N when measuring $R_{21}$ as a function of other quantities. This technique is summarized in Section \ref{sec:Physical} and described in more detail by \citet{Schruba2011}, \citet{JDonaire2017}, and \citet{Cormier2018}. In Figures\,\ref{fig:stack_co21} and \ref{fig:stack_co21}, we show  one application, the stacked \mbox{CO(1-0)} and \mbox{CO(2-1)} brightness temperature as a function of galactocentric radius in NGC~0628. This illustrates the procedure used to stack other galaxies and to stack by other quantities. It also highlights some of the uncertainty associated with the lowest brightness temperature bins, an issue raised in the main text.

Both figures shows that stacked by the local {\sc Hi} velocity produced high signal to noise, coherent spectra out to ${\sim} 10$~kpc. The grey region shows the integration area used to determine the integrated brightness temperature. This clearly corresponds to all real astrophysical signal out to 9~kpc, with the algorithm used to identify the line width doing a good job.

%The fit is used to determine the bounds over which we integrate the intensity. The range is set such that everything above 1\% of the peak is included. The shaded area consequently illustrates the range over which we integrate. 

Outside 9~kpc, we begin to see some breakdowns in the approach. In the tenth radial bin, which covers $r_{\rm gal} =9{-}10$~kpc, the CO(1-0) spectrum includes a second, fainter, \mbox{CO(1-0)} emission peak to the right of the main emission line.  This second peak appears displaced by approximately 100 km~s$^{-1}$ from the main peak. This could represent a noise spike, a problem with the {\sc Hi} velocity field, a problem with baseline subtraction, or real signal. No analogous feature appears in the CO(2-1) spectrum. In this case, we manually extended the integration range to cover the additional emission line, but the profile becomes uncertain in this bin. This uncertainty is higher than the statistical uncertainty. In the next panel, we see that by $10{-}11$~kpc, uncertainties in the baseline produce large ``ripples'' in the spectrum for both CO(1-0) and CO(2-1). Though formally the S/N of the data remain high (there is extended emission over a large velocity range), these results remain uncertain.

These sorts of systematic uncertainties tend to crop up in the outer parts of the binned data. These breakdowns in the stacking procedure and baseline uncertainties contribute to some of the scatter in profiles at low intensity but are not necessarily reflected in the statistical scatter.

\begin{table*}
    
    \caption{{ Signal to Noise Ratio for the radial stack bins (see Figure \ref{fig:stack_co10} for an example showing the individual radial bins for one galaxy). The median of a given bin over the nine galaxies is taken, as well as the 5 and 95 percentiles. Each radial bin has a width of 1 kpc.}}
    \label{tab:SNR_stacks}
    \centering
    \begin{tabular}{c | c c c c c c c c c c c c } \hline \hline
         Center radial bin [kpc] & 0.5 & 1.5 & 2.5 & 3.5 & 4.5 & 5.5 & 6.5 & 7.5 & 8.5 & 9.5 & 10.5 & 11.5\\ \hline
         Median SNR & 142 & 110 & 103 &  96 & 89 & 67 & 54 & 40 & 28 & 9.5&4.1&10.1\\
         5 perc. & 54 & 60 & 62 & 59 & 63 & 49 & 23 & 20 & 11 & 6.5 & 0.8 & 0.8\\
         95 perc. & 323 & 215 & 157 &148 & 97 & 94 & 99 &  78 & 59 & 47 & 36 & 28\\ \hline
    \end{tabular}
    
\end{table*}
%otherwise, the integrated brightness temperature would be too low (and consequently the CO line ratio too high). 

\begin{figure*}
    \centering
    \includegraphics[width = \textwidth]{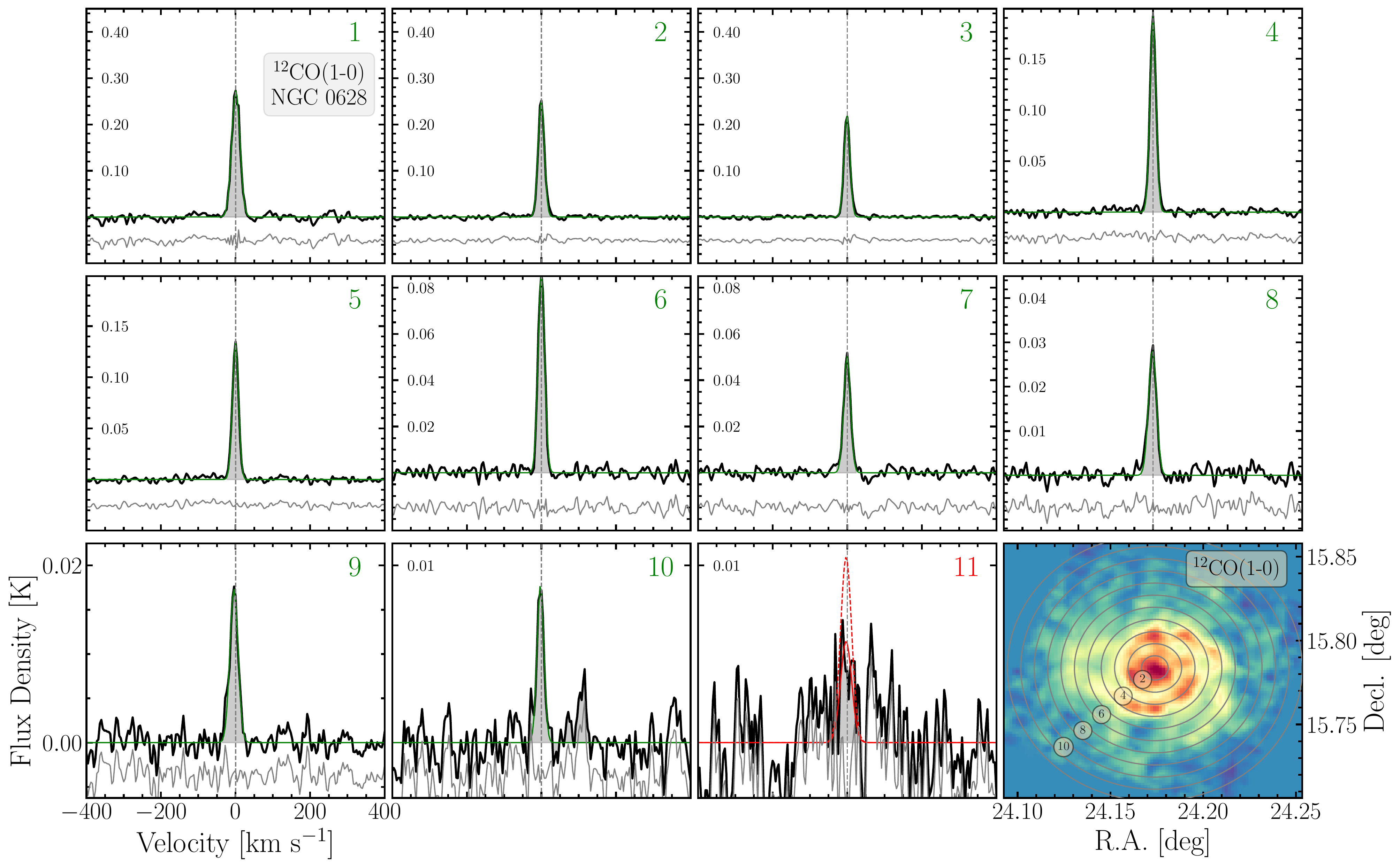}
    \caption{\textbf{Illustration of spectral stacking.} Spectra of CO(1--0) brightness temperature stacked as function of galactocentric radius in NGC 0628. The outer radius, in kpc, of each kpc-wide ring is given in the top right corner, with the color indicating whether the line peak has S/N $>5$ (green) or not (red). Shaded area shows the region over which we integrate to determine the line flux.
    The light gray dotted line indicates $v=0$\,{km\,s$^{-1}$} position. The bottom right panel illustrated the rings used for the stack.}
    \label{fig:stack_co10}
\end{figure*}
\begin{figure*}
    \centering
    \includegraphics[width = \textwidth]{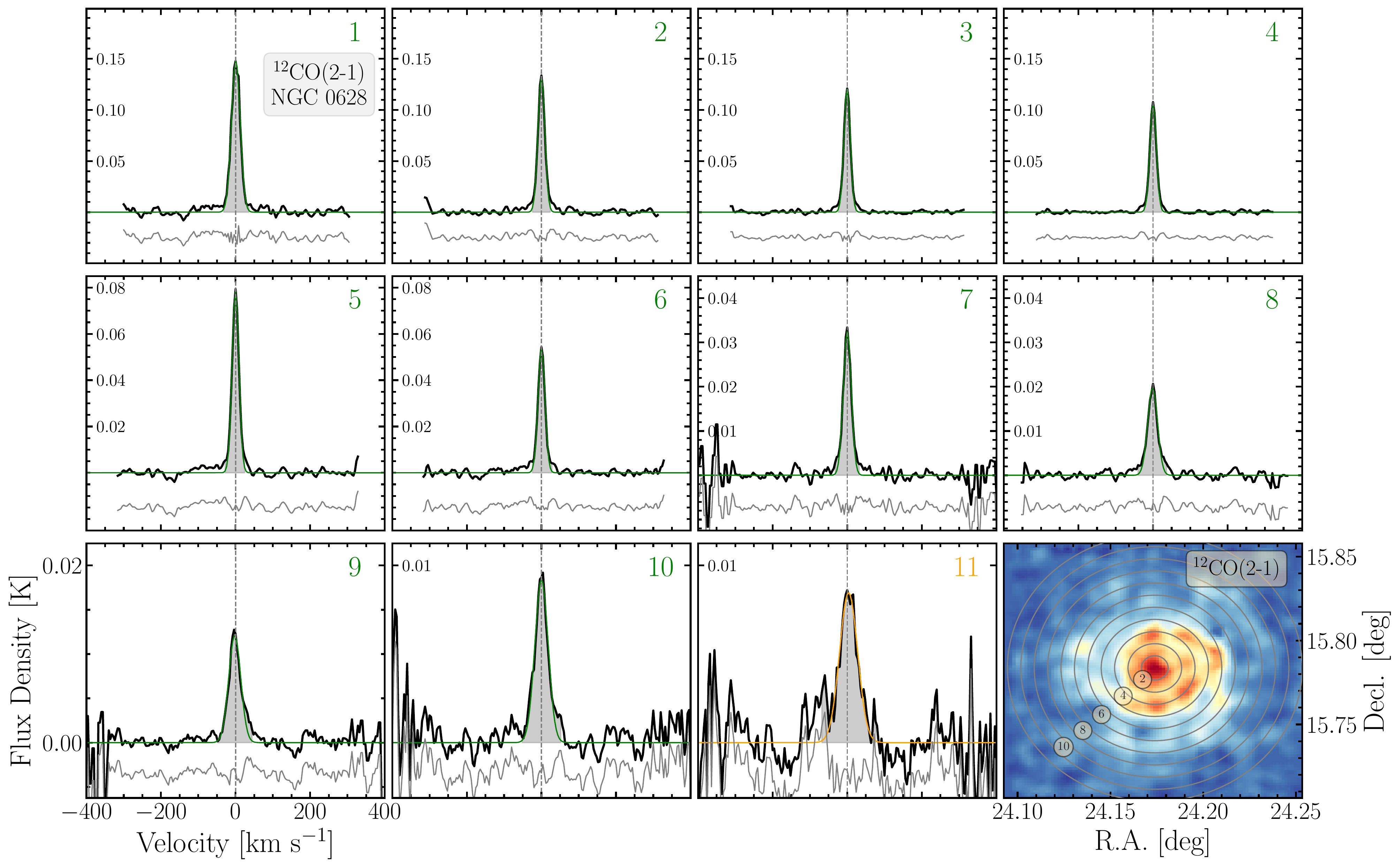}
    \caption{\textbf{Illustration of spectral stacking --- continued.} As Figure\,\ref{fig:stack_co10} but now showing the CO(2-1) line stacked in bins of galactocentric radius. }
    \label{fig:stack_co21}
\end{figure*}

\section{Azimuthal \texorpdfstring{$R_{21}$}{Lg}  Variation in NGC 5194}
\label{app:comp_NRO_EMIR}
\begin{figure*}
    \centering
    \includegraphics[width = \textwidth]{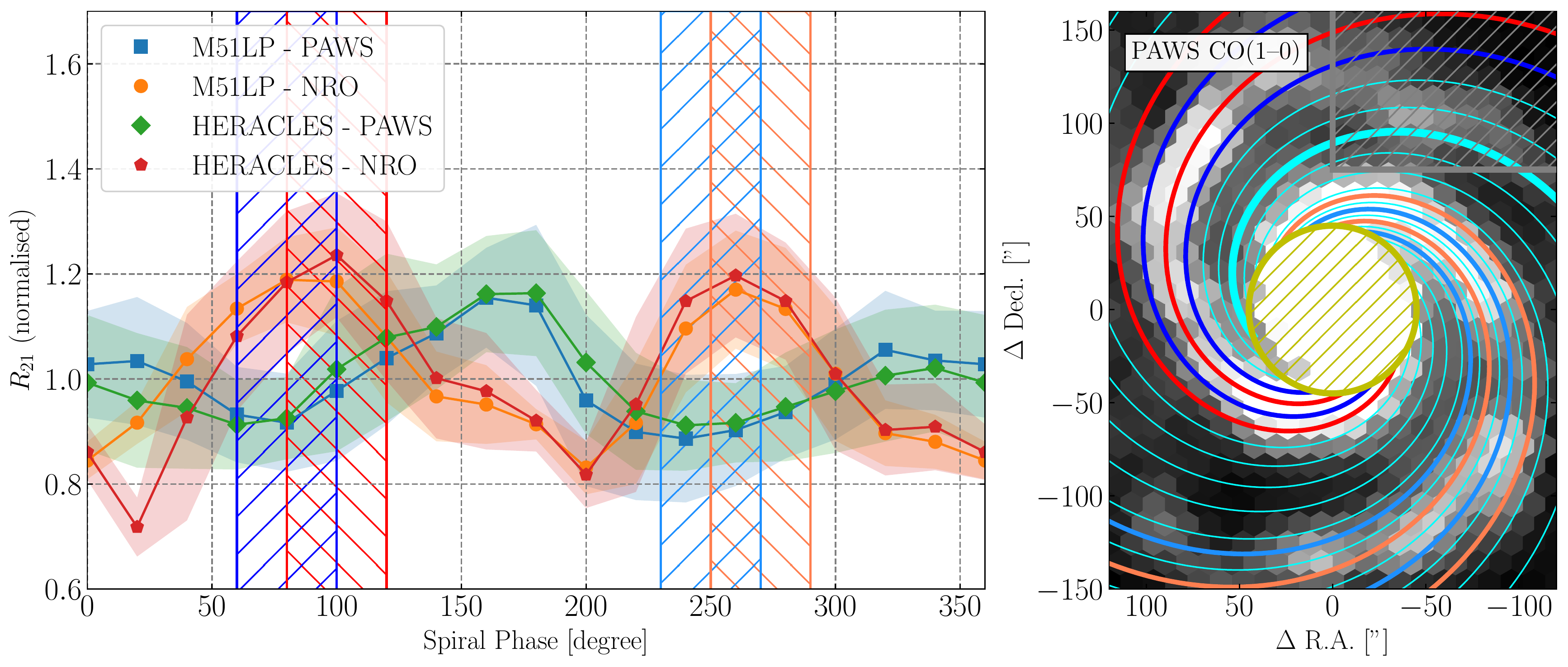}
    \caption{{\bf Normalized CO Line Ratio as function of spiral phase.} We follow the spiral phase analysis of the integrated CO line brightness temperature ratio, as described in \citet{Koda2012}, to study the azimuthal variation. (\textit{Left Panel}) Here we show all four permutations of the two different CO data sets for both transitions. Each line ratio is normalized by the corresponding mean. The shaded area shows the standard deviation of the line ratio binned by spiral phase. The blue,   hashed band shows the molecular arm and the red,  hashed band indicates the trailing star-forming arm, as provided by \citet{Koda2012}. It is evident, that upon changing the CO(1-0) data set (PAWS vs NRO), the trend of the line ratio changes, while changing the CO(2-1) data sets (M51 LP vs HERACLES), has no effect. (\textit{Right Panel}) The CO(1-0) brightness temperature map showing the PAWS data. The spiral phases are plotted in increments of 20$^\circ$. Points within the central  { hashed} inner 45$\arcsec$ area as well as the  {hashed} region in the north--west are excluded from the spiral phase bins. We applied a ${\rm S/N}=10$ threshold.}
    \label{fig:arm_interarm}
\end{figure*}

\begin{figure*}
    \centering
    \includegraphics[width = 0.95\textwidth]{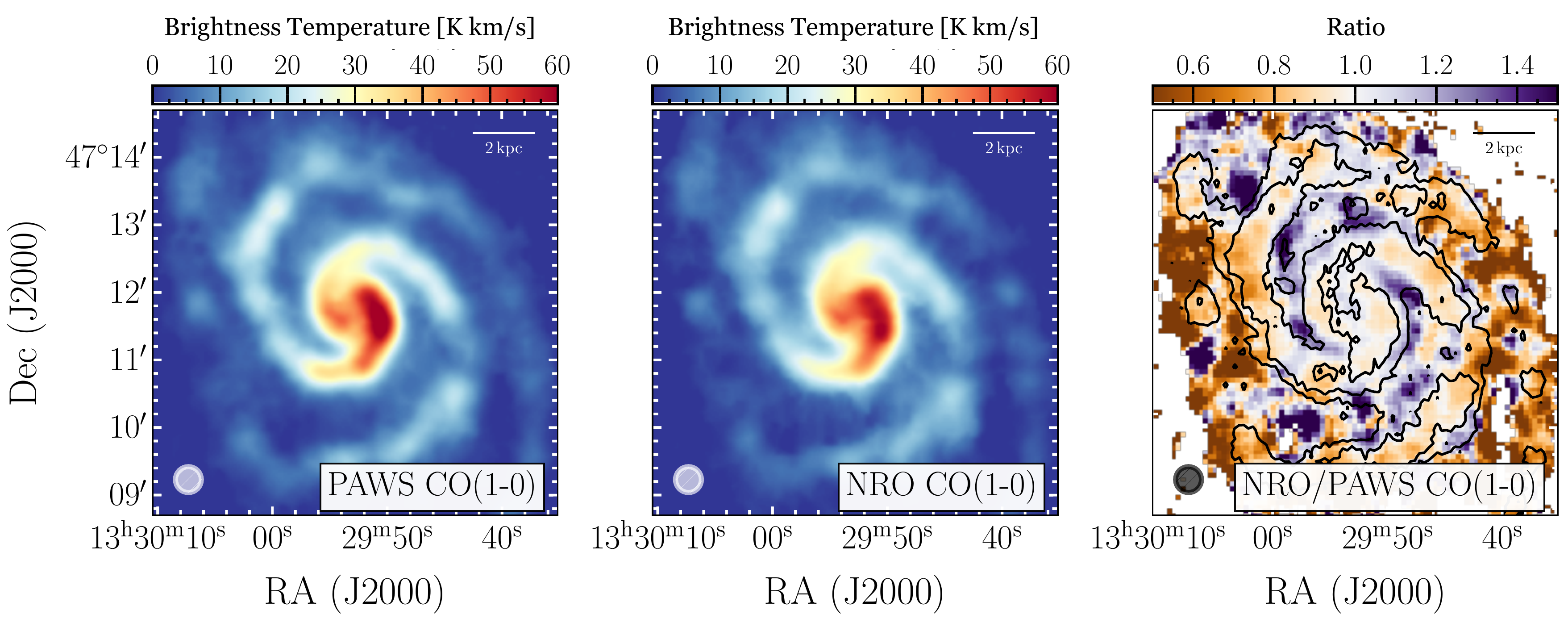}
    \caption{ {\bf CO(1-0) line brightness temperature maps.} Side-by-side comparison of (\textit{left}) the PAWS CO(1-0) line brightness temperature map \citep{Pety2013} and (\textit{right}) the NRO 45-m map \citep[from][]{Koda2011}. Both show similar line brightness temperature values, but the NRO map shows a weaker contrast between arm and interarm region (especially at the position of the outer arms), which we believe causes the discrepancy in the azimuthal line brightness temperature trend discussed in Section \ref{variations} and Appendix \ref{app:comp_NRO_EMIR}. 
    Both maps were created by integrating the full cube over the same masked velocity range. {(\textit{right}) The ratio of the CO(1-0) Line of the NRO map and the PAWS map. The contours are drawn at ${\rm S/N} = 20,40,100$ of the PAWS CO(1-0) map. The discrepancy is strong in the interarm region, where the ratio is clearly ${>}1$.}}
    \label{fig:two_maps}
\end{figure*}

{ In this study, we investigated the spatial variation of the CO line brightness temperature ratio across the full discs of the galaxies. The source NGC 5194 is not only unique in the sense that is shows strong differences between the arm and interarm regions, but the trend we find disagrees with a previous study by \cite{Koda2012}. We find larger $R_{21}$ values (${\sim}0.9{-}1$) in the interarm region compared to the molecular arm region (${\sim} 0.7{-}0.8$). This trend stands in contrast to the one reported in \cite{Koda2012}. Here, we investigate the origin of the discrepancy. As we used different observations than the previous study, it is essential to determine, which data set causes the discrepancy. In particular, we used CO(1-0) observations from the PAWS survey \citep{Pety2013} and CO(2-1) observations from the M51 Large Program (J.~S.\ den Brok et al., in prep.). \cite{Koda2012} on the other hand used CO(1-0) observations from NRO \citep{Koda2011} and CO(2-1) observations from HERACLES \citep{Leroy2009}.

To properly analyze the azimuthal variation, we followed the same procedure as described in \cite{Koda2012} to determine the variation of $R_{21}$ as a function of the spiral phase. Figure \ref{fig:arm_interarm} shows (left panel) the result, where we looked at the line ratio using all possible permutations of the CO(1-0) (PAWS \& NRO) with CO(2-1) (M51 LP \& HERACLES) data sets. {All observations were convolved to a common resolution of 24$\arcsec$.} The red and blue hashed regions indicate the location of the spiral arm, as given in \cite{Koda2012} (see also right panel of Figure \ref{fig:arm_interarm}; molecular arm: blue, $\psi = 60^\circ-100^\circ$ and $230^\circ-270^\circ$; leading edge: red, $\psi = 80^\circ-120^\circ$ and $250^\circ-290^\circ$). Note that the \mbox{$y$-axis} shows the normalized line ratio. It is evident from this plot, that the discrepancy is caused by the use of a different CO(1-0) data set. Substituting the \mbox{CO(2-1)} HERACLES data with the M51 LP observations does not change the azimuthal trend at all. Figure \ref{fig:two_maps} shows the two different \mbox{CO(1-0)} maps side by side. The left panel shows the PAWS brightness temperature map, while the map on the right illustrates the NRO map. {The NRO map has a native resolution of 19.7\arcsec. For the comparison, we convolved it to the resolution of the PAWS CO(1-0) map at 24\arcsec. We integrated both the NRO and PAWS cube over the same velocity range.}
Already from visual inspection, it is evident that the contrast between arm and interarm is higher in the PAWS than in the NRO map (especially for the position of the outer arms). 
{{The discrepancy can be caused by inproper error beam handling, different or unstable $T_{\rm sys}$, variable S/N, or scanning artifacts. To investigate the exact cause for the discrepancy is beyond the scope of this project.}}
%In J.~S.\ den Brok et al. (in prep) we study the precise difference in more detail and we will be able to give a more in-depth analysis.  
%A more in-depth analysis is deferred to J.~S.~den Brok et al. (in prep.) which will compare multiple transitions from various CO isotopologues.
}

% Don't change these lines
\bsp	% typesetting comment
\label{lastpage}
\end{document}

% End of mnras_template.tex